\documentclass[aps,reprint,nofootinbib]{revtex4-1}
\usepackage{graphicx,amsfonts,amsmath,amssymb}
\usepackage{bm}
\usepackage{url}
\makeatletter
\g@addto@macro{\UrlBreaks}{\UrlOrds}
\makeatother

\begin{document}
\title{How to (Un-) Quantum Mechanics}
\date{\today}
\author{C. Baumgarten}
\email{christian-baumgarten@gmx.net}
\noaffiliation

\def\begeq{\begin{equation}}
\def\endeq{\end{equation}}
\def\bquo{\begin{quotation}}
\def\equo{\end{quotation}}
\def\begary{\begeq\begin{array}}
\def\endary{\end{array}\endeq}
\def\bmtx{\left(\begin{array}}
\def\emtx{\end{array}\right)}
\newcommand{\myarray}[1]{\begin{equation}\begin{split}#1\end{split}\end{equation}}
\def\3{\ss}
\def\d{\partial}
\def\h{\eta}
\def\w{\omega}
\def\W{\Omega}
\def\s{\sigma}
\def\eps{\varepsilon}
\def\e{\epsilon}
\def\a{\alpha}
\def\b{\beta}
\def\g{\gamma}
\def\y{\gamma}
\def\d{\partial}
\def\S{{\Sigma}}

\begin{abstract}
When compared to quantum mechanics, classical mechanics is often depicted 
in a specific metaphysical flavor: spatio-temporal realism or a 
Newtonian ``background'' is presented as an intrinsic fundamental classical 
presumption.
However, the Hamiltonian formulation of classical analytical mechanics
is based on abstract {\it generalized} coordinates and momenta: It is a 
mathematical rather than a meta-physical framework.

If the metaphysical assumptions ascribed to classical mechanics are dropped, 
then there exists a presentation in which little of the purported difference 
between quantum and classical mechanics remains. This presentation allows to 
derive the mathematics of relativistic quantum mechanics on the basis of an
abstract Hamiltonian phase space. It is argued that a spatio-temporal
description is not a condition {\it for} but a consequence
{\it of} objectivity. It requires no postulates.

This is achieved by evading spatial notions in the beginning and by assuming
nothing but time translation invariance.
\end{abstract}

\keywords{Quantum Mechanics, Hamiltonian mechanics, Lorentz transformation, Electrodynamics, Dirac equation}
\maketitle

\section{Introduction}

Uncountable articles and books have been written on the interpretation of 
quantum theory and most share a number of assertions, briefly summarized as follows:
Firstly, yes, quantum mechanics is by far the most precise and successful theory ever 
formulated and secondly, no, there is no general agreement on what it tells us 
about the world. Thirdly, it is usually asserted, that classical mechanics (CM) 
is intuitive and clear while quantum mechanics (QM) is counter-intuitive, weird 
and somehow raises questions like ``is the moon there when nobody looks?''~\cite{Mermin85},
questions that no layman should expect to be raised by physicists. But the common
sense in the physics curriculum is that the success of QM provides evidence
that ``Nature is absurd''~\cite{FeynmanQED}. This bewildering conclusion is
repeated over and over in the literature.

Even if there is no agreement concerning the interpretation of QM, there seems
to be consensus that QM and CM are very different. We doubt this narrative.
As we shall argue, most of the mathematical formalism of QM can be obtained from 
classical notions in a straightforward manner, without new axioms or postulates.
Our view contrasts with the {\it standard presentation of QM} (SPQM), which 
stresses the profound mathematical differences between CM and QM. But
closer inspection reveals that there are few. 

This article is dedicated to those physicists who share the author's impression,
that both, CM as well as QM, are just {\it presented} in a way that suggests
deep and profound mathematical differences and who are flabbergasted when they
discover that the real difference between CM and QM is not {\it mathematical}
but roots mostly in the prevailing narrative.

\subsection{What is Quantum in Quantum Mechanics?}

A typical assertion of the SPQM is the following:
``In fact, the basic concepts such as observable, ensemble, state, or yes-no 
measurement, employed in the "usual" interpretation of quantum mechanics, 
are themselves not explainable by known pretheories.''~\cite{Ludwig}
There is a subtle but important difference between the claim that 
some fact or formalism ``is not explained'' (yet) and the claim that some 
fact or formalism ``is not explainable''. But since an ``explanation'' is 
the reduction of a new (element of a) theory to established elements,
there is more historical contingency in physics than usually 
acknowledged~\cite{Cushing}. 

Since ``pretheories'' are those theories that do not require anything
but ``classical'' concepts, it can be precisely defined what has to be
regarded as genuinely ``quantum'' -- namely those facts or formalism 
that can not be derived using the classical mathematical methods of physics. 
Hence if, in what follows, we claim that some equation or statement is not 
``quantum'', then it is not intended to say that it does not belong to the 
techniques usually employed when ``doing'' QM, but that it can indeed be 
derived on the basis of classical mathematical concepts. {\it Therefore}, 
according to this definition, it {\it can not be considered} ``quantum''.

Axioms are only required for things that can not be established otherwise,
and the reader may judge for himself, how much ``quantumness'' eventually 
remains that requires an ``axiomatic'' foundation~\footnote{
Historians of physics lately called into question whether the categories
of classical vs. quantum physics is adequate~\cite{Gooday2013}.
}.

\section{Setup}

\subsection{Is Classical Physics Based on Mass Points?}
\label{sec_SEQ}

The SPQM puts lots of emphasis on the claim that classical particles
are point-like and that the ``position'' of a ``particle'' 
is well-defined in classical mechanics. There are two issues
with this. Firstly, classical physicists were, contrary to the
usual textbook narrative, not at all committed to point 
particles~\cite{vlh_paper}, but secondly, {\it if} a classical particle
would have been defined to be point-like, then it is not the
position {\it of} a point-particle that is well defined; a point
particle {\it is} little more than a position. That is, not the 
position of the particle is well defined, but a particle is, 
apart from some ``attached'' properties of mass, charge and momentum, 
nothing but {\it a position in space}.
However, this narrative is historically untenable. It is a fairy tail.
Physicists of the ``classical times'' (i.e. before 1900), mostly 
avoided claims about the exact form of particles. The representation of
a ``particle'' by its momentum and position was typically understood as
an approximation and the ``position'' represents the center of mass~\cite{vlh_paper}.

For long the existence of atoms, i.e. the granular structure of matter,
was regarded as an unproven hypothesis.
Boltzmann, for example, argued in the introduction to his {\it lectures on 
gas theory}, published 1896, in support of the atomic hypothesis and leaves 
no doubt that he considers this hypothesis to be confirmed by many physical 
and chemical findings. But he made no claims, neither of metaphysical nor 
other nature, which would have suggested an {\it ontological} interpretation 
of material particles (atoms or molecules) by mathematical points.
Instead he wrote:
\bquo
Furthermore there is the age-old view,
that the material bodies do not fill out the space they occupy continuously 
in a mathematical sense, but that they consist of small discrete particles
which can, due to there small size, not be distinguished by our senses.
Philosophical considerations support this view.
~\footnote{
Original (German): ``Hierzu kommt nun die uralte Ansicht, dass die K\"orper
den von ihnen eingenommenen Raum nicht im mathematischen Sinne 
continuirlich erf\"ullen, sondern aus discreten, wegen ihrer
Kleinheit einzeln für die Sinne vollkommen unwahrnehmbaren K\"orperchen,
den Molek\"ulen, bestehen. F\"ur diese Ansicht sprechen philosophische Gr\"unde.''
~\cite{Boltzmann}.
}
\equo
There is no doubt that Boltzmann is a proponent of classical physics.
Nonetheless he does not speak of particles as ``mathematical points''
but only of ``small bodies''.

He defended a corpuscular view with a surprising statement concerning
the continuous filling of space by these ``small bodies'':
\bquo
Furthermore one would need to presume the partial differential equation 
governing the behavior of such hypothetical continuum as foundational~\footnote{Original (German): 
``Ferner muss man bei Annahme eines Continuums die partiellen Differentialgleichungen 
f\"ur das Verhalten desselben als das urspr\"unglich gegebene 
auffassen.''~\cite{Boltzmann}.}.
\equo
As we shall argue, Boltzmann's considerations anticipate the need for a
Schr\"odinger type equation. Nonetheless they seem to ignore the fact that the
problem of continuity does not change with scale: The question whether
a macroscopic body fills out space continuously (or not) does not substantially
differ from the question whether some (sub-) atomic ``particle'' fills out
a tiny amount of space continuously (or not), {\it unless} the (sub-) atomic
particle is envisioned to be a mathematical point.

Most classical physicists interpreted the mathematical point, not as an 
ontological commitment but as an approximation to be used for reasons
of mathematical convenience\footnote{
  Several examples for classical (19th century) textbooks on analytical
  mechanics supporting this account are given in Ref.~\cite{vlh_paper}.
Most authors of the 19th century refused to take the point particle literally
due to the many ``nonphysical features.
For instance the well-known problem of infinite self-energy
of point charges: A charged sphere of radius $r$ has an electric 
self-energy $E_0$ of approximately
\begeq
E_0={q^2\over 4\,\pi\,\eps_0\,r}
\endeq
so that in the limit $r\to 0$ the self-energy becomes infinite.}.

Nonetheless, in the context of quantum theory, one is frequently told
that~\footnote{We found this sentence in a book of Sean Carroll~\cite{Carroll},
  but this way of describing ``the situation'' can be found almost everywhere
  and expresses Born's ``statistical'' interpretation of the wave function,
  for which he received a Nobel prize.}:
\bquo
if we think an electron wave function is a diffuse cloud centered on the
nucleus, when we actually look at it we don't see such a cloud, we see a
point-like particle at some particular location.
\equo
It is highly questionable, that some physicist has ever ``seen''
a point particle. What actually {\it has} been observed are orbitals
as they pop out of Schr\"odinger's atomic theory\cite{Humphreys1999,Zuo1999,Stodolna2013,Villeneuve2017}.
But though it is an evidently absurd claim that point particles could
possibly be observed, orthodox scholars instead denied that orbitals
could possibly be observed~\cite{Scerri2000}. 
Remarkably, it has even been claimed that~\cite{Scerri2001}
\bquo
[...] if these claims [of observed orbitals] were to be sustained it would imply an outright
refutation of quantum mechanics.
\equo
This is a surprising conclusion since orbitals, one might naively think,
arise from solutions of the Schr\"odinger equation and are hence an
invention of QM. How can the observation of orbitals then be ``an outright
refutation of quantum mechanics''? The mentioned debate could have ended
with the reply of Zuo {\it et al} in which they admit that the ``correct'' term
would have been ``electron density''~\cite{Zuo2001}. But even this does not
fully satisfy the dogma, since quantum theorists insist that orbitals
``don't exist''~\cite{Labarca2010}. They don't exist because the SPQM
tells us that we ``see'' point particles whenever we ``look'' at them.
It remains unclear whether this statement contains more than the tautological
claim that the measurement of a position gives, as a result, a position.

According to the SPQM, the squared wave function $\psi^\star\psi$ describes
just the probability density to find ``the particle''. This barely makes sense
unless the ``real thing'' is but a point particle.
Nonetheless it has been pointed out by Sebens, that the interpretation
of $\psi^\star\psi$ as charge density is compatible with quantum chemistry
and QFT~\cite{Sebens}:
\bquo
[..] we examined some of the reasons why it is appealing to think of electron
charge as spread out in the way Schr\"odinger proposed. [...]
Although quantum chemists regularly treat wave functions as describing spread-out
distributions of charge, scholars working on the foundations of quantum mechanics
rarely explicitly include such charge densities in the ontological precise formulations
of quantum mechanics that they propose. [...] Here I have argued that their
fit with quantum chemistry is a point in their favor. When we move to quantum
field theory, I think the case for a spread-out electron charge density is
particularly strong as the theory can be viewed as describing
quantum super-positions of classical field states where electron charge is spread-out.
\equo
The SPQM suggests that classical physics is about point particles and that
this picture was (in some way) more plausible, reasonable and easier to
swallow than quantum mechanics. But the historical truth is that few 
physicists of the ``classical'' era were satisfied with the idea of point
particles. The narrative that there was a satisfying coherent physical
worldview prior to the scientific revolutions of the twentieth century,
is simply false. It is a lie. There never was a satisfying classical theory 
of matter.

There are, of course, reasons why the point particle notion is important
within the story of the SPQM.
Without this narrative it is difficult to make sense of Born's rule.
If the density $\psi^\star\psi$ has to be interpreted as the probability
to find a particle ``at some position'', then the particle must basically
be given by that position. Since how could it possibly be found
``at some position'' if it had a volume? But then, if the particle
is a mathematical point, how can we understand, that many textbooks suggest to 
think of wave functions as point particles which are ``smeared out'' over a
volume? Since one has to spread out the point particle in space only if
it was assumed to be point-like in the beginning.

Struggling with the interpretations of QM, Sean Carroll suggested that~\cite{Carroll}
\bquo
The way to break out of our classical intuition is to truly abandon the idea that
the electron has some particular location.
\equo
This statement is probably correct, but it seems that it implies mostly to
question the notion of point particles. But we disagree with the first part
of the statement: in fact the point particle never was a result of ``classical intuition''~\cite{vlh_paper}.
Carroll does not explicitly refuse point particles and prefers to instead 
speak of ``super-positions''. It remains unclear to us why should it 
be reasonable to regard a distribution as a ``superposition of point particles'' 
and not simply as one distributed object~\footnote{
Stephen N. Lyle was more courageous: ``The point particle
approximation has been extraordinarily successful. But [...] we might
understand physics better by knowing what can be done with
spatially extended particles.''\cite{Lyle}.}? Why do we speak of a ``spread
out'' point charge while classical analytical mechanics legitimizes the
``mathematical point'' by referring to the ``center of gravity'', i.e. as a
mathematical idealization of a matter distribution\cite{vlh_paper}? Isn't it
obvious that the latter makes perfect sense while the former is circular?
What is the sense of saying we ``find'' a particle at a certain position
and why would it be different from saying: We measure the particle's
(average) position: We measure a position and we obtain a position. 
In order to obtain a distribution, one might combine many position 
measurements. And in quantum mechanics as well as in classical mechanics, 
repeated measurements of the same continuous quantity rarely yield a single
infinitely precise number: repeated measurements usually provide
results that are (or can be) described by a probability distribution.
Of course, the SPQM claims to know, that the ``classical'' explanation for
such a distribution is given by the limited precision of measurement results
while it is claimed to be an intrinsic feature of nature according to 
quantum mechanics. But what difference does this really make? 

\subsection{Newton's mechanics includes no (causal) theory of motion}

It has been widely unnoticed, but Newton's mechanics suggests
a description of motion of material objects, but not a description
of motion as a causal process. Newton's theory of gravitation has
been criticized for incorporating action-at-a-distance and it has been
generally understood that a physical ``distortion'' in the form of
a physical (gravitational, electromagnetic) field, must be described
by partial differential equations in order to allow for a causal 
transmission through space. But why should the motion of matter not
be described as a causal process - just as the transmission of any 
other physical effect? The acausal nature of motion is emphasized
in Newtonian mechanics by the ``first axiom''. It has been suggested
to regard ``inertia'' as the cause for straight rectilinear motion.
But if ``inertia'' has to be regarded as the cause of straight
rectilinear motion, why then did Newton introduce straight
rectilinear motion by an axiom (or, actually, by two!)?

In his book on optics, Newton depicts his (personal) view of
particles~\cite{NewtonOpticks}:
\bquo
    [...], it seems probable to me, that God in the
    Beginning form'd Matter in solid, massy, hard,
    impenetrable, movable Particles, of such Sizes
    and Figures, and with such other Properties, and
    in such Proportion to Space, as most conduced to
    the End for which he form'd them; and that these
    primitive Particles being Solids, are incomparably 
    harder than any porous Bodies compounded of them;
    even so very hard, as never to wear or break in pieces;
    no ordinary Power being able to divide what God
    himself made one in the first Creation.
\equo
He imagined that a particle would have to be
described by a density distribution of infinite rigidity,
i.e. a frozen (time-less) ``object''. A Newtonian
particle at rest would be described by a two-valued density
$\rho(\vec x)$ (zero or one) and a moving particle by
\begeq
\rho(\vec x,t)=\rho(\vec x-\vec v\cdot t)
\endeq
such that the motion of the density $\rho$ is the solution,
in the sense of Boltzmann's quote above, of a partial
differential equation, namely of the wave-equation
\begeq
\left({\d^2\over\d t^2}-v^2\,\vec\nabla^2\right)\rho(\vec x,t)=0
\endeq
with the ``trivial'' dispersion relation $\w=v\,k$. But there are
some issues related to this. Firstly, a density is defined
as a positive definite quantity, while wave motion
is mostly associated with oscillations which can be positive and
negative. This problem could be overcome by defining the density
by the squared amplitude of a wave $\rho=\psi^2$ with
\begeq
\left({\d^2\over\d t^2}-v^2\,\vec\nabla^2\right)\psi(\vec x,t)=0
\endeq
But secondly, this kind of homogenuous wave equation is 
exclusively known in cases in which the phase-velocity $v$ is 
a property of the medium. It therefore requires an explanation,
if we wish to identify the phase-velocity with the velocity of 
a particle.

In Newtonian physics there is no physical or mathematical or
logical connection between a particle's velocity and wave motion.
Furthermore, if a rigid Newtonian particle is considered in it's 
rest frame, it is completely static and timeless, incompressible and
un-deformable. The infinite hardness suggested by Newton implies that
any mechanical force acting on the one side of a Newtonian particle
would have to be transmitted to the other side {\it instantaneously},
thus implying an infinite speed of sound within the volume occupied by
the particle. The particle would not only occupy some space, the
suggested properties would effectively {\it negate} the physical existence
of the occupied space: Cause and effect would have to be transmitted
instantaneously within it as if the occupied space would not exist.

Hence it is difficult to imagine how one could formulate a sound
causal theory of motion based on Newton's concept of absolutely
hard and eternal diamond-like particles. Apparently one has to
admit that local causal transmission of physical effects always
requires partial differential equations and if those are linear,
then motion is {\it always} related to wave-like transport.
This is true for fundamental fields like the electromagnetic or
gravitational fields, for pressure waves in liquids and
sound waves in air: Local distortions travel as waves through
space. Only the motion of material bodies is attributed to the
opaque and problematic notion of ``inertia''.

Hence it seems that classical Newtonian mechanics, the flagship
of the causal worldview, offers neither a sound causal theory for
the motion of matter nor for the ``occupation'' of space by
material particles. As we shall see next, Schr\"odinger's Equation
offers a solution to both problems.

\subsection{Schr\"odinger's Equation as a Classical Theory of Motion}
\label{sec_motion}

\bquo
The problem is not the interpretation of quantum mechanics. That’s getting
things just backwards. The problem is the interpretation of classical
mechanics.\\
\mbox{}\hfill {\it -- Sidney Coleman}~\cite{Coleman}.
\equo

A partial differential equations which allows to describe the motion of
a continuous matter distribution is well-known as Schr\"odinger's equation.
This can be derived on two pages~\cite{seq_paper}, but we shall discuss
this in some more depth here. We generalize Newton's hard-core particles
and presume that a ``particle'' can be described by a positive definite
normalizable density function $\rho(\vec x,t)\ge 0$
\begeq
\int\,\rho(t,\vec x)\,d^3x=1\,.
\endeq
However, we drop all other assumptions that Newton thought were necessary.

In order to implement the condition of positive (semi-) definiteness, i.e. $\rho\ge 0$, 
one choses a positive semi-definite expression for $\rho$. One option
is to introduce auxiliary functions $\psi(t,\vec x)$ such that ($n\in\mathbb{N},\,n\ge 1$):
\begeq
\rho(t,\vec x)=\sum_i\,\psi_i^{2\,n}(t,\vec x)\,,
\endeq
or, using complex numbers for later convenience, by
\begeq
\rho(t,\vec x)=\psi^\star\psi\,.
\label{eq_BornRule}
\endeq
As Boltzmann remarked, if matter is distributed, then it is required
to derive the distribution from a partial differential equation (PDE).
However, it is not required that the PDE involves the density directly. 
The requirement that the density must be positive (semi-) definite, 
concerns the {\it possible solutions} of the PDE. 
Since $\rho$ must be positive semidefinite, it can be represented by a
quadratic expression.

The density must be normalizable, which implies that the auxiliary function
$\psi$ must be $L^2$-integrable. Hence it has a $L^2$-norm and is,
by construction, a member (vector) in some Hilbert space. Hence Hilbert space,
quite obviously, requires no ``axiomatic'' foundation. 

This means that the use of Hilbert spaces in physics is {\it as such} not
``quantum'': Any positive semi-definite and normalizable density function
can be represented by the squared modulus of a member of a Hilbert space.
It is simply a math fact that the Fourier transform of $\psi(t,\vec x)$ exists
and is well-behaved:
\begeq
\psi(t,\vec x)\propto\int\,\tilde\psi(\w,\vec k)\,\exp{[-i\,(\w\,t-\vec k\cdot\vec x)]}\,d^3k\,d\w\,.
\endeq
The Fourier transform is a bijective unitary change of coordinate (``address'') space:
Changes of the density in Fourier space have consequences in physical space and vice
versa. In the verbiage of the SPQM, the ``particle'' is represented by a
``wave packet''. But this tells us nothing substantial beyond a mere math fact. 
Nonetheless, the decomposition into an ensemble of partial waves enables us
to proceed further, since waves (as we argued) allow for a causal description
of the transport of matter.

It has been proven that few very basic conditions, mostly causality
(for instance that cause preceeds effect), imply the existence of
a dispersion relation~\cite{Toll,Nussenzweig1960,Nussenzweig1972,Dethe2019}:
if the motion of matter is a causal process in space and time, then a dispersion
relation must exist. Furthermore, we may use another math fact about ``wave packets'',
namely that their velocity, the so-called ``group velocity'' $\vec v_{gr}$,
is (in linear approximation) given by~\cite{Hamilton,Brillouin,Whitham,Pain}:
\begeq
\vec v_{gr}={d\over dt}\langle \vec x\rangle=\vec\nabla_{\vec k}\,\w(\vec k)\,.
\endeq
As well-known from classical Hamiltonian mechanics, the velocity
of some particle is the momentum gradient of the Hamiltonian function, which is 
- in case of a free particle - a function of momentum only, i.e. 
${\cal H}(\vec p)$:
\begeq
\dot{\vec x}=\vec\nabla_{\vec p}\,{\cal H}(\vec p)\,.
\endeq
The equality of these velocities is then little more than
a condition of consistency:
\begeq
\vec\nabla_{\vec p}\,{\cal H}(\vec p)=\vec\nabla_{\vec k}\,\w(\vec k)\,.
\label{eq_wpd}
\endeq
We are hence lead to the math fact that ${\cal H}$ and $\w$ as well as their
arguments may only differ by a scaling constant and an additive function
of coordinates, i.e. a function independent of $\vec k$ or $\vec p$.
Let these additive functions be $\phi(\vec x,t)$ and $\vec A(\vec x,t)$,
respectively. They depend exclusively on the canonical coordinates
(i.e. position in space) and time:
\myarray{
\vec k&\propto\vec p+\varepsilon\,\vec A\\
w(\vec k)&\propto{\cal H}(\vec p)+\varepsilon\,\phi\\
}
with some constant factor $\varepsilon$. Without any physical assumption 
(beyond the classical ones: continuity, finiteness, causality), by pure
mathematical means, one arrives at a point where the de Broglie 
relations ${\cal E}\propto\w$ and $\vec p\propto\vec k$ pop out.

As shown in Ref.~\cite{seq_paper}, the Schr\"odinger equation is obtained 
from here in few steps: If the proportionality constant is denoted by
$\hbar$, then
\begeq
\psi(t,\vec x)\propto\int\,\tilde\psi({\cal E},\vec p)\,\exp{[-i/\hbar\,({\cal E}\,t-\vec p\cdot\vec x)]}\,d^3k\,d{\cal E}\,,
\endeq
so that the ``canonical quantization'' conditions ${\cal E}\to\,i\,\hbar\,\d_t$ 
and $p_j\to\,-i\,\hbar\,\d_j$ directly follow. By the use of the Newtonian 
energy-momentum-relation (EMR) ${\cal E}={\vec p^2\over 2\,m}$ one obtains 
Schr\"odinger's equation for a free particle~\cite{seq_paper}.
Even without any reference to Newton's EMR we can guess that ${\cal H}$ must
- due to isotropy of space - be an even function in $\vec p$ and hence
the second order is most significant at small momentum values.

It is true that the mass is, in this derivation, merely a particle-dependent
parameter and does not refer to specific ``volume elements'' of the
distribution. Hence this equation does not align with a Newtonian space-time
picture, as Heisenberg rightly noted.
But the success of Schr\"odinger's wave mechanics suggests that
Schr\"odinger's equation is -- at least a first (non-relativistic)
approximation -- of the partial differential equation that
Boltzmann considered necessary and foundational for embedding extended
particles continuously into space. It describes the general
properties of motion of a conserved spatial density distributions under the
assumption that this motion is a {\it causal process}. One can derive it
from genuine classical physical principles like the mentioned conditions
of continuity, conservation and causality\footnote{See also Ref.~\cite{GossonHiley}.}.
If there is magic in it, then it is the magic of (Fourier's) mathematics.
It has been argued elsewhere that it is not classical mechanics which is
committed to an ontological notion of the point particle~\cite{vlh_paper}.
It is the SPQM which sticks to this alleged ``classical'' notion.

E.T. Jaynes criticised that\cite{Jaynes6}
\bquo
Because of their empirical origins, QM and QED are not physical theories at 
all. In contrast, Newtonian celestial mechanics, Relativity, and Mendelian 
genetics are physical theories, because their mathematics was developed by 
reasoning out the consequences of clearly stated physical principles which 
constrained the possibilities. To this day we have no constraining principle 
from which one can deduce the mathematics of QM and QED; in every new 
situation we must appeal once again to empirical evidence to tell us how we
must choose our mathematics in order to get the right answers.
In other words, the mathematical system of present quantum theory is, like 
that of epicycles, unconstrained by any physical principles. Those who have 
not perceived this have pointed to its empirical success to justify a claim 
that all phenomena must be described in terms of Hilbert spaces, energy levels, etc. 
This claim (and the gratuitous addition that it must be interpreted physically 
in a particular manner) have captured the minds of physicists for over sixty 
years. And for those same sixty years, all efforts to get at the nonlinear
'chromosomes and DNA' underlying that linear mathematics have been deprecated 
and opposed by those practical men who, being concerned only with phenomenology, 
find in the present formalism all they need.
\equo
As we have shown with the above derivation, no new but only old and well-known
principles are required, enriched by some 19th and 20th century's mathematical
insights. What is indeed required, is the classical continuity assumption,
which is, at least in ``spirit'', the exact opposite of typical assertions
of the SPQM which depicts quantum theory as a theory of discontinuities.

But discontinuity and discreteness have not been introduced by Max Planck,
they were always there. Newton's postulate, quoted above, is a postulate
of ``quantization'', namely a follow-up of the ancient idea of indivisible
atoms: Newton's suggests ``quantization'' of matter in the form of particles,
while Schr\"odinger's equation does (at first sight) the exact opposite:
It introduces a continuous distribution.

As we have shown above, the letter $h$ designates, from a logical point of
view, merely a scaling factor for the conversion of units rather than a
``quantum of action''. In the words of Zeh~\cite{Zeh2009}:
\bquo
In this formulation of quantum theory by means of wave functions, Planck's
constant is {\it not} used primarily to define discrete quantities ("quanta"),
but rather as a scaling parameter, required to replace canonical momenta and
energies by wave lengths and frequencies, respectively, -- just as time is
replaced by length by means of the velocity of light in the theory of
relativity.
\equo
Hermann Weyl explained as early as 1930 that the significance of the two
constants $\hbar$ and $c$ is, from a logical viewpoint, equivalent -- they
are both scaling constants~\cite{Weyl1930}:
\bquo
The constants $c$ and $h$, the velocity of light and the quantum of action,
have caused some trouble. The insight into the significance of these
constants, obtained by the theory of relativity on the one hand and quantum
theory on the other, is most forcibly expressed by the fact that they do not
occur in the laws of Nature in a thoroughly systematic development of these
theories.
\equo
This means that both constants are ``theoretically'' non-entities: beyond
the conversion of units there is no theoretical legitimization for
them to exist. According to John P. Ralston~\cite{RalstonHBar}:
\bquo
Planck’s constant was introduced as a fundamental unit in
the early history of quantum mechanics. We find a modern approach
where Planck’s constant is absent: it is unobservable except as a
constant of human convention. 
\equo
Planck's alleged ``discovery'' of quantized radiation was, in his own
words, a consequence of a mathematical method (or ``trick'') to
facilitate probablity calculus\cite{PlanckBook}:
\bquo
Um nun die Beziehung (9) auf den vorliegenden Fall anzuwenden,
dachte ich mir ein Gebilde, bestehend aus einer sehr gro\3en Anzahl N
von gleichartigen Oszillatoren, und suchte die Wahrscheinlichkeit zu
berechnen, da\3 dies Gebilde die vorgegebene Energie $U_N$ besitzt.
Da nun eine Wahrscheinlichkeitsgr\"o\3e nur durch Abz\"ahlung gefunden
werden kann, so war es vor allem notwendig, die Energie $U_N$ als eine
Summe von diskreten, einander gleichen Elementen $\eps$ anzusehen,
deren Anzahl durch die ebenfalls sehr gro\3e Zahl $P$ bezeichnet sein
m\"oge~\footnote{Translation by the author:
  ``In order to apply relation (9) to the problem at hand, I imagined a
  huge ensemble of $N$ identical oscillators and sought a way to compute
  the probability that the ensemble had the total energy $U_N$. Since
  a probability can only be found by enumeration it was necessary to
  interpret $U_N$ as a sum of discrete elements $\eps$, the number of
  which to be designated by $P$.''
}.
\equo
In other words: it was by no means intended or expected to be the
discovery of a new or fundamental physical principle, as it is widely
depicted today. And the notion of discontinuity was neither Planck's
personal paradigm nor his guiding principle: it was a kind of collateral
damage of the use of calculus, unintended {\it as such} and widely
ignored by Planck's contemporaries.

The discreteness of atomic transition frequencies is not due
to ``quantized action'', it emerges -- quite similar to discrete classical
electromagnetic cavity modes -- from the boundary conditions that the solutions
of some wave equation, here Schr\"odinger's, have to fullfill. Unbound ``states'' 
have continuous frequency spectra. The revolutionary content of Planck's hypothesis
is not ``discreteness'', but the equivalence of energy and frequency at the
fundamental level and hence the direct connection of mechanics with the
requirements of causality. It can also be understood as a replacement of
Newtonian by Hamiltonian notions~\cite{vlh_paper}.
In other words~\cite{QFOQS2014}:
\bquo
Main message: A photon is a pulse of the classical electromagnetic field.
Discreteness is an illusion produced by detectors.
\equo
Despite the fact that the SPQM boldly denies the principle of continuity,
it makes use of it whenever needed. And it is frequently needed, since
strictly speaking, if the assumption of continuity is dropped, then the
fundamental technique of physical theorizing, namely to solve (partial)
differential equations, looses it's foundation and meaning. And {\it some}
continuity assumption is even required just to solve Schr\"odinger's
equation - for instance in the typical textbook example of the step 
potential~\footnote{See for instance Ref.~\cite{Branson79}.
There is no doubt that potential steps are absent in nature, but they are
often used as textbook examples due to their mathematical simplicity.}.

Hence the insight that it is not discontinuity but continuity which allows
for a straight derivation of Schr\"odinger's equation, might be a first step
towards the unscrambling of the quantum omelette~\cite{Jaynes4}:
\bquo
Our present QM formalism is a peculiar mixture describing in part laws of Nature 
and in part incomplete human information about Nature-all scrambled up together 
by Bohr into an omelette that nobody has seen how to unscramble. Yet we think 
that the unscrambling is a prerequisite for any further advance in basic physical 
theory [...].
\equo
But it requires a significant change of perspective, away from metaphysical
prejudices to a sober interpretation of the (surprizingly successful)
mathematical formalism.

\subsection{Natura non facit saltus: Continuity and Signal Analysis}

David Tong expressed this view as follows~\cite{Tong}:
\bquo
The early attempts to understand
quantum theory, most notably by Danish
physicist Niels Bohr, placed discreteness
at its heart. But Bohr’s was not the final word.
Erwin Schr\"odinger developed an alternative approach
to quantum theory based on the idea of waves in 1925.
The equation that he formulated to describe how these
waves evolve contains only continuous quantities -- no integers.
Yet when you solve the Schr\"odinger equation for a specific system,
a little bit of mathematical magic happens. Take the hydrogen atom:
the electron orbits the proton at very specific distances.
These fixed orbits translate into the spectrum of the atom.
The atom is analogous to an organ pipe, which produces a
discrete series of notes even though the air
movement is continuous. At least as far as the atom is concerned,
the lesson is clear: God did not make the integers. He made
continuous numbers, and the rest is the
work of the Schr\"odinger equation.

In other words, integers are not inputs
of the theory, as Bohr thought. They are
outputs. The integers are an example of
what physicists call an emergent quantity.
In this view, the term ``quantum mechanics'' is a misnomer.
Deep down, the theory is not quantum. In systems such as the
hydrogen atom, the processes described by
the theory mold discreteness from underlying continuity.
\equo

In order to take yet another look at the problem of point particles and other 
hypothetical discontinuities, let us reconsider the conditions that wave 
functions (or any other ``signal'') must fulfill to be physically possible
and meaningful. Usually signals are understood as quantities changing in time,
but since signals are also transmitted through space, spatial distributions can 
(and must) be treated on equal footing: In the following
we use the variables $x$ and $k$, but one might as well replace them
with $t$ and $\w$.

In the previous section we argued that any non-negative normalizable density $\rho$ 
can be expressed by the square of some auxiliary function $\rho=\psi^\star\,\psi$
and that this already implies that $\psi$ is $L^2$-normalizable and
hence a ``vector'' in Hilbert space. Furthermore it follows that $\psi$ 
has a well-defined Fourier transform. But does it suffice to know that
{\it some} Fourier spectrum exists or are their further conditions?
Let us next take a closer look on the properties of the Fourier
transforms in cases where $\psi$ or $\rho$ (or their derivatives)
are not differentiable.

What is at stake is the question whether or not the fundamental
assumption underlying the SPQM, the idea of discontinuity, is
mathematically, physically and logically defendable.
Consulting an electrical engineer one would probably be told 
that strictly instantaneous signal jumps don't exist in any real
world system or measurement. And physicists should know the reason why:
No matter how high the sampling rate of an oscilloscope is (or will ever be), 
it will never be possible to confirm that some signal ``jump'' was instantaneous --
because this would require a measuring device with infinite bandwidth,
as we shall demonstrate in the following.

The Fourier transform $F(k)$ or some function $f(x)$ is given by:
\begeq
F(k)={1\over\sqrt{2\,\pi}}\,\int\limits_{-\infty}^\infty\,dx\,f(x)\,\exp{(-i\,k\,x)}\,.
\label{eq_ft0}
\endeq
Integration by parts gives:
\myarray{
F(k)&={1\over\sqrt{2\,\pi}}\,\int\limits_{-\infty}^\infty\,dx\,f'(x)\,{\exp{(-i\,k\,x)}\over -i\,k}\\
    &-{1\over\sqrt{2\,\pi}}\,\left.f(x)\,{\exp{(-i\,k\,x)}\over -i\,k}\,\right\vert_{-\infty}^\infty\\
}
Since $f(x)$ is normalizable, it must vanish at infinity, so that the
last terms is identically zero. It follows that
\begeq
-i\,k\,F(k)={1\over\sqrt{2\,\pi}}\,\int\limits_{-\infty}^\infty\,dx\,f'(x)\,\exp{(-i\,k\,x)}\,.
\endeq
The replacement of a derivative in coordinate space by the unit imaginary and
wave-number of Fourier space is standard in engineering science. It is not
``quantum'', it is ``Fourier''.

Then, if $f(x)$ has one (or several) discontinuities (steps), for instance if 
the pulse shape is assumed to be sharply rectangular or if the considered
signal is supposed to have some kind of instantaneous discontinuity, then the
derivative is the sum of a regular smooth function $\tilde f'$ and one (or
more) Dirac-deltas:
\begeq
f'(x)=\tilde f'(x)+\sum\limits_j\,\eps_j\,\delta(x-x_j)\,.
\endeq
Hence with
\myarray{
-i\,k\,F(k)&={1\over\sqrt{2\,\pi}}\,\int\limits_{-\infty}^\infty\,dx\,\tilde f'(x)\,\exp{(-i\,k\,x)}\\
&+{1\over\sqrt{2\,\pi}}\,\sum\limits_j\,\eps_j\,\exp{(-i\,k\,x_j)}\\
}
the Fourier spectrum $F(k)$ always has a component proportional to $1/k$. 
The ``energy'' contained in the pulse $f(x)$
is, in signalling theory, measured by
\begeq
E_x=\int\limits_{-\infty}^\infty\,dx\,f^\star(x)\,f(x)
\endeq
which is, according to Parseval's theorem, identical to the ``energy''
of the Fourier spectrum~\footnote{In quantum theory the square has a different
interpretation (i.e. number of particles), but is conserved as well
and has the same mathematical form.}:
\begeq
E_k=\int\limits_{-\infty}^\infty\,dk\,F^\star(k)\,F(k)
\endeq
However, even if the ``energy'' of the spectrum is finite (the wave function
normalized to $E_k=1$) {\it and} $F(k)$ has a nonvanishing term proportional to $1/k$,
then the root-mean-square (rms) bandwidth (i.e. momentum spread in QM) 
is infinite:
\begeq
\langle k^2\rangle=\int\limits_{-\infty}^\infty\,dk\,k^2\,\vert F(k)\vert^2=\infty
\endeq
As derived above, whenever steplike discontinuities are present,
$\vert F(k)\vert^2$ must contain a non-vanishing term ${\eps^2\over k^2}$.
Hence any such discontinuity implies an infinite root-mean-square (rms)
width of the spectrum, vulgo: an infinite bandwidth.

This explains why the presumption of {\it almost} instantaneous (i.e. fast)
change substantially differs from {\it really} instantaneous change:
We can imagine the physical existance of arbitrarily high frequencies,
but an infinitely high frequency is physically meaningless.
And so is the idea of instantaneous change on the fundamental level
of physical reality:
not only is it impossible to confirm such hypothesis experimentally,
it can not even be rendered meaningful. According to a certain reading
of the Copenhagen doctrine, assumptions that can not be rendered
meaningful in experimental terms, must be dropped as ``metaphysical''.
There is no reason why this would not also hold for the assumption of
``fundamental'' discontinuity.

The discovery of a physical scaling constant is
undoubtful a strong indication of fundamentally new physics. However,
the conclusion that Planck's constant provides evidence of a fundamental
discontinuity in nature, is not compulsory. It is not even specifically
nearby, since the proportionality of energy and frequency is known
in classical mechanics as the ``adiabatic invariance'' of the phase
space volume~\cite{Heat}.
Hence an equation of the form $E=h\,\nu$ where $h$ has the unit of action,
is {\it as such} not beyond classical mechanics. It becomes ``non-classical''
only by the fact that $h$ is a fixed scaling constant on the fundamental
level. With the existence of a natural constant with a physical unit
it becomes possible and nearby to express the respective quantity in
units of this natural constant -- velocities in units of $c$ and action
in units of $h$ (or $\hbar$). 

If a function $f(x)$ is continuous, but the derivative is not, then
we obtain a delta function with the second derivative and hence the
second moment is defined, but the fourth is infinite (and so on).
As a general rule we have\cite{Cohen1995}:
\myarray{
\langle k^n\rangle&=\int\limits_{-\infty}^\infty\,dk\,F^\star(k)\,k^n\,F(k)\\
&={(-i)^n}\,\int\limits_{-\infty}^\infty\,dx\,f^\star(x){d^n\over dx^n}\,f(x)\,,
\label{eq_Cohen}
}
which is identical to the corresponding quantum expressions and 
{\it as such} nothing but a math fact.

Hence the condition for the finiteness of the $n$-th moment in Fourier space 
depends on the finiteness of the $n$-th derivative in coordinate space. This
must be understood to recognize the implications of the postulate of 
instantaneous (time-less) quantum leaps -- or like-wise of point particles:
The presumption of the reality of instantaneous change at some stage,
no matter temporal or spatial, inevitably leads to more infinities downstream.
But it is quantum mechanics itself which tells us that infinite spectral
bandwidth implies infinite momentum spread and/or infinite energy spread,
respectively, which is in conflict with the laws of thermodynamics.

While the uncertainty relations of the SPQM merely tell us that 
$\Delta k\,\Delta x\ge 1/2$, signal analysis suggests that 
\myarray{
\Delta k<\infty\\
\Delta x<\infty\\
}
and hence
\begeq
1/2<\Delta k\,\Delta x<\infty
\endeq
must hold as well.

In his book ``Der Teil und das Ganze''~\footnote{The English title is {\it Physics
    and Beyond}.}, published 1969, Heisenberg reported that he {\it committed himself
  to believe} that nature is fundamentally discontinuous~\cite{HTG}.
The belief in these discontinuities prevailed over Schr\"odinger's belief
that nature must be fundamentally continuous. In collaboration with Bohr and
supported by the Copenhagen coterie, Heisenberg managed to engrave this
metaphysical prejudice deeply into the curriculum. But as we just showed,
fundamental discontinuities can not be tested experimentally.
Hence they remain a matter of metaphysical prejudice or {\it belief}.
But after what has been said, it should not suprize that recent experimental
findings indicate that the commitment to fundamental discontinuity in the form
of ``instantaneous'' quantum jumps is in conflict not only with mathematical
and physical logic but with experimental reality as well~\cite{NoJumps}. 

Once one has identified the principles upon which Schr\"odinger's equation
is {\it really} build, one recognizes that and why it has indeed a flavour of
inevitability as asserted (though never proven) by Bohr\cite{Beller}.
Schr\"odinger's equation is the basis for a universal (classical) theory 
of motion of continuously distributed matter. And since we derived it from
classical notions, Schr\"odinger's equation is {\it as such} not ``quantum''.

Since Schr\"odinger's equation is the non-relativistic approximation
of Dirac's equation, un-quantum mechanics has to proceed further.  
The algebraic constraints we shall obtain on this path will even allow to
discuss the dimensionality of space-time. This of course requires a longer
chain of arguments~\cite{qed_paper,osc_paper}.

\subsection{No Commandments}

Victor Stenger wrote that
``most laypeople think of the laws of physics as something like the Ten
Commandments -- rules governing the behavior of matter imposed by some great
lawgiver in the sky. However, no stone tablet has ever been found upon which 
such laws were either naturally or supernaturally inscribed''~\cite{Stenger}.
The question at hand is whether it is thinkable to construct physics as a purely 
deductive theory, i.e. whether or not physical laws are (all or at least some) 
logically inevitable. And if this is not considered possible, then there is
an obligation to explain where the residual laws of physics are supposed to
come from. Dirac described it as follows~\cite{Dirac81}:
\bquo
Some physicists may be happy just to have a set of working rules leading 
to results in agreement with observation. They may think that this is the goal 
of physics. But it is not enough. One wants to understand how Nature works. 
There is strong reason to believe that Nature works according to mathematical 
laws. All the substantial progress of science supports this view.
\equo

Commandment-like-laws (CLLs) can be descriptive, obtained from a fit to
experimental data and as such they are not only legitimate but unavoidable.
But such CLLs are prelimenary, at least with respect to their presentation.
Descriptive CLLs can be regarded as pieces of a puzzle, implying 
that even if it is yet unknown {\it how} the bits can be combined, one 
expects that eventually all CLLs ought to fit seamless into a 
more encompassing deductive approach. As soon as a general theory is 
available, we expect that it enables to derive equations which eventually
fit to the data. The probably best known example for this process of 
unification are Maxwell's equations. One must emphazise that it is the
scientific method itself which requires to presume that the presentation
of CLLs must eventually be modified once a larger theoretical framework
has been established. This is sometimes called ``Ockham's razor''~\cite{Ockham}. 

To refuse commandments therefore serves as our main guiding principle. We are
aware that many, maybe most, physicists and philosphers of science hold that
physical theories without assumptions are impossible. And this is of course
correct, since any physical theory must (for instance) presume the existence
of a physical world to be described. The term ``physical world'' includes further
presumptions, for instance that the status of a physical system at time $t_0$
constrains the possibilities for the status at time $t>t_0$ {\it smoothly}.
The mathematical expression for this rule is, that
\begeq
\psi(t+\delta t)=\psi(t)+\d_t\psi(t)\,\delta t+\dots
\endeq
where $\psi$ is the vector of numbers used to describe a physical system.
In the limit $t\to t_0$ the behavior is then linear. Whether one prefers to
regard this as a metaphysical or as an epistemological premise, is a different
debate. One might prefer the epistemological argument because of it's higher
degree of certainty: It is the core of the scientific enterprize to never
release the premise of cause and reason -- even if this commitment seems
unreasonable at first sight: In pre-scientific times, many things appeared
{\it unexplainable}, which nonetheless turned out to be scientifically
explainable as science progressed.
Progress in science was rarely initiated or supported by the believe
that some fact or relation is not possibly explainable, or in some sense
fundamentally non-causal.

The classical expression for continuity is {\it natura non facit saltus},
i.e. that nature does not perform jumps, but that any change is smooth and
continuous~\footnote{Many physicists and philosphers of science
(including some founding fathers of QM) believe that the ``discovery'' of
quantum theory falsified this assumption~\cite{Weinert,Capellmann} and
this believe in ``fundamental'' discontinuities became a pillar of the SPQM.
However, many physicists and philosophers vehemently opposed this conclusion,
for instance Einstein, Schr\"odinger and many others~\cite{Zeh1993,Zeh2009}.}.
As we have shown above, Schr\"odinger's equation can be derived solely on
the basis of exactly those classical assumptions that the SPQM seemingly
abandoned. Without them, Schr\"odinger's equation can, as asserted by Messiah, only
be postulated but it can't even be generally solved\footnote{
  Even in the case of (hypothetically) discontinuous potentials,
  Schr\"odinger's equation can only be solved if the wavefunction and
  it's first derivative are presumed to be continuous\cite{Pade2}.
}.

In the derivation of Schr\"odinger's
equation~\cite{seq_paper}, we presumed that an elementary ``object'' can be described 
by a density distribution in $3$-dimensional space~\footnote{To be more precise:
the dimensionality does not enter directly. But we shall show that the most
general linear Hamiltonian provides arguments for the $3$-dimensionality of
space. See also Ref.~\cite{qed_paper, osc_paper}.}. 
This and some elementary ideas of Hamiltonian mechanics are used in the 
derivation. However, also classical mechanics can, by the use of Hamiltonian 
theory, be presented in a way that allows for an almost assumption-free, 
purely analytical, approach. It is less well known than one should expect, 
but there exists a ``theorem due to Lie and Koenigs on the reduction of any 
system of ordinary differential equations to the Hamiltonian
form.''~\cite{Whittaker}. 
This means that ``any system of ordinary differential equations'' can be cast
into Hamiltonian form. Or, in other words, one can apply Hamiltonian dynamics
whenever a closed physical system can be described by {\it some} system of
ordinary differential equations. Likewise -- according to the Lie-K\"onigs-Theorem --
any closed physical system can be described by a conservation law of the 
Hamiltonian form. This is called {\it Hamiltonization}~\cite{Kerner,Santilli,Perlick}.

\section{Setup}

In a preceeding paper we argued that the only logical possibility to 
elude commandments is to derive the ``laws of physics'' from a definition 
of what is essentially meant by ``objective physical reality'', 
a world composed of {\it real physical objects} (RPOs)~\cite{fit_paper}.
Stenger expressed this idea as follows\cite{Stenger2007}:``If the models of
physics are to describe observations based on an objective
reality, then those models cannot depend on the point of view of the
observer. This suggests a principle of point-of-view invariance that is
equivalent to the principle of covariance when applied to space-time.''

We shall argue in this essay, that it is possible to derive a general
Hamiltonian in Fourier space, required to obtain Dirac's equation from
nothing but a definition of physicality -- from a single fundamental
constraint which serves as a ``reality condition''. 
This constraint is not even exceptionally deep or profound. It is simple, 
evident, well-known and straightforward. Gerard t'Hooft expressed it as 
follows~\cite{ToE2005}:
``[...] in particular string theorists expect that the ultimate
laws of physics will contain a kind of logic that is even more
mysterious and alien than that of quantum mechanics. I, however, will 
only be content if the logic is found to be completely straightforward.''
In this quote t'Hooft described his expectation concerning a final theory.
We do not (and can not) claim to have the final theory at hand, but we
do claim that the described kind of logic already exists and allows to
develop the basics of quantum mechanics in a straightforward way: it
is the logic of Hamiltonian methods~\cite{vlh_paper}.

The question of a definition of {\it physicality} is also raised
by multiverse theories that have been suggested, for instance by 
Tegmark, who conjectured the possibility of many different 
physical worlds, not only different {\it copies} or {\it versions} 
of the same kind of world, but worlds that obey very different physical 
laws~\cite{Tegmark98,Tegmark}. If not only the known physical world, 
but many different worlds are conjectured, then these worlds 
must have something in common that allows to call all of them ``physical''.
Hence a simple and general definition of physicality is required, 
something that must hold in any thinkable physical world, even in 
hypothetical worlds which might, according to Tegmark's idea, be ruled by 
different and alien physical laws, that might have, for instance, 
a different number of spatial dimensions or, who knows, no spatial 
dimensions at all. Of course, if Tegmark was right, then a final
deductive approach would be impossible~\cite{fit_paper}.

\subsection{Symmetries and Quantities}

Physics creates models of (parts of) reality. These models allow to ``simulate
reality'', or, in simple cases, to directly calculate results.
Hence it is arguable, that, whatever is physical in a world, should allow 
for a description by {\it algorithms} that predict (probabilities for
possible) evolutions of physical {\it quantities} in {\it time}. 
Hence the basis of a physical model of reality is a (possibly very long) 
list of quantities $\psi(\tau)$, that depend on time $\tau$. This is the 
raw material for a general physical model of any thinkable physical reality. 

One of the first facts children learn about real objects is 
called {\it object permanence}~\cite{objperm}, namely 
{\it that the moon is still there, even if nobody looks}. 
Object permanence does not seem to be general enough to serve as 
the desired constraint that defines physicality, because macroscopic 
objects can be disassembled and destroyed, one can break tea cups and 
burn wooden chairs. 
But matter can be manipulated only within specific constraints. Objects 
are made of other objects. Insofar as one can disassemble objects, they
can be destroyed, but chemistry found that the amount of matter remains 
unchanged, even if objects are burned. And even though it is theorectially
possible to destroy all individual microscopic objects (particles) that a 
macroscopic thing is made of by annihilation with a perfect copy made of 
anti-matter -- still (to our best knowledge) all energy remains.
This, eventually, is an insight that is as simple as it is not trivial.

The impossibility for a {\it perpetuum mobile} that produces net energy 
is {\it the} fundamental constraint for any known closed physical system or 
process. According to Einstein ``The most satisfactory situation is evidently 
to be found in cases where the new fundamental hypotheses are suggested by 
the world of experience itself.
The hypothesis of the nonexistence of perpetual motion as a basis for
thermodynamics affords such an example for a fundamental hypothesis suggested
by experience''~\cite{EinsteinPR}.

As Einstein rightly remarked, this principle is suggested by experience, 
but once it's depth has been recognized, physicists understood that 
it has the strength and status of a definition of physical realness 
itself. If a theory fails to provide conservation of energy then it is 
{\it unphysical by definition}. But if it is possible to define un-physicality 
on the basis of a conservation ``law'', why then should it not be possible 
to also define physicality this way?

But what exactly is a {\it conservation law}?
Emmy Noether, in 1918, discovered the math fact~\cite{Noether}, as Stenger puts
it, ``[...] that coordinate independence was more than just a constraint on the 
mathematical form of 
physical laws. She proved that some of the most important physics principles 
are, in fact, nothing more than tautologies that follow from space-time 
coordinate independence: energy conservation arises from time translation 
invariance, linear momentum conservation comes from space translation
invariance, and angular momentum conservation is a consequence of space rotation
invariance. These conserved quantities were simply the mathematical generators of the
corresponding symmetry transformation.''~\cite{Stenger}
Hence it is a math fact that a conservation law is nothing but a continuous 
symmetry, the generator of which is a conserved quantity. 

The concepts of {\it energy} as well as of {\it action} can only
be defined on the basis of an already elaborated physical theory~\footnote{
It is known to be a non-trivial problem to find non-circular definitions 
of the central notions (mass, energy, force) within classical physics.}.
This would of course be a theory of the known physical world and
not necessarily valid in any hypothetical physical world. How should one 
{\it know} a priori whether these notions are releveant and meaningful
in any thinkable physical world? Therefore, if arbitrary physical worlds
are considered, these notions are too specific to be used from the start.

However, it is not required to specify the type of the conserved quantity. 
It suffices to formally refer to {\it some} positive 
definite constant of motion (PDCOM) which serves as a measure
of object permanence, because it is known that it is not the object itself 
which is permanent, but some abstract quantity that objects are ``charged with''. 
We can anticipate that this quantity will turn out to be a possible measure 
of the amount of substance. This requires no postulate: If correct, then
the {\it physical meaning} of the conserved quantity should be a 
{\it mathematical} consequence of its conservation. Planck expressed it
this way~\cite{Planck1908}:
\bquo
[...] von dem Gedanken ausgehend, dass der Begriff der 
Energie seine Bedeutung f\"ur die Physik erst durch das 
Princip gewinnt, welches ihn enth\"alt~\footnote{
Translation by the author: ``[...] starting from the idea
that the concept of energy receives it's physical meaning
only through the principle that contains it [the principle
of conservation of energy].''}
\equo

Since we promote herewith
the idea to start the formulation of a general theory of physics from the 
mathematical form alone, it follows that the ``type'' or ``meaning'' of a 
physical quantity has to be understood entirely from its structural relation 
to other quantities. One can identify the conserved quantity with
{\it mass} or {\it energy}, if it has the same formal relationships to other 
quantities as already known from established physics: The physical meaning is a 
consequence of the mathematical form, not vice versa. A quantity $x$ that is
invariant under any transformation and has a relationship with other quantities
$x_i$ such that $x=\sqrt{x_0^2-x_1^2-x_2^2-x_3^2}$, can be identified as mass,
if $x_0$ transforms like an energy and ${x_1,x_2,x_3}$ like the components
of the momentum vector. In other words: We do not postulate the notion of
mass or energy apriori, we recognize the mathematical form and identify the
meaning aposteriori. There are arguments supporting the view that this
might be the only feasible strategy for any {\it final} theory~\cite{fit_paper}.

If a no-assumption-approach allows for any conclusion about the nature of the 
conserved quantity, then it must emerge from the {\it form} of the equations or 
symmetries. Only if known mathematical structures emerge, if the form of the
equations {\it suggests} a specific interpretation, then the
{\it principle of sufficient reason} (PSR) authorizes to map quantities of
the theory to known physical observables, i.e. to propose an interpretation.

Hence the very idea of a real physical object logically requires at least one 
positive definite constant of motion (PDCOM). Little more than this will be 
used to derive quite a number of the basic ``laws'' of physics. 
Without commandments. 

\subsection{Time}

There is no a priori reason to introduce more than a single symmetry, 
namely constancy in time, aka {\it permanence}. Or, to be more precise:
If we would presume more than one single symmetry, then one would need
to specify how many and why not one more or less. 

Time is a primary quantity that remains basically ``undefined''. 
Stanley Goldberg wrote that ``Either you know or you don't know what I 
mean when I use a phrase like "time passes."''~\cite{Goldberg}. 
The ``dimension'' of time is different from spatial dimensions insofar 
as it is {\it unique}. One can discuss the dimensionality of physical space 
and one can, within the classical framework, imagine physical worlds with 
more or less than $3$, maybe even zero, spatial dimensions. But it is 
questionable if it is possible to imagine a physical world without the unique 
dimension of time: ``[...] only one true integer may occur in 
all of physics. The laws of physics refer to one dimension of time''~\cite{Tong}. 

Hamilton wrote that ``the notion or intuition of order in time is not 
less but more deep-seated in the human mind, than the notion of intuition 
of order in space; and a mathematical science may be founded on the former, 
as pure and as demonstrative as the science founded on the latter. There is 
something mysterious and transcendent involved in the idea of Time; but there 
is also something definite and clear: and while Metaphysicians meditate on 
the one, Mathematicians may reason from the other.''~\cite{APT}. 

We shall show that, even though Hamilton's idea to derive algebra as {\it the} 
science of pure time failed~\cite{Ohrstrom}, there is nonetheless {\it an} 
algebra of pure (aka {\it proper}) time.
It reveals the possibility to formulate (central parts of) physics as a 
tautology~\cite{fit_paper}. This has absolutely no negative connotation,
as Goldberg explained: 
``Different branches of mathematics have different rules but in all branches, 
since the rules are predetermined, the conclusion is actually a restatement, 
in a new form, of the premises. Mathematics, like all formal logic, is
tautological. That is not to say that it is uninteresting or that it doesn't 
contain many surprises.``~\cite{Goldberg}
If it is possible to derive (essential parts of) physics from a definition
of physicality, then the result is a tautology in Goldberg's sense. Of course
this means that any analytical theory is a tautology in  Goldberg's sense.

\subsection{Reason}
\label{sec_reason}

Generality is maintained by presuming nothing specific, neither 
about the ``nature'' of the dynamical variables nor about the ``nature'' of 
the conserved quantity. This attitude has been summarized by Hamilton 
under the name {\it principle of sufficient reason} (PSR): 
``Infer nothing without a ground or reason.''~\cite{PSR}. 
In this form it might also be called the {\it principle of insufficient 
reason}~\footnote{Ariel Caticha identified the principle of insufficient 
reason in Quantum mechanics~\cite{Caticha}.}.
Of course, the PSR eventually contains little more than Stenger's claim that
there are no commandments. 

The PSR has a bias towards symmetry since nothingness (the void) is the most 
symmetric state: To assume nothing specific about a number of things or quantities 
has to be understood as assuming no asymmetry and the PSR forbids to introduce 
asymmetries, distinctions and classifications without a ground or reason.

Hence the raw material for the simplest physical object contains a yet unknown
number $\nu$ of dynamical variables $\psi=(\psi_1,\dots,\psi_\nu)$ (quantities)
that depend on a time, the evolution of which is constrained by a 
PDCOM ${\cal H}(\psi)={\cal H}_0=\rm{const}$. 

With the condition that ${\cal H}(\psi)$ is a constant of motion, it is implied 
that $\psi$ itself does not contain any other constant, i.e. all variables 
in the list $\psi$ depend on time so that no linear combination of the elements 
of $\psi$ may presumed to be constant.

\subsection{Structure of the Paper}

In Sec.~\ref{sec_lom} we shall firstly show that in any linear classical dynamical 
system, which can be derived from the assumed PDCOM, the number of true dynamical 
variables is even, i.e. the variables come in pairs and secondly that one can 
always describe the dynamics, after an appropriate change of variables, by 
Hamilton's laws of motion. This is a ``light version'' of the theorem of Lie
and K\"onigs, which provides evidence that Hamiltonian dynamics has maximal 
generality and is not negotiable. If the SPQM suggests that it requires 
modifications then we shall show that this is wrong.

In Sec.~\ref{sec_phasespace} we introduce the phase space distribution as 
the fundamental mathematical representation of physical objects. We show 
that it suffices to consider the simplest possible description of phase space 
distributions, namely the matrix of second moments (sloppily called
auto-correlation matrix), to derive Heisenberg's equation of motion for
operators. We show that it can be made ``quantum'' merely by notation.

In Sec.~\ref{sec_algebras} we shall derive the basic algebras of phase
space, namely the algebra of (skew-) Hamiltonian matrices. We explain 
the necessity to describe stable phase space distributions by second 
and higher even moments, and why this implies that it is impossible 
to measure $\psi$ directly.

In Sec.~\ref{sec_clifford} we explain the meaning and the role of Clifford
algebras in low-dimensional phase spaces. We critically review the specific 
notational convention concerning the use of complex numbers in QM in 
general and specifically in Dirac's theory. We explain the general 
conditions that Hamiltonian physics imposes on the dimensionality of 
phase spaces.

In Sec.\ref{sec_fff} we use simple group-theoretical considerations that,
when applied to the Dirac algebra, suggest an interpretation in terms
of relativistic electrodynamics. We show that this interpretation directly 
yields the Lorentz force law, the Lorentz transformations and the 
relativistic energy-momentum-relation. We demonstrate that this framework
also enables to derive the Zeeman effect, the spin, and the physics of 
adiabatic high frequency transitions (Breit-Rabi-model).

In Sec.~\ref{sec_uqm} we explain what is meant by ``un-quantization'': Since
the so-called ``canonical quantization'' can be derived and explained on 
the basis of classical notions, we simultaneously un-quantize QM in the 
sense explained above and ``quantize'' classical mechanics.
Then we briefly discuss Born's rule and explain why classicality is compatible with 
background independence but nonetheless leads {\it with necessity} to 
$3+1$-dimensional geometrical notions.

\section{The ``Law'' of Motion}
\label{sec_lom}

Eq.~\ref{eq_wpd} raises the question, how to obtain the Hamiltonian function 
${\cal H}$ (or ``dispersion relation'') of the wave ensemble in Fourier space.  
We shall omit specific assumptions as there is a straightforward method to 
construct the required Hamiltonian ``from first principles'': the unavoidable 
presumption that a conserved quantity (represented by a ``Hamiltonian'') exists 
is the only ``first principle'' required to proceed.

Our inventory so far consists of a number $\nu$ of dynamical variables $\psi$,
subject to change in time $\tau$ and a PDCOM ${\cal H}(\psi)$.
With the prelimenary simplifying assumption~\footnote{
The difference between simplifying assumptions and some substantial assumptions
is the following: A simplifying assumption includes no assertion
about reality. It is introduced only to simplify the math and implies that
the assumption will be dropped in a later stage for a more general treatment.} 
that ${\cal H}$ does not explicitely depend on time ${\d{\cal H}\over\d\tau}=0$, 
the physicality constraint can be formulated as follows: 
\begeq
\dot{\cal H}=\sum\limits_{k=1}^\nu\,{\d{\cal H}\over \d\psi_k}\,\dot\psi_k=0
\label{eq_com0}
\endeq
where the overdot indicates the temporal derivative. Note that we do not
claim to know {\it why} there should be a physical world with physical objects
in it. We only claim that {\it if} a world exists that can be called {\it physical} 
and {\it if} this world allows for a theory without commandments, then Eq.~\ref{eq_com0}
must hold. Eq.~\ref{eq_com0} can be written in vectorial notation as
\begeq
(\nabla_\psi{\cal H})\cdot\dot\psi=0
\label{eq_com1}
\endeq
where the ``$\cdot$'' indicates a scalar product and the overdot the time derivative. 
The general solution is given by
\begeq
\dot\psi={\bf J}\,(\nabla_\psi{\cal H})
\label{eq_lom0}
\endeq
with some arbitrary $\nu\times\nu$ skew-symmetric matrix ${\bf J}$. 
Inserted into Eq.~\ref{eq_com1} the condition for constancy of ${\cal H}$
is fulfilled -- by the {\it skew-symmetry} of {\bf J} {\it alone}.

It is a math fact that if $\lambda$ is an eigenvalue of the real square 
skew-symmetric matrix ${\bf J}$, then $-\lambda$ is also an eigenvalue.
Hence any skew-symmetric matrix of size $\nu\times\nu$ with odd $\nu$
has at least one vanishing eigenvalue.
A vanishing eigenvalue indicates the existence of a (hidden) constant in $\psi$.
Since this was excluded by definition and the PSR, ${\bf J}$ must have full
rank and hence $\nu=2\,n$ is an even integer or can be reduced to an even
integer by an appropriate coordinate transformation. In both cases we can
restrict ourselves to an even number of ``true'' dynamical variables without
loss of generality: In any physical world, the number of dynamical variables 
that are required to describe an RPO, is even.

We use the prelimenary simplifying assumption that ${\cal H}(\psi)$ can 
be written as a Taylor series of $\psi$ and initially concentrate on the 
terms of lowest order. For this case of small oscillations one may skip 
higher than quadratic terms and translate by $\psi_0$ such that linear
terms vanish, without loss of generality. The constant term 
can be excluded as trivial. Then ${\cal H}(\psi)$ can be written as
\begeq
{\cal H}(\psi)=\frac{1}{2}\,\psi^T\,{\bf A}\,\psi
\label{eq_H2}
\endeq
with a positive definite symmetric~\footnote{Skew-symmetric components 
don't contribute and are therefore irrelevant.} matrix ${\bf A}$ of size
$2\,n\times 2\,n$. The linearized {\it law of motion} (LOM) Eq.~\ref{eq_lom0} 
then becomes 
\begeq
\dot\psi={\bf J}\,{\bf A}\,\psi\,,
\label{eq_lom1}
\endeq
where $\psi$ is a vector of $2\,n$ components. 

According to a theorem of linear algebra for every non-singular skew-symmetric matrix 
${\bf J}$ of size $2n\times 2n$ there exists a non-singular matrix ${\bf Q}$ such
that~\cite{MHO}:
\begeq
{\bf Q}^T\,{\bf J}\,{\bf Q}=\textrm{diag}(\lambda_0\,\eta_0,\lambda_1\,\eta_0,\dots,\lambda_n\,\eta_0)
\label{eq_strucdef}
\endeq
where $\lambda_k$ are real non-zero constants (the modulus of two eigenvalues) and
\begeq
\eta_0=\bmtx{cc}0&1\\-1&0\emtx\,.
\endeq
Since there is no reason to assume anything specific about the eigenvalues,
beyond being non-zero, the PSR recommends the most symmetric case, 
i.e. the modulus $\vert\lambda_k\vert$ equals unity~\footnote{For $n=1$ 
there is only a single pair of eigenvalues $\pm\lambda$ which gives just a 
factor in Eq.~\ref{eq_lom1} and can therefore be dropped by the use of 
the suitable time unit.
The commitment to the PSR does not permit to introduce different 
eigenvalues for $n>1$ without reason.}. 
In this case ${\bf Q}$ is an orthogonal transformation so that
\begeq
{\bf Q}^T\,{\bf J}\,{\bf Q}={\bf 1}_n\otimes\eta_0\equiv\y_0
\label{eq_strucdef1}
\endeq
Note that this transformation is only required to obtain 
the symplectic unit matrix in a simple form. This firstly shows 
that ${\bf J}^2=\y_0^2=-{\bf 1}$ and secondly that the dynamical
variables can always be regarded as canonical pairs $q_i$ and $p_i$.
One may write $\psi=(q_1,p_1,q_2,p_2,\dots,q_n,p_n)^T$.
A single canonical pair represents 
the smallest thinkable dynamical system with a PDCOM and is called 
a degree of freedom (DOF).

Hence Eq.\ref{eq_lom0} can then be written, without loss of generality,
in the form of Hamilton's equations of motion
\begary{rcl}
\dot q&=&{\d{\cal H}\over\d p}\\
\dot p&=&-{\d{\cal H}\over\d q}\\
\label{eq_lom3}
\endary
or, using the linear approximation (Eq.~\ref{eq_H2}):
\begeq
\dot\psi=\y_0\,{\bf A}\,\psi={\bf H}\,\psi\,,
\label{eq_lom2}
\endeq
where $\y_0\,{\bf A}$ has been replaced by a single matrix ${\bf H}$.
Since nothing but physicality is assumed, Hamilton's equations 
of motion must pop up in any {\it thinkable} physical reality
without commandments.

Aristotelian physics is based on a fundamental distinction between
objects in motion and objects at rest. Newton's revolution rests, 
from a certain perspective, on a different fundamental distinction,
namely on the distinction between accelerated motion and rectilinear
motion. Hamiltonian physics replaces this with the distinction between
those quantities that are constant and those that (can) vary in time.
The former are called ``constants of motion'' while the latter are the
``dynamical variables''. This distinction is, in our view, the most 
fundamental physical distinction that we can think of and 
qualifies Hamiltonian dynamics as foundational for all of physics.

The matrix $\y_0$ is the so-called {\it symplectic unit matrix} (SUM).
A matrix that can be written in the form ${\bf H}=\y_0\,{\bf A}$
is called {\it Hamiltonian}. The transpose of a Hamiltonian matrix is
\begeq
{\bf H}^T={\bf A}\,\y_0^T=\y_0\,{\bf H}\,\y_0\,.
\label{eq_symplex}
\endeq
We define the ``adjunct spinor'' $\bar\psi=\psi^T\y_0^T$ so that the 
Hamiltonian (Eq.~\ref{eq_H2}) can be written as 
\begeq
{\cal H}=\frac{1}{2}\,\psi^T\,{\bf A}\,\psi=\frac{1}{2}\,\bar\psi\,{\bf H}\,\psi\,,
\endeq
since $\y_0^T\,\y_0={\bf 1}$.

We stress again that no substantial assumptions were made to arrive at 
Hamilton's equations of motion (EQOM) and no substantial assumption about 
the meaning of $q_i$ and $p_i$ are implied by the notation. 
The use of the symbols ``$q$'' and ``$p$'' is simply a convention 
of Hamiltonian theory. These quantities represent arbitrary pairs of 
conjugate dynamical variables. And we stress again that they are 
``classical'' in the sense that $q\,p-p\,q=0$.

Time, an arbitrary number of dynamical variables, a constant of motion 
and the PSR are the only required ingredients for the concept of a 
$2\,n$-dimensional phase space. Hence the concept of phase space has 
no intrinsic connection to spatial coordinates or mechanical momenta, 
but is purely abstract. It is the basis of any physical world.

\section{Phase Space}
\label{sec_phasespace}

Almost all classical presentations of quantum mechanics 
as given by Born and Heisenberg~\cite{Born1924} as well as by 
Schr\"odinger~\cite{Schroed1926a,Schroed1926b,Schroed1926c,Schroed1926d}, Dirac~\cite{Dirac0} or von 
Neumann~\cite{JvN1932}, emphasize the Hamiltonian nature of 
Quantum theory. Even if the SPQM postulates that in QM the 
classical Poisson brackets have to be replaced by the 
commutator of conjugate operators, Birkhoff and von Neumann 
wrote that there
``[...] is one concept which quantum theory shares alike with 
classical mechanics and classical electrodynamics. This is the 
concept of a mathematical "phase-space."''~\cite{Neumann}.

But if CM and QM share the concept of phase space, then the purported
fundamental differences between CM and QM must be due to the 
interpretation, due to the assumed relation between phase space 
variables $q$ amd $p$ and measurable (``observable'') quantities. 
Since $q$ and $p$ are classically interpreted as a coordinate value
and the corresponding canonical momentum, then it is no surprize that 
the canonical pair has a different meaning in quantum theory. 
But nonetheless, in 1978, Dirac interpreted spinor components in
exactly the same way that we propose here~\cite{Dirac78}:
``These new degrees of freedom are to be associated here with certain 
dynamical variables $(q_1,p_1)$ and $(q_2,p_2)$ to be thought of as 
corresponding to two independent linear harmonic oscillators''.
As we shall demonstrate, this is indeed the main difference 
between CM and QM. But it is not mathematical, it is interpretational.

In presentations of QM, CM is often reduced to mass point dynamics, 
which implies a direct identification of the dynamical variables 
(i.e. elements of $\psi$) to measurable positions and mechanical momenta. 
But analytical mechanics, as formulated by Lagrange, Hamilton, Jacobi 
and others, is a set of abstract mathematical principles that underlie 
{\it any type of} dynamical system. There is no law in classical physics 
that limits the applicability of these concepts to ``mass points''
~\footnote{Effectively, the applicability of Hamiltonian concepts is not even
limited to physics.}.
Specifically books on quantum theory leave the impression as if
classical physics must inevitably be concerned with mass points.
But if one takes it to the test, 19th century textbooks on analytical
mechanics put few emphasis on a point-like nature of particles and if
they do, then they mostly refer to it as a convenient approximation, not as
an ontological model. The claim that classical physics cannot interpret
particles by any other notion but mass-points mainly serves the purpose 
to underline the revolutionary character of quantum theory. If one
refuses to accept the SPQM account of classicality, much of the
quantum thimblerig collapses.

The methods of Hamilton and Lagrange can be (and are) applied in many if 
not all fields of physics and any kind of dynamical variables, also in 
those cases that do not refer to a spatio-temporal description in the 
first place. A well known example is the Langrangian formulation of
electrodynamics~\cite{Jackson}. But also LC - circuits~\cite{LCHamilton}
or ray optics~\cite{Optics} and statistical physics~\cite{Statmech} make
use of the {\it classical} methods of Hamilton and Lagrange.
The theory of canonical transformations, which allows for any kind of
transformation that preserves the Hamiltonian equations of motion,
is the core concept of analytical mechanics and it is incompatible
with a reduction of CM to mass point dynamics.

The variable list $\psi$ formally represents a ``coordinate''
in some $2\,n$-dimensional phase space but this does neither imply nor 
suggest a specific interpretation of $\psi$. Then real physical objects 
(RPOs) (or systems, respectively) are, in the first place, inhabitants of 
phase spaces.
As a single classical mass point makes no tangible object in ``physical'' 
space, a single coordinate in phase space makes no sensible object as well.
But beyond this naive (but mathematically correct) intuition, any solution
$\psi(\tau)$ of Eq.~\ref{eq_lom1} is a singular solution, a ``trajectory''.
It is well-known in classical analytical mechanics however, that the knowledge
of single trajectories alone does not suffice to establish {\it strong
  stability}: It is a rigorous result of analytical dynamics that a solution
$\psi(\tau)$ is strongly stable only if a non-zero vicinity of the solution
is (strongly) stable as well~\cite{Arnold}. Hence, if we aim to describe real
physical objects by the methods of analytical dynamics, {\it point particles}
are useless: the mathematical possibility of their existence (i.e. that they
are a solution of the equations of motion) is necessary but not sufficient 
to decide whether they could exist with non-vanishing probability in the real
world. Hence some kind of distribution is required to replace single orbits,
metaphorically speaking one can think of such a distribution as
a droplet in phase space~\footnote{
Liouville's theorem states that Hamiltonian motion in phase 
space corresponds to the flow of an incompressible fluid~\cite{Heat}.}.

Since a general distribution $\rho(\psi)$ in phase space implies 
an infinite amount of information~\footnote{
That the ``quantum state'' contains infinite information is the content
of a Hardy's theorem~\cite{Hardy,JenningsLeifer}.
}, it is required to drastically reduce the complexity of the description.
A common way to describe distributions is to evaluate their respective
``size'' in either direction. 
The most common method to measure the size of a distribution is the use
of (the square-root of) second moments~\footnote{
Mathematically it is well-known that reasonable distributions 
with finite moments are completely and uniquely determined by their moments.}.
The matrix of the most important second moments $\Sigma$, for instance,
is given by
\begeq
\Sigma_{ij}\equiv\langle\psi_i\psi_j\rangle=\langle\psi\psi^T\rangle\,,
\endeq
where the embracing angles indicate some (yet unspecified) average~\footnote{
We leave aside subtleties of possible interpretations of how to obtain 
and understand this average. At this point it suffices to agree that one can
average over an ensemble of phase space points or some phase space volume. 
If the system is presumed to be ergodic, the average might also be obtained 
by integration over time.}.
Without loss of generality, one can write this as a matrix
product of some $2\,n\times m$ matrix ${\bf K}$ of the form 
\begeq
\Sigma_{ij}={\bf K}\,{\bf K}^T
\label{eq_sigma}
\endeq
where $m\ge 2\,n$~\cite{osc_paper}.
 
From Eq.~\ref{eq_lom2} one finds the (linearized) equation of 
motion of the autocorrelation matrix:
\begary{rcl}
\dot\Sigma&=&\langle\dot\psi\,\psi^T+\psi\,\dot\psi^T\rangle\\
          &=&\langle{\bf H}\,\psi\,\psi^T\rangle+\langle\psi\,\psi^T\,{\bf H}^T\rangle\\
          &=&{\bf H}\,\Sigma+\Sigma\,{\bf H}^T\\
\label{eq_enveq0}
\endary
Multiplication from the right of both sides with $\y_0^T$ gives:
\begeq
\dot\Sigma\y_0^T={\bf H}\,\Sigma\,\y_0^T+\Sigma\,{\bf H}^T\,\y_0^T
\label{eq_enveq1}
\endeq
Now we define another Hamiltonian matrix ${\bf S}$ by ${\bf S}\equiv\Sigma\y_0^T$
and with $\y_0^T=-\y_0$, $\y_0^T\y_0={\bf 1}$ and Eq.~\ref{eq_symplex}. Then one 
obtains
\begeq
{\bf \dot S}={\bf H}\,{\bf S}-{\bf S}\,{\bf H}\equiv[{\bf H},{\bf S}]\,,
\label{eq_enveq}
\endeq
which is known as Heisenberg's equation of motion for operators~\footnote{
We shall use the term {\it auto-correlation matrix} not only for $\Sigma$ 
but also for ${\bf S}=\Sigma\,\y_0^T$.}.

Eq.~\ref{eq_enveq} is still missing the quantum look and feel, namely the 
unit imaginary and $\hbar$, which are both absent.
But it is wrong to think that these factors are valid indicators for
the quantumness of equations. We prove this by simply introducing 
them from void.

Since all variables in $\psi$ are treated equally they must all have the 
same unit. According to Eq.~\ref{eq_lom2} the elements of the Hamiltonian
matrix ${\bf H}$ have the unit of frequency~\footnote{The autocorrelation 
matrix ${\bf S}$ can, up to this point, be given an arbitrary unit.}.
No one can prevent us from giving ${\bf H}$ the unit of energy by multiplication
with some conversion factor $\hbar$ of the dimension of action. We then 
obtain with ${\bf\tilde H}=\hbar\,{\bf H}$:
\begeq
{\bf \dot S}=\frac{1}{\hbar}\,[{\bf\tilde H},{\bf S}]
\label{eq_Heisenberg0}
\endeq
Any Hamiltonian matrix that represents stable dynamical systems has 
purely imaginary eigenvalues. Furthermore, if $\lambda$ is an 
eigenvalue of a Hamiltonian matrix, then $-\lambda$, as well the complex 
conjugates $\pm\bar\lambda$ are also eigenvalues~\cite{MHO}.
Since ${\bf\tilde H}$ is by definition a stable non-degenerate Hamiltonian 
matrix, it can be written as
\begeq
{\bf\tilde H}={\bf E}\,\rm{Diag}(i\e_1,-i\e_1,i\e_2,-i\e_2,\dots,-i\e_n)\,{\bf E}^{-1}
\endeq
where ${\bf E}$ is the matrix of eigenvectors and $\e_i=\hbar\w_i$ are 
real energy eigenvalues. 
We introduce another matrix ${\bf\breve H}$ by multiplication with $-i$:
\begeq
{\bf\breve H}=-i\,{\bf\tilde H}={\bf E}\,\rm{Diag}(\e_1,-\e_1,\e_2,-\e_2,\dots,\e_n,-\e_n)\,{\bf E}^{-1}
\endeq
which has now real energy eigenvalues so that the unit imaginary appears 
explicitely:
\begeq
{\bf \dot S}=\frac{i}{\hbar}\,[{\bf\breve H},{\bf S}]
\label{eq_Heisenberg1}
\endeq
The introduction of factors that otherwise cancel out can not add anything 
physical to an equation~\footnote{The idea that $\hbar$ represents
a quantum of action in the sense that classical mechanics is restored 
in the limit of $\hbar\to 0$, has been falsified experimentally~\cite{Hu}.}.
Therefore Heisenberg's operator equation is {\it as such} not quantum: 
we just derived it from classical Hamiltonian mechanics.
Without further assumptions it can be regarded as the linearized equations
of motion of second moments in a Hamiltonian phase space.
Furthermore Eq.~\ref{eq_enveq} proves that commutators are
just an algebraic result of {\it considering} the evolution of 
second moments in time. Hence the appearance of (anti-) commutators is, 
as such, not quantum either.

\subsection{Unitary vs. Symplectic Motion}

The matrix ${\bf H}$ is Hamiltonian (not Hermitian), therefore 
it generates symplectic (and not unitary) evolution in time. 
However, symplectic motion is more general than unitary motion since firstly, 
unitary motion is always symplectic~\footnote{See App.~\ref{sec_unisym}.}
but secondly, symplectic motion allows for complex eigenvalues~\footnote{
Complex here means {\it truely} complex, points in the complex plane 
that are neither on the real nor on the imaginary axis.} 
which are excluded in unitary motion. But no law of the universe 
and no commandment forbids complex eigenvalues. Such a law would be 
superfluous anyway as complex eigenvalues and stability are
incompatible. Complex eigenvalues may appear (for a limited time) in 
nature, for instance in case of resonance~\cite{Gamov,MG73,GC1}, but they 
are incompatible with long-term stability. 

Bender and others have shown that unitarity is not universally required, 
not even in the case of real eigenvalues~\cite{Bender98,Bender03,Bender07,NonHermitian}.
The corresponding postulate of QM is therefore not universally valid 
and can be dropped. 

This is a simple math fact and requires no postulates: 
the description of stable states (of motion) {\it requires} the corresponding 
mathematical form of eigenvalues. If one insists on the unit imaginary as an 
indispensable QM factor, then the eigenvalues of a stable system must be real. 
In stable symplectic motion, which is considered to be classical, the unit 
imaginary is not written explicitely and the eigenvalues of stable motion are 
purely imaginary.
In the former case one uses the unit imaginary  {\it explicitely} and writes 
the frequency as $\w={\cal E}/\hbar$, in the latter case the unit imaginary 
is used {\it implicitely} as the eigenvalues have the form $\pm\,i\,\w$ 
with a real valued frequency $\w$. But neither nature nor mathematical logic
cares much about notational conventions: Neither the explicite use of the unit 
imaginary nor unitary evolution are {\it as such} quantum.

The derivation of Eq.~\ref{eq_enveq} from a PDCOM (the Hamiltonian) 
suffices to generate a structure preserving symplectic framework in 
which ``probability current conservation'' pops up automatically, 
since symplectic motion is known to conserve the occupied volume of 
phase space as we shall see in the next section.

\section{Phase-Space Algebra}
\label{sec_algebras}

Let us mention some math facts about Hamiltonian matrices and symplectic 
motion that are known, but maybe not well-known. We begin with the fact 
that Eq.~\ref{eq_enveq} constitutes a so-called {\it Lax pair}. 
As Peter Lax has shown, if a pair ${\bf S}$ and ${\bf H}$ of operators
obeys Eq.~\ref{eq_enveq}, then the trace of any power of ${\bf S}$ 
is a constant of motion~\cite{Lax}:
\begeq
\rm{Tr}({\bf S}^k)=\rm{const}\,,
\label{eq_COM}
\endeq
for all $k\in\mathbb{N}$. This also holds for non-linear operators.
Within our approach, both matrices are by definition the product of the 
skew-symmetric SUM $\y_0$ and a symmetric positive definite matrix ${\bf A}$. 
According to linear algebra they have a vanishing trace and it can be shown
that all odd powers of ${\bf H}$ and ${\bf S}$ share this property:
\begeq
\rm{Tr}({\bf S}^{2k+1})=0\,,
\label{eq_odd_trace}
\endeq
for $k\in\mathbb{N}$, so that only the even powers in Eq.~\ref{eq_COM}
are ``non-trivial'' constants of motion (COMs). It has been shown elsewhere
that the eigenvalues of the autocorrelation matrix are a measure of the 
occupied phase space volume~\cite{Dragt,Wolski}. Again: This is a math fact
and requires no postulate.
Furthermore, in statistical mechanics, a phase space density is as close 
as can be to a probability density.

As already mentioned, both matrices, the driving matrix ${\bf H}$ and the
matrix ${\bf S}$, have the same structure, namely both are 
{\it Hamiltonian} (Eq.~\ref{eq_symplex}).
It is well known that such matrices are generators of symplectic motion
by the fact that the solution of Eq.~\ref{eq_lom2} is the
matrix exponential
\begeq
\psi(\tau)=\exp{({\bf H}\,\tau)}\,\psi(0)={\bf M}(\tau)\,\psi(0)\,.
\label{eq_sol1}
\endeq
It is straightforward to show that a symplectic matrix ${\bf M}=\exp{({\bf
    F}\tau)}$ holds\footnote{
A matrix ${\bf N}$ is skew-symplectic if ${\bf N}\,\y_0\,{\bf N}^T=-\y_0\,.$
}:
\begeq
{\bf M}\,\y_0\,{\bf M}^T=\y_0\,.
\label{eq_symplectic}
\endeq
One finds after few steps
\begeq
{\bf S}(\tau)={\bf M}(\tau)\,{\bf S}(0)\,{\bf M}^{-1}(\tau)={\bf M}(\tau)\,{\bf S}(0)\,{\bf M}(-\tau)\,.
\label{eq_sol4}
\endeq
The result of Hamiltonian evolution in time, the result of motion, is a 
symplectic similarity transformation (SST). And since similarity
transformations do not change eigenvalues, the eigenvalues are COMs.

\subsection{Eigenvectors and Eigenvalues}
\label{sec_eigen}

It is yet another math fact that commuting matrices share a system of eigenvectors. 
According to Eq.~\ref{eq_enveq} the matrix ${\bf S}$ (and hence the second 
moments) is constant, iff ${\bf H}$ and ${\bf S}$ commute. Only diagonal 
matrices always commute, so that commuting matrices must have the same 
matrix of eigenvectors ${\bf E}$:
\begary{rcl}
{\bf D}_S\,{\bf D}_F&=&\,{\bf D}_F\,{\bf D}_S\\
({\bf E}\,{\bf S}\,{\bf E}^{-1})\,({\bf E}\,{\bf H}\,{\bf E}^{-1})&=&({\bf E}\,{\bf H}\,{\bf E}^{-1})\,({\bf E}\,{\bf S}\,{\bf E}^{-1})\\
{\bf S}\,{\bf H}&=&{\bf H}\,{\bf S}\\
\label{eq_diag}
\endary
where ${\bf D}_S=\rm{Diag}(i\,\eps_1,\dots,-i\,\eps_n)$ is the diagonal matrix 
containing the eigenvalues of ${\bf S}$. 

{\it Therefore} eigen-vectors and -values must play an important role
in any physical world. This is a consequence of constructing observables 
from second moments on the basis of Hamiltonian mechanics. Again this requires no
postulates, neither quantum nor otherwise.
Oscillatory systems have eigenvalues -- the frequencies -- and eigenvectors~\footnote{
Note that ${\bf S}$ commutes with analytical functions of ${\bf H}$: if it
commutes with ${\bf H}$, it commutes with ${\bf M}=\exp{({\bf H}\tau)}$ as well.}.
It is a math fact that strongly stable systems must have purely imaginary 
eigenvalues and complex eigenvectors. And since the eigenvalues come in
pairs (or quadruples, if complex), the eigenvectors also come in complex
conjugate pairs. This is a math fact about (classical) coupled oscillating 
systems, subject to linear Hamiltonian motion. The use of Eigenvalues and
Eigenvectors in the description of physical systems is therefore {\it as such}
not quantum. 

\subsection{Symplectic Motion is Structure Preserving}

Since similarity transformations do not change eigenvalues,
this also holds for linear symplectic motion, i.e. SSTs.
SSTs also preserve the structure of Hamilton's equations of motion, 
i.e. the form of the matrix $\y_0$. With respect to an RPO
it is specifically the {\it dynamical structure} which determines the 
properties, or the {\it type} of these objects. The fact that evolution 
in time is a SST, guarantees that Hamiltonian matrices remain Hamiltonian.
The exponential of a Hamiltonian matrix is symplectic and the logarithm
of a symplectic matrix is Hamiltonian~\cite{MHO}.
A skew-Hamiltonian matrix ${\bf C}$ is a product of the SUM $\y_0$ and an 
arbitrary skew-symmetric matrix ${\bf B}=-{\bf B}^T$:
\begeq
{\bf C}=\y_0\,{\bf B}
\endeq
such that
\begeq
{\bf C}^T=-\y_0\,{\bf C}\,\y_0\,.
\label{eq_cosymplex}
\endeq
Accordingly the number of linear independent elements $\nu_s$ in a Hamiltonian 
matrix of size $2\,n\times 2\,n$ is
\begeq
\nu_s=n\,(2\,n+1)
\label{eq_nsym}
\endeq
and in a skew-Hamiltonian matrix it is $\nu_c$:
\begeq
\nu_c=n\,(2\,n-1)
\label{eq_ncosym}
\endeq
 
It is a straightforward exercise to show that the (anti-) commutators of 
Hamiltonian (${\bf S}_i$) and skew-Hamiltonian (${\bf C}_j$) matrices
have (anti-) commutators of the following type:
\begary{ccc}
\left.\begin{array}{c}
{\bf S}_1\,{\bf S}_2-{\bf S}_2\,{\bf S}_1\\
{\bf C}_1\,{\bf C}_2-{\bf C}_2\,{\bf C}_1\\
{\bf C}\,{\bf S}+{\bf S}\,{\bf C}\\
{\bf S}^{2\,n+1}\\
\end{array}\right\} & \Rightarrow & \mathrm{Hamiltonian}\\&&\\
\left.\begin{array}{c}
{\bf S}_1\,{\bf S}_2+{\bf S}_2\,{\bf S}_1\\
{\bf C}_1\,{\bf C}_2+{\bf C}_2\,{\bf C}_1\\
{\bf C}\,{\bf S}-{\bf S}\,{\bf C}\\
{\bf S}^{2\,n}\\
{\bf C}^n\\
\end{array}\right\} & \Rightarrow & \mathrm{skew-Hamiltonian}\\
\label{eq_cosy_algebra}
\endary
Note that the unit matrix ${\bf 1}$ is skew-Hamiltonian.
It is remarkable that it is possible to derive the complex
structure~\ref{eq_cosy_algebra} from nothing but algebraic
symmetry arguments, i.e. from a pure algebra of time~\footnote{
  Even though this ``algebra of time'' allows to obtain
  the Dirac matrices as we shall show below starting with
  Sec.~\ref{sec_hca} which contain a representation of
  Hamilton's quaternions, there is no evidence that Hamilton
  had anticipated this specific course of reasoning. Nonetheless
  it is remarkable in how many aspects this reasoning is based
  on and/or related to Hamilton's work.
}.

\subsection{Observables and Generators}
\label{sec_cosy}

Quantum mechanics is not the first physical theory that requires a reflection
about the meaning of a measurement. Also special relativity is a theory that 
struggles with the meaning of time and length measurements. 

There is a great pusillanimity in physics to use notions like ``existence'' or
``reality''. However the assumption that something does (or can) exist
(some body or field or force or dynamical variable) is implied by any
physical theory. All physical theories must, for instance, presume that
some ``observable'' quantity exists and that observables
can be measured. This in turn implies the existence of some direct or
indirect {\it reference} quantity.

Our considerations are based on a definition of physical realness and 
are therefore incompatible with claims that $\psi$ is somehow ``unreal''. 
In the contrary, $\psi$ was the only ontic ``thing'' we presume at all. 
This view is confirmed by the no-go-theorem of Pusey, Barrett and Rudolph 
in which the authors claim that 
``if a quantum state merely represents information about
the real physical state of a system, then experimental predictions are
obtained which contradict quantum theory''~\cite{PBR}. 
There is a minority report of physicists who do not subscribe a dogma of 
unreality. Roger Penrose, for instance, wrote ``if we are to believe that 
any one thing is in the quantum formalism is 'actually' real,[...], then I 
think it has to be the wavefunction [...]''~\footnote{Sect. 20.6 in
  Ref.~\cite{Penrose}.}, or Lev Vaidman: 
``The only fundamental physical ontology is the quantum wave 
function''~\cite{Vaidman}. 

But even if we defy to regard $\psi$ as somehow unreal, it can not
be denied that the physical meaning of $\psi$ is not self-evident.
So far we did not consider the physical unit of the variables $\psi$. 
What type of quantity do these variables represent? We shall stay agnostic
and answer that ``since it' s part of nature, we don' t really
know.''~\cite{RalstonBook}. But this situation is not new to physics:
Also classical notions like energy or force can not be properly defined
in such a way that anyone could answer ``what {\it is} energy after
all''~\cite{LopesCoelho2021}?

We can only stress again that {\it formally} $\psi$ is a coordinate in 
some phase space. Classical phase space coordinates have no fixed units, 
only the product of canonical pairs is - in the conventional system of
units - fixed to the unit of action, or angular momentum, respectively. 

Hence we can say that a $2\,n$-dim. volume ($n\ge 1$) of phase space may
have a unit, single coordinate values, i.e. elements of $\psi$ don't.
From the PSR it follows that both, the canonical coordinates and momenta
forming $\psi$, have the same unit.

Though one might formally say that $\sqrt{\hbar}$ would be the nearby
unit for a phase space coordinate, this has little practical value. 
A unit requires not only a name and a symbol. For a direct measurement
it is necessary to have a reference artifact that has a constant 
property of the same type: a certain weight, length, clock frequency or 
voltage. But since the variables $\psi$ are supposed to be fundamental, 
how should such an artifact emerge from a more fundamental level, if there
is none, {\it by definition}? 

Furthermore, $\psi$ is - by construction - a list of dynamical {\it variables} 
in the literal sense. The definition stipulates that none of these variables 
(and no linear combination thereof) can be considered a non-zero constant. 
Hence there is no constant reference and {\it therefore} $\psi$ can not be 
directly measured~\cite{qed_paper,osc_paper}. 
Only {\it available} constant quantities can provide a reference, i.e. 
second or higher even moments like the Hamiltonian ${\cal H}(\psi)$.
Linear Hamiltonian theory is based on a quadratic form, the Hamiltonian, 
which is a PDCOM by construction and provides the reference for all 
second (and possibly higher even) moments. 
Hence the use of second (or higher even) moments and correlations to describe 
the phase space distribution is not only a convenient and natural choice,
within our considerations it is the {\it only possible} choice.

There is no need to {\it postulate} that $\psi$ can not be measured:
unless someone presents a solution to the reference problem, we 
doubt that it has a solution. Mermin asserted that ``[...] the proper 
subject of physics [are] correlations and only correlations''~\cite{Mermin}.
Here is a logical reason why this assertion makes perfect sense: at least
two variables are required to obtain correlations.

Humans are inhabitants of a physical world and have the perspective of
insiders. One can not prevent anyone from considering the possibility 
that some supernatural being, some 'Maxwellian demon', might have a 
different perspective and is in the posession of a reference that enables 
to ``measure'' the elements of $\psi$ directly. But from {\it within} 
a physical world, a direct measurement is hardly possible, if $\psi$
is indeed fundamental. 
Some schools of philosophy deny the possibility to presume existence of 
unmeasurable entities. But we do not suggest that the {\it entities} are 
unmeasurable, we just doubt that the {\it values} of the variables in 
$\psi$ can be directly measured.

If one regards this as a reason to exclude the wave-function from 
classical physics, then classicality would have to be limited to 
observable physical quantities as well.
However, such a limitation of classicality would be historically and 
logically untenable:
Newton based the fundament of his theory on the existence of something
unmeasurable, namely absolute space. Furthermore he suggested a 
corpuscular theory for light, before any such corpuscle 
was experimentally detected. Boltzmann used atoms in the kinetic theory of 
gases before there was sufficient evidence that atoms exist at all. 
A sober view of physics reveals that there are plenty of entities which
can not be observed ``directly''. This is due to the very method of 
physics: Physics proceeds by presenting simple but not directly
observable {\it common} causes. If the cause would be directly observable,
it would not need a theory for it.

The reference problem explains many of the difficulties to understand 
and accept un-quantum physics for inhabitants of physical worlds, if these 
inhabitants share the prejudice that a physical theory must be based on 
{\it measurable} quantities alone. This also holds for human beings.

The matrix of second moments ${\bf S}=\Sigma\y_0^T$, and the spinor $\psi$ 
have, at first sight, very different LOMs. In contrast to the dynamical variables 
in $\psi$, that can {\it by construction} not be constant, the variables 
in ${\bf S}$ (the ``observables'') are constant, if ${\bf S}$ and ${\bf H}$ 
commute. Otherwise they (for instance) oscillate with some frequency and 
amplitude~\footnote{The branch of physics that best illustrates the relation 
between $\psi$ and ${\bf S}$ (or $\Sigma$, respectively) is probably accelerator
physics. Using the notions of charged particle optics, $\psi$ corresponds 
to individual particle positions (position not in space, but in phase space)
described in the local co-moving frame and the matrix $\Sigma$ corresponds
to the matrix of second moments used to
describe the {\it envelope} of a beam~\cite{rdm_paper,geo_paper}.}.
The Hamiltonian as a PDCOM is available as a reference quantity so 
that the correlations of ${\bf S}$ can always be measured. Hence there
are (at least) two different levels of reality, the ``spinor'' $\psi$ and 
its auto-correlation matrix ${\bf S}$, but only the latter can be directly
observed.

The matrix ${\bf S}$ is, like ${\bf H}$, a Hamiltonian matrix. 
Since skew-Hamiltonian matrix components do not contribute to the Hamiltonian, 
they cannot be (linear) generators of possible evolutions in time. Correspondingly the
autocorrelation matrix $\Sigma$ is symmetric. It follows that skew-Hamiltonian 
matrices have zero ``expectation'' values~\cite{qed_paper}. This means 
that there are further ``parameters'' emergent in the theory that necessarily 
vanish and are in this sense ``unmeasurable'' or ``hidden''. 

Classical statistical mechanics tells us that a phase space density is 
constant if (and only if) it is exclusively a function of COMs. 
Since only even moments can mathematically generate COMs, a stable
phase space density is an even function of $\psi$: $\rho(\psi)=\rho(-\psi)$.
Classical statistical mechanics is concerned with many DOFs
and in this case, only positive definite values do not cancel 
by averaging over some (thermal) ensemble, namely the known PDCOM, 
so that eventually in this case one finds $\rho=\rho({\cal H})$. 
The Boltzmann distribution $\rho({\cal H})\propto\exp{(-\beta\,{\cal H})}$ 
is such a case and corresponds, using Eq.~\ref{eq_H2}, to a multivariate 
normal distribution in $\psi$, up to a normalization. 

The constraint that only Hamiltonian ``operators'', parameters of the
Hamiltonian matrix, represent observables, might be regarded as the true 
origin of the Hermiticity condition for complex Dirac matrices~\footnote{
Given by $\y_0\,\y_\mu\,\y_0=\y_\mu^\dagger$~\cite{Pal}. In the
standard presentation of the Dirac electron theory, the adjunct spinor
is defined as $\bar\psi\equiv \psi^\dagger\y_0$, but the multiplication
with $\y_0$ is, to my knowledge, rarely explained.}.
We believe that our ``classical'' approach is clearer, straightforward and more 
stringent. Our analysis suggests that {\it fundamental} dynamical quantities 
{\it can not be observables} and that, instead, second (or higher 
even) moments are required to obtain observables~\footnote{
Therefore our approach implies a denial of the possibility that classical 
Newtonian metaphysics could be fundamental at all.}.
Then of course, some strange effects concerning the statistical properties of
observables are unavoidable~\footnote{
D.N. Klyshko considered that many if not all ``quantum paradoxes'' have 
a common origin, namely the ``failure to find a solution to a certain 
moments' problem''~\cite{Klyshko}. 
}. 

\subsection{The Pauli Matrices from Hamiltonian Symmetries}

Hamiltonian canonical pairs (HCP) are the basic elements of dynamical system.
If one considers a single HCP, the matrices $\y_0$, ${\bf H}$, ${\bf S}$ and ${\bf M}$ 
are of size $2\times 2$. Consider an arbitrary real $2\times 2$ matrix ${\bf K}$:
\begeq
{\bf K}=\bmtx{cc}a&b\\c&d\emtx
\endeq
The ``parametrization'' of ${\bf K}$ by individual matrix elements is the
simplest but it does not fit to the structural properties of linear 
Hamiltonian theory. Any square matrix of size $2\,n\times 2\,n$ can be 
expressed by the sum of a Hamiltonian and a skew-Hamiltonian matrix. 
We would like to find a set of parameters such that 
each parameter belongs to one and only one of those. Since Hamiltonian
matrices have zero trace, we can easily identify a skew-Hamiltonian part
as a multiple of the unit matrix $\eta_3={\bf 1}_2$:
\begeq
{\bf K}=\bmtx{cc}a&b\\d&-a\emtx+c\,\bmtx{cc}1&0\\0&1\emtx\,.
\endeq
Since the definition of Hamiltonian matrices uses matrix transposition,
it is required to distinguish between the purely symmetric, 
purely skew-symmetric, and diagonal matrix elements. We thus arrive at
\begeq
{\bf K}=s_0\,\eta_0+s_1\,\eta_1+s_2\,\eta_2+c\,\eta_3\,,
\endeq
where 
\begary{rclp{5mm}rcl}
\eta_0&=&\bmtx{cc}0&1\\-1&0\emtx&&\eta_1&=&\bmtx{cc}0&1\\1&0\emtx\\
\eta_2&=&\bmtx{cc}1&0\\0&-1\emtx&&\eta_3&=&\bmtx{cc}1&0\\0&1\emtx={\bf 1}\\
\endary
are the real Pauli matrices (RPMs). $\eta_0=(\y_0)_{2\times2}$ is the SUM
for a single DOF. Hence the RPMs provide a parameterization which perfectly
matches the symmetries relevant to Hamiltonian dynamics:
\begeq
{\bf K}=\bmtx{cc}c+s_2&s_0+s_1\\s_1-s_0&c-s_2\emtx\,.
\endeq
An analysis of the properties of these matrices reveals that
\begary{rclp{5mm}rcl}
(\eta_0)^2&=&-{\bf 1}&&(\eta_1)^2&=&{\bf 1}\\
(\eta_2)^2&=&{\bf 1}&&(\eta_3)^2&=&{\bf 1}\\
\endary
Furthermore one finds that these three non-trivial matrices mutually
anti-commute, i.e. for $i,j\in\,[0,1,2]$:
\begeq
\eta_i\eta_j+\eta_j\eta_i=2\,\rm{Diag}(-1,1,1)\,.
\endeq
All individual RPMs are either symmetric or skew-symmetric, they
either pairwise commute or anti-commute, they square to $\pm{\bf 1}_2$ and
they are either Hamiltonian or skew-Hamiltonian, symplectic
(Eq.~\ref{eq_symplectic}) or skew-symplectic (Eq.~\ref{eq_symplectic}).
Their trace vanishes except for the unit matrix.
The relevant symmetries of the Pauli algebra are given by
\begary{rcl}
\eta_i\,\eta_j&=&\pm\,\eta_j\,\eta_i\\
\eta_i^2&=&\pm\,{\bf 1}\\
\eta_i&=&\pm\eta_i^T\\
\rm{Tr}(\eta_i)&=&0\,\,\,{\rm unless }\,\eta_i={\bf 1}\\
\label{eq_rpmsym}
\endary
Note the math fact that skew-symmetric matrices $\eta_i$ square to 
$-{\bf 1}$ while symmetric matrices $\eta_i$ square to ${\bf 1}$~\cite{osc_paper}. 
The signature (the sign of the trace of the square) of the Pauli 
matrices corresponds to their symmetry under matrix transposition.

The type of transformation that these matrices generate (Eq.~\ref{eq_sol1}) 
is the matrix version of Euler's formula $e^{i\phi}=\cos{\phi}+i\,\sin{\phi}$:
\begary{rccp{5mm}rcr}
\exp{(\eta_i\,\phi)}&=&\cos{\phi}+\eta_i\,\sin{\phi}&\rm{for}&\eta_i^2&=&-{\bf 1}\\
\exp{(\eta_i\,\phi)}&=&\cosh{\phi}+\eta_i\,\sinh{\phi}&\rm{for}&\eta_i^2&=&+{\bf 1}\\
\label{eq_euler}
\endary
Formally trigonometric functions belong to rotations while the hyperbolic functions 
belong to boosts~\cite{Gourgoulhon,lt_paper}. 
Hence Eq.~\ref{eq_euler} suggests that it is thinkable to derive 
the Lorentz transformations directly from the Hamiltonian algebra:
Without presuming the notion of Minkowski space-time, the Hamiltonian algebra
of proper time automatically allows to derive the mathematical means to
describe space-times of the Minkowski type~\cite{lt_paper}.

Note that only the transformation matrix for rotations is symplectic {\it and} 
orthogonal, while for boosts it is only symplectic. The matrix algebra of a single 
DOF is the real Pauli algebra. Since we derived the significance of the real Pauli 
matrices (RPMS) from {\it classical Hamiltonian theory}, also the Pauli algebra can
- as such - not be quantum. Instead it is obtained by a Hamiltonian constraint
and the analysis of algebraic symmetries.

\subsection{The Kronecker Product and Hamiltonian Clifford Algebras}
\label{sec_hca}

Two methods to generalize the Pauli algebra are possible:
One can either {\it add} more DOFs and analyze the properties of
Hamiltonian systems with two, three, four DOF and so on, {\it or}
one may use a multiplicative approach based on the Kronecker product. 
The next system, constructed from an additive approach has two DOF and
requires the use of the real $4\times 4$ matrices, i.e. the real Dirac 
algebra. Three DOF would require $6\times 6$-matrices and one can 
anticipate that the natural symmetries inherited from the real Pauli 
matrices will be broken.

The multiplicative generalization is based on Kronecker (or tensor) products. 
The Kronecker product of two Pauli matrices ${\bf A}=\{a_{ij}\}$ and ${\bf
  B}=\{b_{kl}\}$ is given by:
\begary{rcl}
{\bf C}&=&{\bf A}\otimes{\bf B}=\bmtx{cc}a_{11}{\bf B}&a_{12}{\bf B}\\a_{21}{\bf B}&a_{22}{\bf B}\emtx\\
&=&\bmtx{cccc}
a_{11}b_{11}&a_{11}b_{21}&a_{12}b_{11}&a_{12}b_{12}\\
a_{11}b_{12}&a_{11}b_{22}&a_{12}b_{21}&a_{12}b_{22}\\
a_{21}b_{11}&a_{21}b_{12}&a_{22}b_{11}&a_{22}b_{12}\\
a_{21}b_{21}&a_{21}b_{22}&a_{22}b_{21}&a_{22}b_{22}\\
\emtx\,,
\endary
i.e. the Kronecker product is a method to systematically write down all 
possible products between all elements of ${\bf A}$ and ${\bf B}$, respectively.
The general rules of Kronecker matrix products are~\cite{MatrixAlgebra}:
\begary{rcl}
({\bf A}\otimes{\bf B})^T&=&{\bf A}^T\otimes{\bf B}^T\\
{\bf A}\,\otimes\,({\bf B}+{\bf C})&=&{\bf A}\,\otimes\,{\bf B}+{\bf A}\,\otimes\,{\bf C}\\
({\bf A}\,\otimes\,{\bf B})\,({\bf C}\,\otimes\,{\bf D})&=&{\bf A}\,{\bf C}\otimes\,{\bf B}\,{\bf D}\\
\mathrm{Tr}({\bf A}\,\otimes\,{\bf B})&=&\mathrm{Tr}({\bf A})\,\mathrm{Tr}({\bf B})\\
({\bf A}\otimes{\bf B})^{-1}&=&{\bf A}^{-1}\otimes{\bf B}^{-1}\\
\label{eq_Kronecker}
\endary
It is straightforward to verify that Kronecker multiplication preserves the 
symmetries of Eq.~\ref{eq_rpmsym}. 

The Kronecker product allows to construct all Clifford algebras (CAs) with real 
matrix representations from the real Pauli matrices, including all Hamiltonian 
Clifford algebras (HCA)~\footnote{As we started out from a conservation law, 
we are {\it specifically} interested in {\it Hamiltonian} Clifford algebras, 
and in {\it Clifford} algebras mainly as they preserve the dynamical 
symmetries of symplectic theory, but we are not specifically interested 
in Clifford algebras {\it as such}.}. 
Obviously there is exactly one algebra that follows both rules, which is 
the only additive {\it and} multiplicative generalization of the Pauli algebra, 
namely the real Dirac algebra.

Before we discuss the Dirac algebra, we shall first give a (very)
brief introduction to CAs as they are usually presented, i.e. without 
reference to Hamiltonian theory, and explain our motivation to
restrict us to Clifford algebras with (irreducible) real matrix representation. 

\section{Clifford Algebras}
\label{sec_clifford}

{\it Mathematically} Clifford algebras can be defined as generated by
a list ${\bf e}_k\,,k\in[0\dots N-1]$ of $N$ pairwise anticommuting elements 
that hold ${\bf e}_k^2=\pm{\bf 1}$. These are called the {\it generators} of the
Clifford algebra. A Clifford algebra that consists of $p$ generators that 
square to ${\bf 1}$ and $q=N-p$ generators that square to $-{\bf 1}$, 
is denoted by $Cl(p,q)$. It has a signature (or ``metric tensor'')
$g_{\mu\nu}=\rm{Diag}(1,\dots,1,-1,\dots,-1)$ with $p$ positive and $q$
negative entries in the diagonal and has dimension $N=p+q$. 

The $N$ generators can be used to obtain new elements by multiplication
since products of two (or more) {\it different} generators ${\bf e}_i{\bf e}_j$ 
are new elements, different from the unit element and from each factor. 
It is easy to see that they square to $\pm 1$ as well. It follows from
combinatorics that there are $\left({N\atop k}\right)$ products of
$k$ generators. These products are called $k$-vectors, so that one has
\begeq
\sum_k \left({N\atop k}\right)=2^N
\endeq
Clifford algebraic unit elements ($k$-vectors) in total. 
Hence $m\times m$ matrix representations require at least
the same number of independent parameters, so that $m^2\ge 2^N$ or $m\ge 2^{N/2}$.
An isomorphism between a given CA and some corresponding matrix algebra
is only possible for even $N$ with $m=2^{N/2}$.

Clifford algebras are used in various branches of physics,
but the Dirac algebra is of special interest since, firstly,
it is the algebraic kernel of QED, secondly, it describes the
coupling between two Hamiltonian canonical pairs,
which places it at the foundation of classical phase space and 
thirdly because it matches the geometry of $3+1$-dimensional 
space-time, electromagnetism and the Lorentz
transformations~\cite{STA,SLAC4237}.

As we elaborated in some detail in Ref.~\cite{lt_paper}, the
real Clifford algebra $Cl(3,1)$ provides the most compact form
to represent the Lorentz covariance of energy, momentum and
the electromagnetic field.
The commutator table of this Dirac algebra determines the 
form of the electromagnetic field tensor~\cite{rdm_paper}.
Furthermore the (anti-) commutation properties enable to naturally 
explain the vector cross product and therefore to describe the 
handedness of space~\cite{lt_paper}.

Certainly CAs are an interesting mathematical topic in their own right
and Hestenes has shown, that it is possible to give a presentation of 
Dirac's theory without any specific matrix representation~\cite{STA}.
From a general mathematical point of view, the question of
a matrix representation can be outsourced into a special branch of 
mathematics called {\it representation theory}, which is then another 
interesting topic in its own right. 
It is a quite common approach to abstract CAs from their respective 
matrix representations. And of course CAs do not need Hamiltonian theory 
to be interesting and useful. They generate geometric spaces even if 
they are not considered in context of a Hamiltonian phase space.

But within the deductive approach of our presentation, the representation
by real matrices is an {\it indispensable} element in which matrix
transposition allows to distinguish Hamiltonian from skew-Hamiltonian elements and to analyze 
their algebra (Eq.~\ref{eq_cosy_algebra}). This is essential for the {\it motivation} 
to consider Clifford algebras {\it at all} in our approach~\footnote{
For a discussion of real CAs in the context of linear Hamiltonian theory 
see~\cite{rdm_paper,geo_paper,osc_paper,lt_paper}. We would have prefered 
to also cite other authors in this context, but haven't found any.
In the standard Refs.~\cite{Chessboard,Lounesto}, for instance, the 
name ``Hamilton'' is mentioned exclusively in context of quaternions.}. 
 
\subsection{The Complex Numbers}
\label{sec_complex}

Consider for instance the case $N=1$ in which we have, besides the unit
element ${\bf 1}$ only one single non-trivial element ${\bf e}_0$, which
either gives the algebra $Cl(0,1)$ or $Cl(1,0)$. In context of Hamiltonian
mechanics, any considered algebra {\it must} contain the symplectic unit
matrix (SUM) to establish Hamilton's equations of motion and hence $Cl(1,0)$
can not be fundamental since it does not allow for a skew-symmetric
unit element. 

Hence, because the SUM squares to $-{\bf 1}$, the corresponding ``Clifford number'' 
(CNs) ${\bf x}$ has the form ${\bf x}=a\,{\bf 1}+b\,{\bf e}_0$. 
The multiplication of two CNs is then
\begary{rcl}
{\bf x}\,{\bf y}&=&(a_x\,{\bf 1}+b_x\,{\bf e}_0)(a_y\,{\bf 1}+b_y\,{\bf
  e}_0)\\
&=&(a_x\,a_y-b_x\,b_y)\,{\bf 1}+2\,(a_x\,b_y+b_x\,a_y)\,{\bf e}_0\,.
\endary
This is the product of two complex numbers, which are hence almost 
identical to the Clifford algebra $Cl(0,1)$. We say ``almost'', because
the theory of complex numbers knows the operation of complex conjugation,
which has no correspondence in $Cl(0,1)$ {\it unless} we refer to a real
matrix representation, where complex conjugation can be naturally obtained
by matrix transposition.
Since ${\bf e}_0=\eta_0$ is the (only) skew-symmetric element and complex 
conjugation is identical to matrix transposition, so that in the conventional 
notation $z=x+i\,y$ is
\begeq
z=\bmtx{cc}x&y\\-y&x\emtx
\endeq
and 
\begeq
z^\star=x-i\,y=z^T=\bmtx{cc}x&-y\\y&x\emtx\,.
\endeq
This means that transposition and complex conjugation can not be properly
distinguished with full logical rigour. One might also substitute matrix
transposition by a multiplication with the signature of the corresponding
Clifford $k$-vectors. But the signature is only uniquely determined, if
exclusively real representations are used.

It also means that the complex numbers are, regarded from our perspective, 
a special case of (the algebra of) $2\times 2$-matrices. It can be shown
(Ref.~\cite{osc_paper}) that the reduction of the real Pauli algebra to 
the algebra of the complex numbers corresponds to the reduction to the 
general LOM of a DOF to the normal form of an harmonic oscillator~\cite{MHO,NormalForms}. 
It is well known in many branches of physics, for instance in accelerator 
physics, that the unit circle in the complex plane is the normal form 
trajectory of the motion of a single DOF so that the complexity of the 
actual state of affairs can be reduced to a single number, namely 
the ``phase advance'' (i.e. time)~\cite{Wiedemann}.

The analysis of normal forms is of course a useful mathematical technique, 
but one should keep in mind, that it describes the system in a
special coordinate system and that coordinate transformations have a 
two-fold meaning. They can be understood as {\it passive} transformations
and, regarded this way, they just concern our mathematical methods to simplify
the description of a given physical problem. But symplectic transformations also 
describe the full space of physical possibilities, of possible evolutions 
in time. This space of possibilities is substantially narrowed if we 
restrict our math to the use of normal forms {\it only}.

\subsection{The Unit Imaginary and the Dirac Equation}

\bquo
Thus a complex Hilbert vector is not a more general kind of 
quantity than a real one. A real Hilbert space is the more 
elementary concept. A complex Hilbert space should be looked 
upon as a real one in which a certain structure is introduced, 
namely a pairing of the coordinates, each pair being then 
considered as a complex number. Changing the phase factors of 
these complex numbers then provides a special kind of rotation 
in the Hilbert space.\\
\mbox{}\hfill{\it -- P.A.M. Dirac}~\cite{Dirac74}
\equo

Foreshadowing what we are going to argue below, let us remark that it is 
not as nearby as often suggested to consider representations ``over'' the 
complex numbers as fundamental~\footnote{See also Ref.~\cite{KMG}:
``[...] complex numbers are not required in order to describe quantum 
mechanical systems and their evolution [...]''.}.
Freeman Dyson wrote~\cite{Dyson}
\bquo
probably all these connections would have been clarified long ago,
if quantum physicists had not been hampered by a prejudice in favor
of complex and against real numbers
\equo
There is reason to suspect that the incautious use of complex (or quaternionic) 
``numbers'' substantially contributes to the scrambling of the ``quantum 
omelette''. To understand this point correctly is indispensable for a successful 
unscrambling. Let us therefore spend a few words on it.

It is part of the logic of quantum theory that the Schr\"odinger equation 
is the non-relativistic {\it approximation} of the Dirac equation~\cite{Messiah2}.
Hence the Dirac equation must be regarded as more fundamental compared to the 
Schr\"odinger equation and is the {\it logical} basis of quantum theory. 
Nonetheless, discussions about the interpretation of QM rarely refer to 
Dirac's theory. As Hestenes noted, ``[it] has long puzzled me is why Dirac 
theory is almost universally ignored in studies on the interpretation of quantum
mechanics, despite the fact that the Dirac equation is widely recognized
as the most fundamental equation in quantum mechanics''~\cite{Hestenes}.

Dirac's theory is required to provide Lorentz covariance, to explain the spin, 
the gyromagnetic ratio and essential parts of the hydrogen spectrum. 
The Dirac equation is the basis of QED and QFT. Furthermore both, special
relativity and quantum mechanics, are required to solve problems
inherited from Maxwell's electrodynamics, the former to provide Lorentz
covariance and the latter to solve the problem of the stability of atoms.
It is true that Schr\"odinger's equation is often more useful than
Dirac's equation in order to solve textbook problems, but it is the wrong
reference when one aims to understand the fundamentals of QM. It is therefore 
annoying that students must first and sometimes exclusively undergo the 
``brainwashing''~\footnote{Murray Gell-Mann is quoted with the following words:
``The fact that an adequate philosophical presentation [of quantum physics]
has been so long delayed is no doubt caused by the fact that Niels Bohr 
brainwashed a whole generation of theorists.''~(See page 253 in Ref.~\cite{Becker}).}
of textbook QM before they have the chance to understand that the 
{\it formalism}, at least the part that can be understood, is fully
classical~\footnote{The author learned about 
the fact that the {\it dynamics} governing the time evolution of quantum theory is 
essentially {\it classical} for the first time in 1998 or 1999, in 
a talk given by John Ralston at DESY in Hamburg. The title of the talk was 
``Spin and the well-dressed Quark''. Like F. Strocchi in 1966, Ralston 
argued that the time-dependent part of Schr\"odinger's equation is identical 
to Hamilton's equations of motion~\cite{Strocchi,Ralston1989}. Many other
authors found classical structures within quantum 
theory~\cite{Heslot,Gray1994,Gosson2001,GossonHiley} and/or vice versa.}. It is an oddity
that most textbooks on quantum theory do not treat the Dirac equation at all -- or 
just briefly as a kind of addendum, typically in the last chapter or in the 
second volume. But quantum theory can not be fully understood using Schr\"odinger's
equation alone.

The Dirac equation, however, {\it allows for} but it does not {\it require} 
the explicite use of the unit imaginary. 
This becomes obvious from the fact that the complex Clifford
algebra $Cl(1,3)$ can be directly replaced with the real Clifford algebra
$Cl(3,1)$, just by letting the $\y$-matrices ``absorb'' the notorious unit imaginary. 
Due to Pauli's fundamental theorem of the Dirac matrices~\cite{Pauli} 
any choice of Dirac matrices that generates the same metric, can be obtained 
by similarity transformations from any other. Hence all representations are 
physically equivalent. Therefore one can always use the (purely imaginary) 
Majorana matrices. If we denote the generating elements of the real algebra 
$Cl(3,1)$ by $\y_\mu$ and those of a complex CA $Cl(1,3)$, using the Majorana 
basis, by $\Gamma_\mu$, then $i\,\Gamma_\mu=\y_u$ and we are done. 
The ``complex'' version:
\begeq
(i\,\Gamma_\mu\d_\mu\pm m)\psi=0
\endeq
becomes
\begeq
(\y_\mu\d_\mu\pm m)\psi=0
\label{eq_direac_real}
\endeq
with purely real matrices $\y_\mu$. Since this is just a question of notation,
it can by no means imply different physics~\footnote{
Of course, the use of $Cl(3,1)$ instead of $Cl(1,3)$ is accompanied
with the use of the of the so-called ``east coast metric'' (ECM) instead
of the ``west coast metric'' (WCM)~\cite{Woit}. But though the WCM is
used more frequently, both choices are physically equivalent within the SPQM.
As mentioned by Woit, Weinberg preferred the ECM in his presentation of
QFT~\cite{Weinberg}.
A third notational convention that writes time as a kind of imaginary 
fourth coordinate in the form of $ds^2=dx_1^2 +dx_2^2+dx_3^2+dx_4^2$ with
$dx_4=i\,c\,dt$ has also been used, for instance by Sommerfeld in 
Ref.~\cite{Sommerfeld} and by Einstein in Ref.~\cite{Essays}.}.

The algebraic form that is obtained in energy-momentum space is an
eigenvalue equation
\begeq
(\y_0\,{\cal E}+p_x\,\y_1+p_y\y_2+p_z\,\y_z\pm i\,m)\psi=0\,,
\endeq
where $\lambda=\pm\,i\,m$ is an eigenvalue of the matrix
\begeq
{\bf H}=\y_0\,{\cal E}+p_x\,\y_1+p_y\y_2+p_z\,\y_3\,,
\endeq
such that a positive mass corresponds to a purely imaginary 
eigenvalue $\lambda$, as required in stable linear Hamiltonian 
systems~\cite{MHO}. Moreover, the matrix ${\bf H}$ is a Hamiltonian
matrix and the Dirac equation is therefore just a special case of
Eq.~\ref{eq_lom2}, hence it is {\it mathematically} but classical 
linear Hamiltonian theory, applied to a new kind of fundamental 
variables $\psi$.

This is the proof that the Dirac equation (in momentum space) is
{\it as such} not quantum, as it can be obtained classically.
It will be discussed in more detail below. The use of the unit imaginary
is pure notational convention and does not make it quantum either.
The use of scaling factor like $\hbar$ 
also can not be quantum~\cite{RalstonHBar}. 
Since the spin is an original result of Dirac's theory, also spin 
is not quantum. Hence, if the more general equation, namely the
relativistic equation of Dirac, is not quantum, then the non-relativistic
approximation, Schr\"odinger's equation, can't be quantum either.

\subsection{Mathematics, Physics and Logic}

In our view it is important to understand that, though physics requires the use
of math (and mathematical logic), mathematics does not require physics and can
not replace physical logic: even if many tools developed in pure mathematics 
turned out to also be useful in physics, they usually have not been {\it designed} 
for this purpose. Physicists have to {\it re-design} and to select the mathematical 
tools that serve their needs best. And they need to address the question of the
underlying logic in a substantially different way. Mathematics is {\it as such} not
a natural science. There is no need, within mathematics, to explain {\it why}
one starts with certain set of assumptions. Mathematical ``assumptions'' are
mostly definitions and they imply no statement about reality. They do not say
what the world is like, they spell out the logical consequences of certain initial 
definitions~\footnote{This was not always acknowledged in history of science. 
Before the ``discovery'' of non-Euklidean geometries it was assumed that geometry 
is {\it by nature} Euklidean. But this is wrong. {\it If} Euklidean geometries
have a special logical significance, then this derives from logical or mathematical 
arguments and can not be argued from {\it presumptions about nature}.}.
Thus, for mathematics, it suffices to present
the axioms (initial assumptions) of a theory. But in physics, it does not suffice
to elaborate what {\it would} follow from ``A'' if ``A'' {\it was} true.
Physics aims for a description of what actually {\it is} the case and why. 
Newtonian physics was falsified by the discovery that its axiomatic basis fails
to provide the correct energy-momentum relation. This is usually attributed to 
the fact that space-time is non-Euklidean, but this does not ``falsify'' Euklidean 
geometry as a mathematical theory~\footnote{Not everyone is troubled by the claim 
that ``space-time'' is usually presented as some kind of ``thing'' with properties, 
properties that are somehow supposed to cause particles to have a specific 
energy-momentum relation. The reasoning presented here allows for a different view
but we will not elaborate on the philosophical consequences.}.

The contemporary presentations of quantum mechanics can be devided roughly into
two major groups. The first groups favours a ``classical'' phenomenological introduction
into QM by the description of core experimental findings. The second one is an
axiomatic approach, often considered to be ``modern''. Both are unsatisfactory.
The phenomenological approach is poor science, because it simply ignores that 
physics is at its core a deductive science and not an unsorted collection of 
(possibly surprizing) phenomena that we try to describe somehow by some mathematical 
patchwork. On the other hand, the axiomatic approach is, due to the poor quality 
of the ``axioms'', not much better: ``[...] the traditional meaning of axiom [is]
a self-evident first proposition, which is neither provable nor in need of a 
proof''~\cite{Pulte}. The claim that a ``physical system'' is represented by a
vector in a separable Hilbert space is, without the detailed explanation given
above, not of that kind. It is neither simple nor basic and certainly
not self-evident.

Of course, one {\it could} formulate an axiom stating that
planetary orbits are conic sections. One {\it can} do that but for what
benefit? Newton's law of gravitation is by far simpler and it explains
{\it why} planetary orbits are conic sections. Equivalently it is poor science
to introduce Hilbert space in an axiomatic way: As we have shown above, the
requirement of positive definiteness and normalizability allow to explain the
prominent role that Hilbert space plays in QM. That is: positive definiteness
(of ``substance'') and finiteness (normalizability) are indeed self-evident
first propositions, which are neither provable nor in need of a proof. And
they allow, combined with the classical assumptions of continuity and causality, 
to derive Schr\"odinger's equation.

The PSR suggests the equivalence of all variables in $\psi$, until it turns out 
to be inevitable to formally break this equivalence by the 
introduction of the skew-symmetric matrix $\y_0$, the symplectic unit matrix. 
The skew-symmetry of $\y_0$ suggested to {\it formally} introduce canonical 
pairs and Hamilton's equations of motion.
This is the reason why Schr\"odinger's non-relativistic equation must be
complex: It needs to implement a canonical pair to generate a constant
of motion~\cite{Strocchi,Ralston1989}. We insisted on real-valued dynamical
variables $\psi$ and PDCOM ${\cal H}$, but this does not imply to abandon the 
use of the complex numbers {\it as such}. We merely argue against the view that
they have a specific prominence for the distinction between the ``classical''
and the ``quantum'' realm.

As stressed before, stable Hamiltonian systems have purely imaginary 
eigenvalues and complex eigenvectors. This is accepted and known from classical 
mechanics and cannot be regarded as an argument for philosophical conclusions
against the reality of the dynamical quantities. What is unacceptable in a 
classical setting, however, is the {\it a priori} use of complex numbers for the 
coordinates $\psi$ or the conserved quantity ${\cal H}$. As we have shown, it is not 
{\it required} by the math of QM: the unit imaginary does not generate quantumness.
We can not prevent anyone from using a suggestive notation, but 
we doubt that the use of a specific notation is {\it physically} relevant.
The only logical constraint for $\psi$ is an even number of variables. 
This alone does not {\it require} the use of complex numbers, even if it 
might be very convenient to use them.

The use of complex numbers for $\psi$ and ${\cal H}$ is not wrong 
{\it as such}, but in the context of Dirac's theory it wrongly suggests 
the physical significance of sixteen matrices, while the restriction 
to the reals enables to properly distinguish between ten Hamiltonian and six 
skew-Hamiltonian components. Also Dirac found and discussed only ten
generators (and not sixteen)~\cite{Dirac63}. The six skew-Hamiltonian elements
appear, within the SPQM, in the theory of the weak interaction (``axial
currents''), but this theory gives no explanation why the weak interaction
is {\it weak}. The distinction between Hamiltonian and skew-Hamiltonian
terms however could explain why axial currents are weaker than vector currents
and why free magnetic monopoles don't exist.

If intended or not, it seems that the complex notation mainly mystifies the 
role of the wave-function. But as Ralston argued, there is little in 
quantum theory that proves the non-reality of the wave-function~\cite{RalstonBook}: 
``Bohr and Heisenberg had made up their minds about a philosophy of unreality 
before the actual quantum theory existed.'' Mara Beller has given a distinguished 
account of the history of the ``making'' of this ``scientific revolution''~\cite{Beller}.

\subsection{Hamiltonian Clifford Algebras}

The usefulness of Clifford algebras in the context of Hamiltonian
theory is due to Eqs.~\ref{eq_cosy_algebra}.
Given we have a set of $N$ anti-commuting matrices $\y_k\,,k\in[1..N-1]$ 
and the SUM $\y_0$ generated by (repeated) Kronecker multiplication of the 
real Pauli matrices, then the matrix system has a dimension $2^N=2^m\times 2^m$
~\footnote{Latin indices $\y_k$ denote a range $k=[1,\dots,N-1]$, greek indices 
$\y_\mu$ a range $\mu=[0,1,\dots,N-1]$.}:
It follows that also all real Dirac matrizes are either Hamiltonian or skew-Hamiltonian: 
\begary{rcl}
\y_0\,\y_k\,\y_0&=&\pm\,\y_0\,(\y_0\,\y_k)\\
               &=&\pm\,\y_k\\
               &=&\pm\,\y_k^T\\
\label{eq_rpmcosy}
\endary
Then any matrix $\y_\mu$ that anti-commutes with $\y_0$, holds 
\begary{rcl}
\y_0\,\y_\mu\,\y_0&=&\y_0\,(-\y_0\,\y_\mu)\\
                 &=&\y_\mu\\
\label{eq_x0}
\endary
(since $\y_0^2=-{\bf 1}$) and is therefore either Hamiltonian and symmetric 
or skew-Hamiltonian and skew-symmetric. This connection between the
different symmetries is of severe importance for the theory of Hamiltonian 
Clifford algebras and has consequences for the general description of $n$ DOF.
It is specifically the mixture of the properties of Clifford algebras and
Hamiltonian constraints (Eq.~\ref{eq_cosy_algebra}), that produces new rich
and complex structures.

As the generators of Clifford algebras all anti-commute (by definition), 
Eq.~\ref{eq_x0} is such a constraint: In systems in which all generators 
of the Clifford algebra are also generators of symplectic motion
(i.e. Hamiltonian), the metric tensor necessarily has the form
$g_{\mu\nu}=\rm{Diag}(1,1,\dots,1,-1)$ and the Clifford algebra has dimension
$Cl(p,q)$ with $q=1$ and $p=N-1$. Hence the formalism reproduces the fact 
that {\it time is unique}.

The use of Hamiltonian {\it Clifford} algebras (and not merely Hamiltonian algebras)
for the parametrization of even moments in phase space provides maximal symmetry with 
respect to the individual variables in $\psi$ as well as with respect to the 
individual HCPs~\footnote{The Hamiltonian algebra of $6\times 6$-matrices as it is
usually associated with the classical motion of particles in ``physical'' 
space, has fewer symmetries.}.
This naturally conforms with the requirement of the PSR to treat all elements on 
equal ground unless there is reason to do otherwise: the matrix representation 
of all $k$-vector elements $\y_A$ of any real Clifford algebra has one (and only one) 
entry of $\pm 1$ in each row and each column while all other elements are identically 
zero.

Bott's periodicity theorem taught us that Hamiltonian Clifford algebras $Cl(p,q)$, 
i.e. CAs with real representations, exist only for~\cite{Okubo}
\begeq
p-q=0,\,1,\,2\,\rm{ mod }\,8\,,
\endeq
which, since $N=p+q$ must be even, reduces in our case to
\begeq
p-q=0,\,2\,\rm{ mod }\,8\,.
\label{eq_bott0}
\endeq
From Eq.~\ref{eq_x0} we derived that, if all generators of $Cl(p,q)$ are
Hamiltonian (i.e. ``observable''), then one has $q=1$ and $p=N-1$ so that
\begeq
N-2=0,\,2\,\rm{ mod }\,8\,.
\label{eq_bott}
\endeq
This establishes a distinction between those dimensionalities that
allow for a representation by Hamiltonian Clifford algebras and those
that do not. The former are listed in Tab.~\ref{tab_dim}. The simplest
algebras are the real Pauli algebra $Cl(1,1)$, and the real Dirac 
algebra $Cl(3,1)$.

According to Eq.~\ref{eq_bott} there are two sequences of HCAs, given by
\begeq
N=8\,m+2
\label{eq_bott1}
\endeq
which we call a HCA of the {\it Pauli type} and 
\begeq
N=8\,m+4
\label{eq_bott2}
\endeq
which we call a HCA of the {\it Dirac type} where $m\in\mathbb{N}$.
\begin{table}
\begin{tabular}{|c||c|c|c|c|c|c|c|}\hline
Type &  m   & 0 & 1 & 2 & 3 & 4 & 5\\\hline
Pauli&$N=8\,m+2$ & 2 & 10 & 18 & 26 & 34 & 42\\\hline
     &p+q & 1+1&9+1&17+1&25+1&33+1 & 41+1\\\hline\hline
Dirac&$N=8\,m+4$ & 4 & 12 & 20 & 28 & 36 & 44\\\hline
     & p+q & 3+1&11+1&19+1&27+1&35+1 & 43+1\\\hline
\end{tabular}
\caption[Possible dimensionalities of HCAs]{
\label{tab_dim}
Possible dimensionalities of HCAs in which all generators
of the Clifford algebra are Hamiltonian matrices and therefore
correspond to non-vanishing (and hence ``observable'') 
auto-correlations of Hamiltonian spinors $\psi$.
}
\end{table}
If the dynamical significance of CAs and hence the requirement
of a real representation is ignored, then CAs can be defined
for practically any dimensionality. In a purely mathematical 
setting, this might be an interesting generalization, but in 
a physical context it is the fast road to dynamical
misconceptions.

In order to understand the logical and dynamical properties of
the individual elements of HCAs of the mentioned dimensionalities, 
it is important to notice that all generators except the SUM $\y_0$
are symmetric real matrices. Since we have no specific argument to 
prefer any of them, it is nearby to consider the role of matrices 
that play a special role by their formal position within the 
CA. Besides the SUM $\y_0$ and the unit matrix ${\bf 1}$, any CA 
has two unique elements, the first being the $N$-vector $\y_\pi$, 
which is the product of all generators 
\begeq
\y_\pi=\prod\limits_{\mu=0}^{N-1}\,\y_\mu
\label{eq_PS}
\endeq
and which is called {\it pseudo-scalar}. 

Since $N$ must be an even integer, the pseudo-scalar anti-commutes with 
all generators (vector elements), it therefore commutes with all 
$2$-vectors and anti-commutes with all $3$-vectors and so forth: 
The pseudo-scalar distinguishes even from odd $k$-vectors. Therefore 
the pseudoscalar of the Dirac algebra induces {\it charge conjugation}, 
namely a change of sign of the bi-vectors only (see Sec.~\ref{sec_rdm} below).

As derived in App.~\ref{sec_PS}, the pseudoscalar of Pauli type
HCAs is Hamiltonian and symmetric while in Dirac type algebras it 
is skew-Hamiltonian and skew-symmetric.

Furthermore, as shown in App.~\ref{sec_PS}, in both, the Pauli type
and the Dirac type algebras, only $k$-vectors with 
$k=1,2,5,6,9,10,\dots$ are Hamiltonian while $k$-vectors with 
$k=3,4,7,8,11,12,\dots$ are skew-Hamiltonian (App.~\ref{sec_PS}).

Another special element is the product $\y_0\,\y_\pi$, which is the
product of all Clifford generators except $\y_0$.
This operator anti-commutes with the SUM and $\y_\pi$, but commutes with 
all other generators of $Cl(N-1,1)$. 
It can hence distinguish between the two types of Clifford generators
and is part of the CPT-theorem~\cite{qed_paper}.
 
\subsection{The real Dirac algebra}
\label{sec_rdm}

Since the smallest dynamical system with some kind of {\it internal} 
dynamics, with interaction, is composed of two HCP and described 
by the Dirac algebra, the Dirac algebra is as fundamental as the
real Pauli algebra.

The usefulness of the real Dirac algebra in {\it classical} Hamiltonian 
theory stems from the fact that it provides a framework to describe the
general (de-) coupling of two HCPs~\cite{rdm_paper,geo_paper,jacobi_paper}.
For systems with more than two canonical pairs, a general block-diagonalization
of stable Hamiltonian matrices can be achieved with a Jacobi type iterative
algorithm: In each step, two DOF are blockdiagonalized based on symplectic 
similarity transformations using the real Hamiltonian Dirac matrices as 
generators~\cite{geo_paper,stat_paper,jacobi_paper}. Hence the real Dirac 
algebra suffices to describe all {\it possible} linear interactions between 
two (and more) HCPs. The (real) Dirac algebra is fundamental 
for the description of dynamical coupling in the same way as $2\times 2$ 
matrices are fundamental to describe rotations in a plane.

In the previous section we derived the conditions for a possible isomorphism
between real matrix reps of CAs $Cl(p,q)=Cl(N-1,1)$ and Hamiltonian algebras. 
Note that the size of the spinor that corresponds to $Cl(N-1,1)$ is 
$2\,n=2^{N/2}$ so that $Cl(9,1)$ corresponds to a spinor of size $2\,n=2^5=32$
and an algebra with $2^N=1024$ elements, $n(2n+1)=528$ Hamiltonian and
$496$ skew-Hamiltonian elements. These numbers alone clearly indicate that
$Cl(9,1)$ can not represent the simplest possible RPO. But there are 
more reasons why a RPO must be composed of two HCPs, which are discussed 
elsewhere~\cite{qed_paper,osc_paper}. Here we restrict us to a short summary 
of the main points: we stress again that the Dirac algebra with 
$4\times 4$-matrices is the minimal size required to represent the 
general case of complex eigenvalues. And there is no fundamental reason for 
nature to exclude those types of dynamical processes that require, maybe 
for a short time, complex eigenvalues; they belong to the full scope of 
possibilities.

Secondly, if the number of variables in the spinor is supposed to 
correspond to the number of variables representing the RPO, then 
\myarray{
2\,n&=N=2^{N/2}\\
N^2&=2^N\\
}
which has only two solutions, namely $N=2$ or $N=4$.
And thirdly, as we shall elaborate now, the system of Clifford 
generators should allow to completely determine the structure of the 
algebra, without ambiguity, and hence provide the basic web of 
physical notions uniquely.

Regarding the real Dirac algebra $Cl(3,1)$, one has the following
unique elements: The SUM $\y_0$, the pseudoscalar (Eq.~\ref{eq_PS})
$\y_\pi=\y_0\y_1\y_2\y_3$ and the product of both ($\y_{10}$).
While there is only a single skew-symmetric element in the real Pauli algebra, 
the Dirac algebra contains six of them. 
By Pauli's fundamental theorem of the Dirac matrices~\cite{Pauli} it is
allowed to select any of the skew-symmetric matrices to represent the SUM $\y_0$. 
Above we have chosen the form ${\bf 1}_2\otimes\eta_0$.
If one choses to use a different skew-symmetric matrix as SUM, 
this is equivalent to a permutation of the order of the elements in $\psi$.
One obtains $\psi=(q_1,q_2,p_1,p_2)^T$ in case of $\y_0=\eta_0\otimes{\bf 1}_2$.
Hence the real Dirac matrices have their meaning {\it relative} to the initial 
choice of the SUM.

Next one has to select one of nine symmetric matrices~\footnote{
The unit matrix can not be a generator of a CA since it commutes with all
others.}, however it is a math fact that $\y_0$ anticommutes only with 
six of them. Hence one has to choose again one of six matrices~\footnote{
We can't tell if ``god throws dices'' or not. Here dices might be an option.} 
and fix it as $\y_1$. This choice is again arbitrary insofar as all choices
give the same physics~\cite{rdm_paper}. But the selection of these two matrices
suffices to decide about the type of all remaining matrices, i.e. whether
they are vectors, bi-vectors and so forth.

According to Eq.~\ref{eq_nsym} and Eq.~\ref{eq_ncosym} there
are $10$ Hamiltonian elements in the Dirac algebra, but up to now we
identified only $4$ of them, namely the generators (called $1$-vectors or simply ``vectors''). 
In the previous section we have shown that in HCAs in which all generators of
the Clifford algebra are Hamiltonian, only $k$-vectors for $k\in[1,2,5,6,9,10,\dots]$ 
are Hamiltonian. Since the highest $k$-vector of the Dirac algebra is 
the pseudo-scalar with $k=4$, a Dirac type Hamiltonian may contain only 
vector and bi-vector elements. We use $\y_{14}=\y_0\y_1\y_2\y_3$ to 
denote the pseudo-scalar.

Since $\y_0$ differs in signature from $\y_1$, $\y_2$ and $\y_3$, bi-vectors
that include $\y_0$ as a factor, are symmetric while those that do not,
are skew-symmetric. Hence the symmetry properties of the bi-vectors force
us to distinguish two groups: the first (symmetric) triple of bi-vectors is 
given by
\begary{rcl}
\y_4&=&\y_0\,\y_1\\
\y_5&=&\y_0\,\y_2\\
\y_6&=&\y_0\,\y_3\,,
\label{eq_biboost_def}
\endary
and the second (skew-symmetric) triple by
\begary{rcl}
\y_7&=&\y_{14}\,\y_4=\y_2\,\y_3\\
\y_8&=&\y_{14}\,\y_5=\y_3\,\y_1\\
\y_9&=&\y_{14}\,\y_6=\y_1\,\y_2\,.
\label{eq_birot_def}
\endary
The skew-Hamiltonian $3$-vector elements are
\begary{cccp{5mm}ccc}
\y_{10}&=&\y_{14}\,\y_0&=&\y_1\,\y_2\,\y_3&&\\
\y_{11}&=&\y_{14}\,\y_1&=&\y_0\,\y_2\,\y_3&&\\
\y_{12}&=&\y_{14}\,\y_2&=&\y_0\,\y_3\,\y_1&&\\
\y_{13}&=&\y_{14}\,\y_3&=&\y_0\,\y_1\,\y_2&&\,.
\label{eq_trivectors_def}
\endary
The last element is the scalar $\y_{15}={\bf 1}$, e.g. the unit matrix.

Any $4\times 4$-matrix ${\bf M}$ can be written as a linear 
combination of the real Dirac matrices (RDMs):
\begeq
{\bf M}=\sum\limits_{k=0}^{15}\,m_k\,\y_k
\endeq
Where a sequential index $k\in[0,\dots,15]$ is used instead of the multi-index 
convention $\y_\mu\y_\nu$. 

This means that the Dirac algebra enables, as the real Pauli algebra, 
for a general re-parametrization of the elements of $4\times 4$-matrices,
suited to symmetries relevant in abstract Hamiltonian dynamics. This is 
usually presented in wrong order: It is true that Dirac introduced his 
matrices with heuristic arguments from the relativistic energy-momentum 
relation. But, as Levy-Leblond remarked, ``the chronological building 
of order of a physical theory, however, rarely coincides with its logical 
structure.``~\cite{LevyLeblond}. Within our no-assumption-approach $3+1$
dimensional space-time follows from the structure of the Hamilton-Dirac-Clifford
algebra and not vice versa.

Since all RDMs besides the unit matrix are orthogonal and have zero 
trace, one obtains the coefficients $m_k$ by
\begeq
m_k=\frac{1}{4}\,\textrm{Tr}({\bf M}\,\y_k^T)\,.
\endeq
The general form of a Hamiltonian matrix ${\bf H}$ that couples two
canonical pairs is a linear combination of ten components,
4 vectors and 6 bi-vectors:
\begary{rcl}
{\bf H}&=&\sum\limits_{k=0}^{9}\,f_k\,\y_k\\
{\bf S}&=&\sum\limits_{k=0}^{9}\,s_k\,\y_k\\
\label{eq_symplexbygammas}
\endary
In order to simplify the calculation one may use Eq.~\ref{eq_euler} 
to analyze the result of a SST (compare Eq.~\ref{eq_sol4}):
\begeq
{\bf S}(\tau)=\exp{(\y_a\,\tau)}\,{\bf S}(0)\,\exp{(-\y_a\,\tau)}\,.
\label{eq_symtrans}
\endeq
so that, using the convention of from Eq.~\ref{eq_symplexbygammas}, the $s_k(\tau)$ 
are functions of $s_k(0)$ and $\tau$. According to Eq.~\ref{eq_euler} the 
matrix exponentials yield, (hyperbolic) trigonometric functions~\footnote{
This holds for exponentials of non-singular matrices. 
The general case is described for instance in Ref.~\cite{exp_paper}.}. 

\section{Function Follows Form}
\label{sec_fff}

We promised that the classical notions of mass, energy and momentum 
would follow from the logic of the imposed dynamical constraint.
Of course it is impossible to provide a logical proof for an 
interpretation. Interpretations can be consistent and plausible, 
but not logical or illogical. The proof that the ten coefficients
of the real Hamiltonian Dirac algebra match to the {\it structure} 
of Dirac's electron theory should not be a big surprise after all. 

As well known in classical mechanics, the generators of canonical 
transformations correspond to physical quantities (``observables''). 
We started with a single PDCOM and apparently this suffices to 
explain the emergence of $10$ quantities, that act as generators 
of SSTs within the Dirac algebra, waiting for an interpretation -- 
four vector components and $3+3$ bi-vector components. The tri-vectors, 
the scalar and pseudo-scalar are skew-Hamiltonian, they do not 
correspond to non-zero correlations, and require -- at this point
-- no interpretation. 

In Clifford algebras with $N=p+q$ being an even number,
one can construct $k$-vectors with even $k$ from products of 
$k$-vectors for $k$ even or odd, but one can not obtain $k$-vectors 
with odd $k$ from products of even $k$-vectors: bi-vectors can be 
obtained multiplicatively from vectors but not vice versa. 
In other words: the even elements, namely the $0$-vectors (scalar), 
the six bi-vectors, and the $4$-vector element (pseudoscalar) form 
the even subgroup. 

It is the structure of the algebra which forces us to distinguish 
between the set of quantities associated with the vector 
elements $(\y_0,\y_k)$ and two sets of $3$ bi-vectors each, 
namely the symmetric elements $\y_4,\y_5,\y_6$ and the skew-symmetric 
elements $\y_7,\y_8,\y_9$. Postulates are, yet again, not required.

This structure suggests to interpret vector components as
representing particles (RPO) and bi-vector components as 
fields: objects are the sources (generators) of fields, fields act 
on objects but can not generate them. 
Hence, it is nearby to interpret the six bi-vectors, $3$ skew-symmetric 
and $3$ symmetric, as generators of the Lorentz transformations. 
And this interpretation matches surprizingly well.

For a detailed account of the Lorentz transformations as they naturally
emerge from $Cl(3,1)$, see Refs.~\cite{lt_paper,rdm_paper,geo_paper}.
Here we just mention the result, namely that the skew-symmetric bi-vectors
$\y_7$, $\y_8$ and $\y_9$ are generators of spatial rotations while the 
symmetric bi-vectors $\y_4$, $\y_5$ and $\y_6$ generate Lorentz boosts 
in the corresponding {\it directions}, both a mathematical consequence of 
Eq.~\ref{eq_euler} or Eq.~\ref{eq_symtrans}, respectively.
This interpretation is completely determined by the structure of the 
Hamiltonian Dirac algebra and by the transformation properties of the
quantities under SSTs.

Since the structure of the real Dirac algebra is the basis of relativistic
electrodynamics, it is reasonable to interpret the parameters accordingly.
The vector parameters of the auto-correlation matrix ${\bf S}$, can
be identified with a four-vector quantity, for instance with energy and
momentum:
\begary{rcl}
s_0            &\equiv&{\cal E}\\
(s_1,s_2,s_3)^T&\equiv&{\vec P}\\
\label{eq_emeq0}
\endary
and the bi-vectors, for instance, with the electromagnetic fields:
\begary{rcll}
(s_4,s_5,s_6)^T&\equiv&{\vec E}\\
(s_7,s_8,s_9)^T&\equiv&{\vec B}\\
\label{eq_emeq1}
\endary
Due to the fact that the {\it electromagnetic fields} appear in this
context as based on pure dynamical notions, we named this interpretation
the {\it electro-mechanical equivalence} (EMEQ) in preceeding papers 
Ref.~\cite{rdm_paper,geo_paper}. In App.~\ref{sec_emeq} it is shown
that this interpretation is consistent with both, the Dirac equation
as well as with Maxwell's equations. In Ref.~\cite{qed_paper} we have
shown how well this interpretation fits to Maxwell's equations.

We collect the vector quantities in the matrix ${\bf P}$
\begeq
{\bf P}={\cal E}\,\y_0+\vec p\,\cdot\,\vec\y
\endeq
and the fields in a second matrix ${\bf F}$
\begeq
{\bf F}=\y_0\,\vec E\,\cdot\,\vec\y+\y_{14}\,\y_0\,\vec B\,\cdot\,\vec\y
\endeq
where the notation using the dot ``$\cdot$'' for the scalar product
is purely formal. Using the pseudo-scalar this can be written as 
\begary{rcl}
{\bf P}&=&({\bf S}+\y_{14}\,{\bf S}\,\y_{14})/2\\
{\bf F}&=&({\bf S}-\y_{14}\,{\bf S}\,\y_{14})/2\\
\endary
Since both, the density matrix and the Hamiltonian matrix have the same
structure, the EMEQ applies to both. 

In isolated equilibrium systems, the Hamiltonian matrix ${\bf H}$ can only 
be a function of quantities produced by the RPO itself: ${\bf H}=f({\bf S})$. 
Consider that $f$ is an analytical function that it can be
written as a Taylor series~\footnote{Then, since ${\bf H}$ and ${\bf S}$ are 
Hamiltonian, only odd terms can contribute, since only odd powers of a
Hamiltonian matrix are Hamiltonian.}. Obviously then ${\bf H}$ and ${\bf S}$
commute and we find from Eq.~\ref{eq_enveq} that ${\bf\dot S}=0$. Even though
the {\it spinor} oscillates and has eigenfrequencies, the {\it observables} 
are ensemble properties which are in this case {\it constant}~\footnote{
In accelerator physics this corresponds to a matched beam ellipsoid in a 
constant focusing channel.}.
In order to obtain a change of observable quantities from Eq.~\ref{eq_enveq}, 
we must add some external Hamiltonian ${\bf H}_x$:
\begary{rcl}
{\bf\dot S}&=&({\bf H}+{\bf H}_x)\,{\bf S}-{\bf S}\,({\bf H}+{\bf H}_x)\\
&=&{\bf H}_x\,{\bf S}-{\bf S}\,{\bf H}_x\\
\endary
Hence, self-interaction is, as long as ${\bf H}=f({\bf S})$ and within this 
linear approximation, {\it unobservable}.

Then, given the RPO is in interaction with external fields, one obtains:
\begary{rcl}
{\bf\dot P}+{\bf\dot F}&=&{\bf F}_x\,({\bf P}+{\bf F})-({\bf P}+{\bf F})\,{\bf F}_x\\
&=&{\bf F}_x\,{\bf P}-{\bf P}\,{\bf F}_x+{\bf F}_x\,{\bf F}-{\bf F}\,{\bf F}_x\\
\endary
It follows from the commutator table of the Dirac algebra, that this can be
splitted into:
\begeq
{\bf\dot P}={\bf F}_x\,{\bf P}-{\bf P}\,{\bf F}_x\,,
\label{eq_lorentzforce}
\endeq
which is the Lorentz force equation as we shall show in Sec.~\ref{sec_lorentz}
and secondly
\begeq
{\bf\dot F}={\bf F}_x\,{\bf F}-{\bf F}\,{\bf F}_x\,,
\label{eq_spin}
\endeq
which describes spin precession (see Sec.~\ref{sec_lorentz} below).

\subsection{Units: The Schwinger Limiting Fields}
\label{sec_units}

In Eq.~\ref{eq_symplexbygammas} different physical quantities like 
electromagnetic fields, energy and momentum are directly added. 
This is allowed if one uses appropriate {\it natural} units. 
Modern physics identified a number of scaling factors, namely 
the ``speed of light'' $c$ for the scale between mass, energy 
and momentum and between electric and magnetic fields, respectively, 
and $\hbar$ for the scale between energy and frequency and 
the unit charge to scale fields relative to mechanical quantites. 
A detailed account of how physical constants are to be understood 
has been given in Ref.~\cite{fit_paper}.

As we shall show below, according to the EMEQ, the eigenvalues of the 
Hamiltonian matrix correspond to the mass of a particle. 
Hence, if the RPO has the mass of the electron $m_e$, then this
scales the electromagnetic fields automatically relative to the so-called 
Schwinger limiting fields $E_S$ and $B_S$, which were first derived by 
Sauter~\cite{Sauter,HeiEu,Schwinger} $E_S$ and $B_S$~\footnote{
These fields are, given in SI-units: 
$E_S={m^2\,c^3\over e\,\hbar}=1.323\,\cdot\,10^{18}\,\rm{V/m}$ and
$B_S={m^2\,c^2\over e\,\hbar}=4.414\,\cdot\,10^{9}\,\rm{T}$. These
values are beyond any technical scale. Only the largest modern pulsed
lasers might allow to generate fields of this strength~\cite{SchwingerLimit}.}. 
The scaling factor between a magnetic field $B$ in SI-units and in
units of frequency is of the order ${e\over m}$ and for electric fields
$E$ of order ${e\over m\,c}$. The fields as they appear here, are scaled 
relative to the properties of the RPO, $e$ and $m$.

\subsection{The Eigenvalues of Dirac's Hamiltonian Matrix}
\label{sec_eigenvalues}

Let us first have a look at the eigenvalues of the Hamiltonian matrix ``operator'' 
that follows the parametrizations Eq.~\ref{eq_emeq0} and Eq.~\ref{eq_emeq1}, 
separately and combined. The trace of a matrix equals the sum of its 
eigenvalues, the trace of the squared matrix equals the sum of the squared 
eigenvalues and so on. 
From Eq.~\ref{eq_odd_trace} we know that the trace of any odd power of some 
Hamiltonian matrix vanishes. Hence only even powers are left, i.e. the second 
and fourth power:
\begary{rcl}
\rm{Tr}({\bf H}^2)&=&\sum\limits_k\lambda_k^2\\
\rm{Tr}({\bf H}^4)&=&\sum\limits_k\lambda_k^4\\
\endary
which allows to compute the eigenfrequencies~\cite{geo_paper,osc_paper,exp_paper}.
The result is given by:
\begary{rcl}
K_1&=&-\mathrm{Tr}({\bf H}^2)/4\\
K_2&=&\mathrm{Tr}({\bf H}^4)/16-K_1^2/4\\
\omega_1 &=&\sqrt{K_1+2\,\sqrt{K_2}}\\
\omega_2 &=&\sqrt{K_1-2\,\sqrt{K_2}}\\
\w_1^2\,\w_2^2&=&K_1^2-4\,K_2=\mathrm{Det}({\bf H})\\
K_1&=&{\cal E}^2+\vec B^2-\vec E^2-\vec P^2\\
K_2&=&({\cal E}\,\vec B+\vec E\times\vec P)^2-(\vec E\cdot\vec B)^2-(\vec P\cdot\vec B)^2\\
\label{eq_eigenfreq}
\endary
Hamiltonian matrices of stable systems have purely imaginary eigenvalues, 
corresponding to real frequencies $\w_i$, so that for stable
systems one has $K_2>0$ and $K_1>2\,\sqrt{K_2}$.

From this we find that $K_2=0$ when $\vec E=\vec B=0$, i.e. for pure vectors (Eq.~\ref{eq_emeq0})
\begeq
\w=\pm\sqrt{{\cal E}^2-\vec P^2}\,,
\label{eq_omega1}
\endeq
and for pure bi-vectors, where ${\cal E}=0=\vec P$, the frequencies are (Eq.~\ref{eq_emeq1})
\begeq
\w=\pm\sqrt{\vec B^2-\vec E^2\pm\,2\,\sqrt{-(\vec E\cdot\vec B)^2}}\,.
\label{eq_omega2}
\endeq
The frequencies are invariants under SSTs and hence are Lorentz scalars, 
i.e. invariant quantities. We therefore know that pure bi-vectors have two 
relativistic invariants, namely $\vec B^2-\vec E^2$ and $\vec E\cdot\vec B$
and we derived this without any reference to Maxwell's equations. Furthermore
we directly {\it know} that a stable bi-vector type oscillation is only
possible if $\vec E\cdot\vec B=0$, since only under this condition one 
obtains real frequencies (aka purely imaginary eigenvalues)~\footnote{
Note that, if one uses the metric of $Cl(1,3)$ instead of $Cl(3,1)$, the
terms representing e.m. fields, receive a factor $i$ and the signs of the
squares are reversed.}.

In the theory of electromagnetic wave propagation one finds that 
$\vec B^2-\vec E^2=0$ so that, if this is inserted into Eq.~\ref{eq_omega2},
apparently electromagnetic fields have no eigen-frequency. This is
generally known to be correct, it nevertheless leads, in our 
approach, to a degenerate matrix ${\bf H}$. This can be understood if we 
consider the frequency of the RPO Eq.~\ref{eq_omega1}, which apparently
provides a constant and invariant frequency, proportional to the mass of 
the RPO. As the frequency of the Dirac spinor that describes
a particle at rest is (up to constant scaling factors $\hbar$ and $c$)
identical to the mass, the time variable $\tau$ must be identified 
with proper time, the time of a ``comoving observer''.
Then it is clear why the electromagnetic bi-vector has a vanishing frequency:
electromagnetic waves, regarded from the perspective of a (hypothetical)
comoving observer, are indeed static. Expressed in language of special 
relativity we would say that we can not transform into the co-moving 
frame of an electromagnetic wave, as it moves with the speed of light.
However, this requires no commandment concerning space-time, but is a 
math fact about symplectic boost transformations. Remarkably, these facts
are known (but widely ignored) since decades~\cite{KimNoz81,HKNL1993}.

\subsection{Special Relativity in a Nutshell}
\label{sec_str}

The analysis of the Dirac algebra provides evidence that an 
RPO is essentially described by the vector type elements that 
are associated with the 4-momentum $({\cal E},{\vec p})$:
\begeq
{\bf H}={\cal E}\,\y_0+\y_1\,p_x+\y_2\,p_y+\y_3\,p_z
\label{eq_RPO}
\endeq
The square of this matrix is 
\begeq
{\bf H}^2=-{\cal E}^2+{\vec p}^2=-m^2
\endeq
so that 
\begeq
\ddot\psi={\bf H}^2\,\psi=-m^2\,\psi\,,
\endeq
i.e. the mass $m$ is proportional to the oscillation frequency, 
an eigenvalue of ${\bf H}$. It is a constant of motion and a scalar,
a $0$-vector of the Clifford algebra. Nonetheless we have no unique 
state of affairs of the RPO. We just selected a ``mass shell''. 
The structure of the Dirac algebra given in the previous section 
suggests that the RPO as described by Eq.~\ref{eq_RPO} is not in
interaction, it is a free ``particle''. 

Is it possible not only to formally derive rotations and boosts, 
i.e. Lorentz transformations 
(Ref.~\cite{lt_paper}), but also a space-time interpretation? This 
requires to switch to the different level of description, to use another 
emergent constant of motion as Hamiltonian. The original Hamiltonian
described the motion of the spinor $\psi$. But spinors are not directly
measurable. In order to (re-) construct classical physics, we need 
equations that involve exclusively observable quantities and in which 
the mass is just a constant ``parameter''.
If we reinterpret the equations of motion for observables 
(Eq.~\ref{eq_enveq}) in a Hamiltonian context,
then one obtains the ``classical'' (relativistic) Hamiltonian of a 
free RPO. The only unique choice for the Hamiltonian of the RPO is the 
parameter ${\cal E}$, which then depends on the vector components of 
the momentum. First we have to introduce a new time variable $t$ that
is conjugate to ${\cal E}$ and it's (proper) time derivative given by
\myarray{
\dot t&={dt\over d\tau}={\d{\cal H}\over\d{\cal E}}={\d\sqrt{{\cal E}^2-\vec p^2}\over\d{\cal E}}\\
&={{\cal E}\over m}=\y\,,
}
so that $dt=\y\,d\tau$ must hold.
Then we use ${\cal E}$ as the new Hamiltonian and obtain the classical relativistic 
energy-momentum-relation (EMR)
\begeq
{\cal E}={\cal H}(\vec p)=\sqrt{m^2+\vec p^2}\,.
\label{eq_EMR}
\endeq
Obviously, besides the constant ``parameter'' $m$, the RPO provides three
additional observables, the components of $\vec p$. When they are interpreted
as the new canonical momenta, then Hamilton's equation of motion for the
(coordinate) velocity $\vec\beta$ is defined by:
\begeq
\vec\beta=\dot{\vec q}=\vec\nabla_{\vec p}{\cal H}(\vec p)={\vec p\over{\cal E}}\,,
\label{eq_velocity} 
\endeq
This is the desired purely Hamiltonian dispersion relation. It derives almost
entirely from the pure algebra of time, i.e. it is an intrinsic feature of
second order moments of Hamiltonian phase spaces.

Inserting Eq.~\ref{eq_velocity} into Eq.~\ref{eq_EMR} gives
\begeq
{\cal E}=\y\,m\,,
\endeq
where $\y\equiv{1\over\sqrt{1-\vec\beta^2}}$ and 
\begeq
\vec p=m\,\y\vec\beta\,.
\endeq
Hence, within our approach, it is just another math fact that the 
velocity $\vert\vec\beta\vert$ is limited to $1$. Therefore the
limiting velocity is not a property of some reified space-time 
but a consequence of an eigenfrequency equation.
The Lorentz transformations are therefore not a {\it consequence} 
of the constancy of the speed of light, but vice versa. It does
not even require the notion of light to find these transformations.
Both, a maximal speed for massive objects as well as the Lorentz 
transformations, emerge from the same Hamiltonian formalism. 
Even the very concept of ``coordinate velocity'' itself can be 
regarded as a result of this formalism.

There is no commandment and no a priori existent space-time
required that determines the energy-momentum relation, but vice versa: 
Minkowski space-time emerges as a consequence of the energy-momentum 
relation (EMR), which is itself a consequence of a classical Hamiltonian 
theory and our elaboration on Hamilton's idea of an algebra of proper time. 
Since $\vert\vec\beta\vert\le 1$, one may write $\beta=\tanh{(\eps)}$,
where $\eps$ is the so-called ``rapidity'', and then one obtains 
$\y=\cosh{(\eps)}$, ${\cal E}=m\,\cosh{(\eps)}$ and 
$p=m\,\sinh{(\eps)}$~\cite{lt_paper}.

Since this approach implements the Lorentz transformations as the fundamental
set of canonical transformations, it is mathematically equivalent with the standard 
presentation of relativity theory (SPRT).
Nonetheless it modifies the SPRT insofar as both, Lorentz transformations
and ``inertial frames'' are notions that require no direct reference to a
predefined space-time {\it at all}. We introduced and explained them in a purely 
abstract Hamiltonian context. This result came out almost {\it automatically}.
But relativity is not the central issue of this article and hence we will not
elaborate in more detail. 
We refer to the introduction in Ref.~\cite{Mundy} and the detailed presentation
in Ref.~\cite{lt_paper}.

\subsection{The Lorentz Force}
\label{sec_lorentz}

The three skew-symmetric Hamiltonian elements $f_7,f_8,f_9$ act as generators 
of rotations in a $3$-dimensional parameter space and are therefore functionally
gyroscopic quantities as for instance magnetic fields or the angular momentum. 
The three symmetric elements $f_4,f_5,f_6$ act as generators of boosts in a
$3+1$-dimensional parameter space and are hence associated with a linear
accelerating quantity like the electric field.
In other words, the parametrization of Hamiltonian dynamics by its natural
symmetries, which guided us towards the use of Clifford algebras, uncovers 
a unique structure and establishes certain transformation classes for otherwise 
uninterpretable elements of the Hamiltonian matrices ${\bf S}$ and ${\bf H}$.
If the physical properties of real physical objects are represented by the 
second moments of ${\bf S}$, then the matrix ${\bf H}$ contains the 
{\it driving terms} of the symplectic motion, which must then be called 
(self-) fields.

Eq.~\ref{eq_lorentzforce} written explicitely using the EMEQ, yields: 
\begary{rcl}
\dot{\cal E}&=&\vec P\cdot\vec E\\
\dot{\vec P}&=&{\cal E}\,\vec E+\vec P\,\times\vec B\\
\label{eq_Lorentzforce0}
\endary
which are the Lorentz force equations formulated in proper time 
$\tau$~\cite{lt_paper,osc_paper,qed_paper,geo_paper}. With
${d\over d\tau}=\y\,{d\over dt}$ one finds (assuming $c=1$):
\begary{rcl}
\y\,{d{\cal E}\over dt}&=&m\,\y\,\vec v\cdot\vec E\\
\y\,{d{\vec P}\over dt}&=&m\,\y\,\vec E+m\,\y\,\vec v\,\times\vec B\\
\endary
and hence
\begary{rcl}
{d{\cal E}\over dt}&=&m\,\vec v\cdot\vec E\\
{d{\vec P}\over dt}&=&m\,\vec E+m\,\vec v\,\times\vec B\\
\endary
As explained in Sec.~\ref{sec_units} also the bi-vector fields, 
like all elements of ${\bf H}$, have a unit of frequency and require 
a re-scaling by $e/m$ to obtain their values in SI units~\footnote{
As we argued in Ref.~\cite{fit_paper}, scaling factors like these
can not be derived logically since they depend on a historical,
hence arbitrary, choice of units. They have to be introduced
in an {\it ad hoc} fashion if equations are to be aligned to the
MKS system. See also Sec.~\ref{sec_fff}.}:
\begary{rcl}
{d{\cal E}\over dt}&=&q\,\vec v\cdot\vec E\\
{d{\vec P}\over dt}&=&q\,\vec E+q\,\vec v\,\times\vec B\\
\label{eq_Lorentzforce}
\endary

The second part, Eq.~\ref{eq_spin}, describes the precession
of the remaining correlations. We denote the internal bi-vectors by
$\vec a$ and $\vec s$ to distinguish them from the external fields
$\vec E$ and $\vec B$, so that 
\begeq
{\bf F}=\y_0\,\vec a\cdot\vec\y+\y_{14}\,\y_0\,\vec s\cdot\vec\y\,,
\endeq
and one obtains (Eq.~\ref{eq_spin}):
\begary{rcl}
\dot{\vec a}&=&\vec a\times\vec B+\vec s\times\vec E\\ 
\dot{\vec s}&=&-\vec a\times\vec E+\vec s\times\vec B\\
\label{eq_eqom_envelope}
\endary
If one uses a complex notation $\vec\sigma=\vec s+i\,\vec a$ 
and $\vec F=\vec B+i\,\vec E$, then 
\begeq
\dot{\vec\sigma}=\vec\sigma\times\vec F
\endeq
which is the equation that describes the precession of the spin in
an external field~\cite{Corben1961}.

\subsection{Electromagnetic Waves and Spin (-flips)}

The Lorentz force that we derived from Eq.~\ref{eq_enveq} using the EMEQ,
refers to static, or at least slowly varying, electromagnetic fields.
The relevant frequency scale is given by the mass of the RPO,
i.e. $0.511\,\rm{MeV}$ in case of electrons. Visible light belongs to
frequencies of order of $\rm{eV}$, i.e. several orders of magnitude
below the typical de-Broglie frequency of electron waves. Still, the wave
length of visible light is in the order of nanometers, while technical
fields, for instance in undulators or spectrometer magnets, vary
with macroscopic ``wave-length'', i.e. order of millimeters up to meters,
again several orders of magnitude larger than the wavelength of light. 
Hence it is legitimate to assume that the variation of the fields is slow
and therefore adiabatic.

How do we treat the case in which (the electromagnetic part of) ${\bf H}$ 
{\it varies}, slowly compared to the de-Broglie wave? 
Of course this depends on the type of variation. For
electro-magnetic waves, we know from Maxwell's equations~\footnote{
We have shown how to derive Maxwell's equations on the basis of this
approach in Ref.~\cite{qed_paper}.} that polarized e.m. waves, as seen
by the RPO can be described by a rotating ``Dreibein''. 
If ${\vec E}(\tau=0)=E\,\vec e_x$ and ${\vec B}(\tau=0)=B\,\vec e_y$, then, 
for an observer in some ``inertial reference frame'', these vectors 
rotate with frequency $\W$ around the $z$-axis. The generator for
rotations around the $z$-axis is $\y_9$. Hence the time dependency 
can be written, according to Eq.~\ref{eq_enveq} as~\footnote{The 
factor $\frac{1}{2}$ is required to generate a spatial rotation 
frequency $\W$ and is a peculiarity of spinors~\cite{lt_paper}.}
\begeq
{\bf\dot F}=\frac{\W}{2}\,(\y_9\,{\bf F}-{\bf F}\,\y_9)\,.
\endeq
It is then possible (see App.~\ref{sec_Tdep}) to represent the time
dependency of ${\bf F}$ by adding the term $\W/2\,\y_9$ to ${\bf F}$,
effectively the same as a magnetic field component $B_z=\W/2$. 
Hence a circular polarized electromagnetic wave can be described in 
this approach by effective electromagnetic terms which give
$\vec E\cdot\vec B=0$ and $\vec B^2-\vec E^2=\W^2/4$, i.e. with an 
additional energy term that is proportional to frequency $\W/2$
(times $\hbar$, in MKS-units), in agreement with Eq.~\ref{eq_omega2}.

\begin{figure}[t]
\includegraphics[width=7cm]{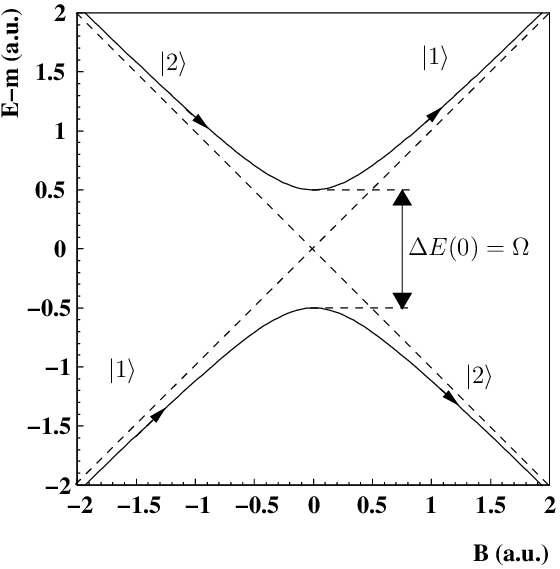}
\caption[Finestructure of the electron]{
Finestructure (energy levels) of an RPO in the presence of a static magnetic 
field $B$ and a circular polarized electromagnetic wave of frequency $\W$.
\label{fig_fs}}
\end{figure}

Assuming that we describe the RPO in its ``rest frame'', then $\vec P=0$ and
the eigenfrequencies, given by Eq.~\ref{eq_eigenfreq}, are:
\begary{rcl}
K_1&=&{\cal E}^2+B^2+\W^2/4\\
K_2&=&\sqrt{{\cal E}^2\,\vec B^2}\\
   &=&{\cal E}^2\,(B^2+\W^2/4)\\
\w&=&\pm\sqrt{{\cal E}^2+B^2+\W^2/4\pm\,2\,{\cal E}\,\sqrt{B^2+\W^2/4}}\\
  &=&\pm\sqrt{({\cal E}\pm\sqrt{B^2+\W^2/4})^2}\\
\endary
so that with $\w=m=\rm{const}$, assumed here to be positive, one finds
two possible eigen-frequencies
\begeq
m={\cal E}\pm\sqrt{B^2+\W^2/4}\,,
\endeq
so that
\begeq
{\cal E}-m=\Delta{\cal E}/2=\pm\sqrt{B^2+\W^2/4}\,.
\label{eq_fs}
\endeq
Even if one can estimate that the amplitude $B$ of the magnetic part of the 
wave is small $B^2\ll\W^2$ and can usually be neglected, it is of course 
possible to increase this field component by some external static field 
without any other change in the calculation. Then the fine-splitting of the 
energy in dependence of $B$ and $\W$ is shown in Fig.~\ref{fig_fs}.
Eqn.~\ref{eq_fs} is, for instance, used to describe the mechanism of adiabatic 
high frequency transitions between spin-states~\cite{Philpott}; as shown in
Fig.~\ref{fig_fs}, if the field $B$ is slowly raised from negative to 
positive values, the rf-field, indicated by $\W$ causes, using ``quantum'' 
language, a mixture of the ``pure'' states $\vert 1\rangle$ and $\vert 2\rangle$
and a level splitting $\W$. When the field, as indicated by the arrows, is raised
even further, an object that was originally in state $\vert 1\rangle$ is
flipped into state $\vert 2\rangle$ and vice versa.

This example provides further insights. Firstly, energy and frequency of
electromagnetic fields have, on the basic level, the exact same meaning and 
secondly, the two oscillator states are mixed around $B=0$ and it is thus 
thinkable to explain un-quantum leaps by resonances~\cite{RalstonBook}. 
Furthermore, without any reference to Hilbert spaces or ``quantum
principles'', we found separate energy levels which are usually postulated 
to be pure ``quantum'' effects. Hence Hilbert space is not quantum but a
convenient mathematical tool to describe the dynamics of eigenstates.
Everything so far suggests that ``quantization'' indeed {\it is} an eigenvalue 
problem as suggested by
Schr\"odinger~\cite{Schroed1926a,Schroed1926b,Schroed1926c,Schroed1926d}, 
and can {\it therefore} be described by classical notions.

It is sometimes claimed that discrete eigenfrequencies or the
``superposition'' of different states of motion with different
frequencies would be weird and could not be understood classically~\cite{Reynolds}.
This is wrong. A classical Hamiltonian system with two DOF has in general two
eigenfrequencies and two (pairs of complex conjugate) eigenvectors. 
The general state of motion is a superposition of these eigenvectors. 
This is purely classical Hamiltonian physics. Literally the same
mathematics emploid to describe adiabatic high frequency transitions
between spin states (as used above) allows to describe the resonant
emittance exchange between two beam planes in accelerators~\cite{Aiba2020,Kallestrup2020}.
Since the latter requires only purely classical physics, then also
the mathematical description of an adiabatic spin-flip is {\it as such}
not quantum.

\section{Same same but different}
\label{sec_dbl}

It is likely the alleged fundamental difference between the classical
and quantum theory which is responsible for the fact that one frequently
faces two versions of the same math in physics, ``classical'' version
and a ``quantum'' version. We mentioned an example of doubled math in
the previous section:
Adiabatic high-frequency transitions between different spin states
are the ``quantum'' version~\cite{Philpott} but the exact same
math used to describe emittance exchange between horizontal and
vertical beam planes is a ``classical'' version~\cite{Aiba2020,Kallestrup2020}.
The fact that quantum mechanics and (classical) beam dynamics are
based on the same mathematical (namely symplectic) structures has
been recognized and emphasized by many researchers before~\cite{DragtPRL,
  Forest,Dattoli1,Dattoli2,Qin2013,Gosson}, but the general fact has
not received much attention.

Another example from beam dynamics is the Riccati-Ermakov equation.
In a quantum physics paper one meets the Riccati-equation~\footnote{
  See for instance Ref.~\cite{CRUZ2015}.}:
\begeq
\ddot\alpha=\w^2(t)\,\alpha={1\over\alpha^3}
\endeq
and in Ref.~\cite{Fedele1998} we find the same equation in context
of accelerator physics (with reference to Ermakov):
\begeq
\ddot E+k_1\,E={I_0^2\over 4\,E^3}
\endeq
And indeed, in accelerator physics the equation stems from the
description of beam envelopes and is, in the presence of an additional
space-charge term, attributed to Kapchinskij and Vladimirskij~\cite{KV1959}.
Therefore it is sometimes called K-V equation~\cite{Scheinker2021}.
Let us emphasise that linear beam dynamics of accelerators is an entirely
classical theory and mostly consists of the statistical description
of ensembles of point particles by Hamiltonian methods.

The duplicity of mathematical relations does not stop at the perhaps
less-well-known equations like the Riccati-Ermakov equation. Also
Heisenberg's ``uncertainty relations'' have been mentioned in this
context. Leon Cohen writes~\cite{Cohen1995}:
\bquo
For signal analysis, the meaning of the uncertainty principle and its importance
have been clearly stated often enough, although a mystery still persists for some.
Right on the mark is the statement by Skolnik: ``The use of the word
'uncertainty' is a misnomer, for there is nothing uncertain about the
'uncertainty relation'.... It states the well-known mathematical fact
that a narrow waveform yields a wide spectrum and a wide waveform yields
a narrow spectrum and both the time waveform and frequency spectrum cannot
be made arbitrarily small simultaneously.''
[...]
If the interpretation is so straightforward in signal analysis, why is it one of the
most profound discoveries in physics? Here is why: In classical physics and in every
day life it seemed clear that we can choose the position and velocity of objects at
will. No one imagined that we cannot place a ball at a given spot with a given
velocity. However, present day physics, quantum mechanics, says precisely that
and it is one of the great discoveries regarding the behavior of matter.
\equo
But these claims of ``great discoveries'' rest not on a rigorously
proven mathematical incompatibility but merely on the assertion that the
wave-function does not represent the ``real thing'' but merely provides a
probability to ``find'' the real thing, namely the imagined point particle,
``at some location'' (whatever precise meaning this saying might have).

It further rests on the claim that one could place a classical object
precisely on some location with a precise velocity. But this assertion,
though frequently repeated, is simply false. In order to place
a classical particle with definite velocity at a definite location, one
would need to have the particle definitely at absolutely zero Kelvin,
which was recognized to be impossible in classical (statistical)
thermodynamics by
\begeq
E={n\over 2}\,k\,T
\endeq
where $n$ is the number of degrees of freedom.
If quantum theory was about ``energy quantization'' then only in the
(classical!) form
\begeq
E_n={n\over 2}\,k\,T_0
\endeq
with some zero-point temperature $T_0$. But this is not what quantum theory
really says, even if the energy levels of the quantum harmonic oscillator
appear to have this form:
\begeq
E_n=(n+{1\over 2})\,\hbar\,\w
\endeq
But note that $\w$ is here a system's property and not a universal quantity.
The core relations of quantum theory are the de-Broglie-relations and they
arise, as we have shown, from a classical theory of motion developed
from classical assumptions such as continuity, causality and finiteness.

It is true that classical physics has been speculatively extrapolated
by philosophically minded people, perhaps in order to clarify it's content.
This phenomenon is well-known from quantum theory too~\cite{Mermin81}:
``We now know that the moon is demonstrably not there when nobody looks''.
But it is likely that the number of physicists which were concerned 
by these speculative extra\-polations was limited at all times.
And as we demonstrate here, the metaphysical baggage has changed far more
than the mathematical content. It is true, that any relativistic equation
is decorated with $c$ and every quantum-equation with $\hbar$, but only
as long as we don't use a unit system in which $c=1$ and $\hbar=1$ hold by
definition.

\section{``Canonical'' (Un-) Quantization}
\label{sec_uqm}

We argued in Sec.~\ref{sec_SEQ} how to obtain the so-called ``canonical
quantization'', i.e. why a spatial derivative represents the momentum operator
and a time derivative an energy operator. It requires almost zero steps.

We introduced the phase space 
density $\rho$ and the matrix of second moments $\Sigma$ of this density. 
It is nearby and well-known in the theory of probability distributions to
use the Fourier transform to represent the moments of a distribution.
In Sec.~\ref{sec_str} we introduced velocities (and hence 
space-coordinates by the option to integrate the velocity $\beta$ over time) 
by the eigenvalue equation. The Fourier transform is mathematically rigorous 
and directly yields the mechanics of waves. However it requires that the
phase space density and the spinor are functions not of time and position
but of energy and momentum: These are the constraint quantities and these
are the quantities which characterize the physical state, while physics
must always presume, from the very beginning, that the abolute time and
the absolute position of some event are physically meaningless: space and
time coordinates are always relative. Furthermore their scales can
{\it de facto} only be derived from other, namely energy-related, scales:
Even the classical electron radius refers to the mass (rest-energy) of the
electron. If this would not be fixed, there wouldn't be a fixed radius either.

The phase space density $\rho({\cal H})$ of some
stable state depends on the constant parameters ${\cal E}$ and ${\bf p}$,
that describe a particle (RPO) so that $\rho({\cal H})=\rho({\cal E},{\bf
  p})$. Equivalently, the eigen-vectors of ${\cal H}$ are, in case of stable
oscillations, also functions of energy and momentum. Hence we define the
four-component spinor $\Psi=\Psi({\cal E},{\bf p})=\psi\sqrt{\rho}$ so that the matrix of 
second moments can be written as~\footnote{See also Ref.~\cite{GLTZ}.}:
\begeq
\Sigma=\int\,\psi\,\psi^T\,\rho(\psi)\,d^4\psi=\int\,\Psi\,\Psi^T\,d^4\psi\,,
\label{eq_rho}
\endeq
which is the so-called ``density matrix'' in the SPQM. 

Hence the spinor $\Psi$ is square integrable and therefore has a Fourier 
transform, which can be written as
\begeq
\tilde\Psi(t,\vec x)\propto\int\,\Psi({\cal E},\vec p)\,\exp{(-i\,{\cal E}\,t+i\,\vec p\cdot\vec x))}\,d^4p\,.
\label{eq_ft}
\endeq
This requires no postulate. It is just the Fourier transform of a 
phase space function and as such not quantum, so that also the operator
rule, the so-called ``canonical quantization'',
\begary{rcl}
\langle{\tilde\Psi^\dagger(t,\vec x)\,\cal E}\,\Psi(t,\vec x)\rangle&=&\langle\Psi^\dagger(t,\vec x)\,i\,\d_t\,\Psi(t,\vec x)\rangle\\
\langle\tilde\Psi^\dagger(t,\vec x)\,\vec p\,\Psi(t,\vec
x)\rangle&=&-\langle\Psi^\dagger(t,\vec x)\,i\,\vec\nabla\,\Psi(t,\vec x)\rangle\\
\endary
often shortly written as
\begary{rcl}
{\cal E}&=&i\,\d_t\\
\vec p&=&-i\,\vec\nabla\\
\label{eq_canonical}
\endary
is not quantum. It is but a special way to compute averages, aka
statistical mechanics. By construction, the parametric space-time, 
represented by $t$ and $\vec x$, matches to the framework of SSTs
that has been developed, if $t$ and $\vec x$ are constructed as
vector components in a Dirac algebra. In this case, the phase of
the Fourier transform  is an invariant quantity, i.e. a scalar. 

Inspection of the (anti-) commutator tables of the Dirac algebra~\cite{rdm_paper} shows, 
that, if ${\bf P}={\cal E}\y_0+\vec p\cdot\vec\y$ is a vector and 
${\bf X}=t\,\y_0+\vec x\cdot\vec\y$ is also a vector, then the 
anticommutator $({\bf P}\,{\bf X}+{\bf X}\,{\bf P})/2$ 
is a scalar. Hence the anti-commutator is but a generalization of the scalar 
(``inner'') product. The commutator is, no big surprise, a generalized vector
(``outer'') product, which suggests the following convention:
\begary{rcl}
{\bf P}\cdot{\bf X}&\equiv&({\bf P}\,{\bf X}+{\bf X}\,{\bf P})/2=(-{\cal
  E}\,t+\vec x\cdot\vec p)\,{\bf 1}\\
{\bf P}\wedge{\bf X}&\equiv&({\bf P}\,{\bf X}-{\bf X}\,{\bf P})/2=(\vec p\times\vec x)\cdot(\y_{14}\y_0\vec\y)\\
{\bf P}\,{\bf X}&=&{\bf P}\cdot{\bf X}+{\bf P}\wedge{\bf X}\\
\endary 
This is just a matter of convenient notation.

The ``physical'' space, defined this way, is no more the reified
fundamental container of everything as in Newtonian physics, or
Minkowski's relativistic reformulation of Newton, but is recognized, 
as it should be, as an entity emerging from the structural properties
of dynamical systems~\cite{BrownPooley}.

However, the Fourier transform requires that energy ${\cal E}$ and 
momentum ${\bf p}$ are real-valued, which excludes resonances. 
Even if we can not elaborate here in detail, but the use of the 
Fourier transform which allows to obtain
a spatio-temporal image of the phase space process, seems unproblematic 
only in specific circumstances, namely in eigenstates of the energy.
 
The general auto-correlation matrix requires $10$ parameters, while the 
Fourier transform uses only $4$ and hence ignores spin. Then there
are variables and correlations that receive no spatio-temporal ``location''
by the formalism, at least not in an obvious form.

Furthermore, Eq.~\ref{eq_emeq0} defines energy
and momentum as (linear combinations of) second moments of $\psi$, i.e.
energy and momentum depend on $\psi$, while the Fourier transform 
is formulated as if $\psi$ was a function of energy and momentum: 
the dependency is reversed. It remains to be elaborated if the dependency 
is bijectiv or not. The eigen-spinors of free Dirac particles are in many 
textbooks expressed as functions of $({\cal E},{\bf p})$, but this 
is not the case if one uses the corresponding eigenstates.
Or, in other words, there might be ``loopholes'' and it is not
far-fetched to assume that these might allow to explain the findings 
that required to introduce the projection postulate. But in any case 
it is clear that the most ``mysterious'' features of QM are, if 
locality is not presumed to be fundamental, merely {\it technical} 
or {\it mathematical} issues. In our presentation these features 
neither suggest nor do they suffice to establish a philosophy of 
unreality. 

Space-time geometry and electromagnetism carry the signature of
the simplest possible description of Hamiltonian interaction. This 
can be taken literally: as shown in ref.~\cite{rdm_paper,geo_paper,jacobi_paper},
the parametrization of a general $4\times 4$ Hamiltonian matrix that
describes the linearized coupling of two DOF by the use of Dirac 
matrices and the EMEQ allows a straightforward analysis and 
transformation to normal forms. Hence the {\it physical} notions of 
the Dirac algebra, provide the {\it mathematical} means to solve 
the general problem of diagonalizing Hamiltonian matrices. This is 
remarkable insofar as usually one expects that math is used to solve 
physical problems and not vice versa.

If physical notions are useful to solve a general math problem,
then the two are apparently isomorphic. All possible terms that are 
allowed by the physicality constraints are parameters of the 
Hamiltonian matrix and have physical significance. Math and physics 
are isomorphic, the theory is saturated. It might be noted by the
way that, within this approach, magnetic mono-poles don't exist. The
apparent symmetry of the Maxwell equations under duality transformations
is broken because the duality rotation is not of Hamiltonian nature~\cite{qed_paper}. 

The Dirac electron is described by a wave-function, i.e. by a charge 
distribution. In App.~\ref{sec_emeq} it is shown that the ``fields'' 
generated by the Dirac current density obey Maxwell's equations. 
This picture appears to be self-consistent. The difference to 
the ``classical'' picture of a 4-current density is a matter of the 
order of the presentation. 

Classical {\it metaphysics} presumes that space-time is more 
fundamental than the matter it ``contains''. According to this view
a charge distribution can be split into infinitesimal parts that are
distinguishable and can be addressed and fully identified by their
positions in space. Accordingly these infinitesimal parts should be able 
to move independently and should repel each other by means of the Coulomb
force. Hence a ``classical'' charge distribution can not be stable 
without further (ad hoc) presumptions. Classical electrodynamics
escapes by postulating ``point charges''. But there is no smooth transition
between these ideas, since the self-energy of such a ``compressed''
charge distribution becomes infinite in the point charge limit. 
Hence classical electromagnetism combined with space-based physics has a 
re-normalization problem. One can conclude that classical (meta-) physics is
neither intuitive nor clear nor free from self-contradictions. Nonetheless few 
physicists and/or philosophers urged for an interpretation. Specifically
when compared to QM, classical physics is presented as if these problems
never existed. The prevalent narrative establishes the myth that classical 
(meta-) physics is coherent, reasonable and close to common sense,
while quantum physics is none of this. The truth is closer to the description
given by Rohrlich~\cite{Rohrlich1996}:
\bquo
In the present paper, I want to point out that an equal lack of rational
explanation can be accorded to the enormous success by great scientists
when they make decisive progress in theory construction in {\it spite} of very
serious objections of their {\it own}. Such objections may be of a conceptual or
of a mathematical nature. Of course, they very often make seminal progress
when there are no such objections or when they are not aware of them, but
we are concerned here with those cases where they are, and where they
proceed despite objections, urged on by their intuition.

As the following examples will show, the greatness of these people lay
exactly in that fact: that they were not deterred by objections no matter
how serious they seem to be. Lesser scientists may not have dared to
proceed in this way. While it is true that most physicists are pragmatists
being concerned primarily with what works, they are often prevented to
pursue an idea by the objections raised against it. The greatness of a
scientist lies in these cases exactly in their ability, to recognize when such
objections should be ignored.
\equo
Maybe the Copenhagen interpretation was equivalently ``great'' in the sense,
that it ``encouraged'' scientists to proceed despite all pending problems.

Rohrlich also mentions the problems related to point particles and point
charges, respectively. In our approach no problems
of this kind can plague the theory, because neither space-time nor
electromagnetic fields are considered fundamental. Instead the space
of spinors, which has been identified as a classical phase space 
formed by two canonical pairs, is considered fundamental and the Lorentz
force does not act in some apriori given Minkowski space-time, but
the notion of space-time emerges together with electromagnetism:
The Lorentz force is due to Heisenberg's equations of motion, which is
little more than an equation of motion for second order moments.
Maybe this picture can only be visualized by people who frequently
deal with phase space ensembles, like for instance accelerator physicists.

But this approach stems directly from Eq.~\ref{eq_wpd}, i.e. from the
very idea that a Hamiltonian is involved and that this Hamiltonian is
identical to the dispersion relation of the wave motion (Eq.~\ref{eq_EMR}).
The way in which the dispersion relation emerges is, in this account, directly
(and exclusively) connected to the algebraic properties of matrices that
have been derived as second moments of spinor-distributions: Prior to
the averaging over phase space, there is neither a momentum nor electro-
magnetic fields. This means that the mechanical momentum of a particle
is only defined as a statistical property of the Schr\"odinger-Dirac
wave-ensemble.

In this approach no issues related to point charges arise, since the
charge density is only ``projected'' into space by the Fourier transform,
but this does not imply that we could split it into pieces and conclude
about the self-interaction by applying Maxwell's equations to these pieces:
Also Maxwell's equations emerge by the same projection~\cite{qed_paper,osc_paper}.
Hence we make no use of spatial notions to begin with and hence no
inconsistencies can arise here. The Lorentz force emerges 
as an {\it effective} force acting only on second moments of a density
distribution in an underlying phase space {\it of different dimension}.
The math used is still classical, but the presumption of space as {\it the}
fundamental notion has been dropped. This does not align with classical
Newtonian (meta-) physics, however, since it is derived from the formal
framework of classical Hamiltonian mechanics, it is legitimate to call
it a classical theory.

In our presentation, classicality is a logical or a mathematical 
but not a metaphysical notion: not space but real physical objects
(RPOs) are fundamental. Without having established real physical
objects first, it seems not sensible to introduce spatial notions.
And again, since Eq.~\ref{eq_enveq} holds for statistical averages
of second moments and only these second moments obey the Lorentz force.
Hence it would be inconsistent to cut an RPO into different ``parts''
and to consider those as 'parts' being independently 'located' in space-time. 
It makes even less sense to presume that they repel each other by a force
that is only established by the second moments of the complete 
phase-space ensemble. This demonstrates that un-quantum physics 
can not be consistently understood, as long as space is regarded as 
fundamental and this is the core of ``how to'' un-quantum mechanics: 
to accept that space, though real, is {\it not fundamental}. 

Furthermore, the description of an electron, which is mostly defined 
by its talent for electromagnetic interaction, should in some way provide 
the mathematical means to explain this talent. The standard presentation
simply postulates that electrons ``carry'' charge. If we aim to avoid 
commandments, this is not satisfying.
In our presentation, the talent for electromagnetic interaction is 
provided by the bi-vector elements of the Hamiltonian matrix parameterized 
by the Dirac algebra. Hence electromagnetism, special relativity and
quantum theory emerge together from a Clifford-algebraic approach to
classical (linear) Hamiltonian theory.

\subsection{Uncertainty Relations}

In his later years, Dirac made a number of remarks
that strongly contrast with the SPQM. For instance, he wrote 
that\cite{Dirac63b} 
\bquo
... this uncertainty relation cannot play a fundamental 
role in a theory in which $\hbar$ itself
is not a fundamental quantity. I think
one can make a safe guess that uncertainty 
relations in their present form will not
survive in the physics of the future.
\equo

Heisenberg's uncertainty relations are, depending on the 
presentation, a consequence of Eq.~\ref{eq_canonical} or 
of the Fourier transform. According to Ralston, they 
describe yet another math fact~\cite{RalstonBook}: 
``No working physicist anywhere views the uncertainty relation
as anything more than an easy math fact. However, the core of
the Copenhagen {\it presentation} puts the items early and
{\it in a particular order to make them appear essential}.''
A certain width of a distribution in momentum space 
determines a minimal width of the Fourier-transformed 
distribution in physical space and vice versa. And yet again, 
since it is a math fact about Fourier transforms, it requires 
no postulates and therefore it cannot be ``quantum''. 

Griffith wrote\cite{Griffith}:
\bquo
This principle is often discussed in terms of measurements of a 
particle's position or momentum, and the difficulty of simultaneously 
measuring both of these quantities. While such discussions are not without
merit [...] -- they tend to put the emphasis in the wrong place, 
suggesting that the inequality somehow arises out of
peculiarities associated with measurements.
\equo
The so-called ``uncertainty relations'' are not due to measurement
uncertainties. Nowhere in the derivation of the uncertainty relations 
is it {\it required} to refer to ``distortions by measurements''. 
There is not even an intrinsic need to speak of ``uncertainties'' 
{\it at all}, since the width of a distribution is not 'uncertain'. 
It is just a non-zero width. Neither do we negate the fact that 
physical states {\it can be} distorted by measurements nor do
we question the fact that measurements possess uncertainty.
We simply do not recognize any necessity to postulate 
some mysterious ``uncertainty principle'' to establish
these findings.

\subsection{The Born Rule}

We used the born rule in the introduction to show that the
Schr\"odinger equation emerges from classical notions. But within
our no-assumption approach, the Born rule has not been established yet.
Born's rule says that $\Psi^\dagger(\vec x,t)\Psi(\vec x,t)$ can be 
regarded as a probability density to ``find'' a particle at position 
$\vec x$ and time $t$~\cite{BornRule}.
Or, in other words, the transfer of a phase space density $\rho(\psi)$
into the parametric space-time, has not yet been sufficiently clarified,
since $\rho$ and not $\Psi^T\Psi=\psi^T\psi\,\rho$ {\it is} - in
our presentation - a density in phase space. So why should the 
latter {\it now} be interpreted as a density in space-time? This 
question is legitimate and nearby and has a simple answer: 
The mathematical principles of spatio-temporal logic require that
a density distribution is normalizable and part of a
$4$-vector that obeys a continuity equation~\footnote{See App.~\ref{sec_realDirac}.}.
This suffices to validate the consistency of the spatio-temporal 
``image'' and Born's rule. Not more and not less. It is not within the 
scope of the continuity equation to guarantee the absence of other,
non-spatial, correlations between physical quantities.

The mathematical correctness of an image, however, does not 
change the fact that the true postal address of the ``components''
of an RPOs is an abstract phase space.
This means that we can take the mathematical form of QM
{\it seriously}: The squared ``amplitude'' merely generates 
the mathematical form of a substance inhabiting space-time.
What we perceive and measure in space-time, is a Fourier
spectrum. The Fourier transform is a reversible unitary
transformation, an isomorphism with respect to the transformed 
properties: Constraints imposed in either space have consequences 
in the other. This justifies to regard the image appearing in 
space-time as {\it real}, but it does not suffice to make it 
{\it fundamental}. 

It is sometimes claimed that QM and specifically the Born rules
requires some kind of non-classical probability theory. 
Leon Cohen has shown that this is just another myth~\cite{Cohen}. 
We found no convincing reason to think otherwise.

\subsection{Space-Time: The Arena of Avatars?}

As explained in the previous section, Minkowski's space-time is,
from the perspective of a single RPO, some kind of a 4-dimensional 
holographic screen, it is metaphysically less but logically more
fundamental than supposed ``classically'' - nonetheless it is more 
than a mere image of reality. For many physicists it seems simply 
not imaginable to weaken local realism~\footnote{
Einstein wrote 1947 in a letter to Born: ``I cannot seriously
believe in it [quantum mechanics] because the theory cannot be 
reconciled with the idea that physics should represent a reality 
in time and space, free from spooky actions at a distance.''~\cite{BELett}
}, but from a logical perspective a merely commandment-based 
(fundamental) space-time concept is not satisfactory. 
Even Einstein, in later years, wrote: 
``Spacetime does not claim existence on its own, but only as a 
structural quality of the field''~\cite{EinsteinRel}.

The experimental tests of Bell's theorem are often interpreted such
that nature has, on the fundamental level, non-local aspects~\cite{Popescu}. 
As Maudlin expressed it~\cite{Maudlin}: ``What Bell proved, and what 
theoretical physics has not yet properly absorbed, is that the physical 
world itself is non-local''~\footnote{Maybe one should not over-emphasize 
the importance of non-locality, though. Since, as elaborated above, 
the space-time description is a fairly direct and immediate consequence 
of our approach. Maybe too many bold claims concerning the meaning 
of quantum theory have been made in the past. Nobody felt the need to 
interpret the physically questionable concept of a ``classical'' point 
charge, so why would we need to interpret now? If the result of the SPQM 
can be summarized as {\it shut up and calculate!}~\cite{ShutUp}, then 
there seems to be no urgent need for an interpretation, premature or not.}.
De Haro and de Regt argued that physical theories without an apriori 
presumed background (space-time) are indeed able, contrary to other claims, 
to ``provide scientific understanding''~\cite{dHdR}, though they
are - of course - difficult to visualize.

Mariani and Truini suggested the basic principle that 
``there is no way of defining spacetime without a preliminary concept 
of interaction''~\cite{MT}. This is exactly what we argued here, 
but we doubt that it requires a principle.
In this work, such a principle does not appear, since the definition
of physicality did not require to specify ``where'' some object is. 
The question ``where'' a thing ``is'', can only have physical meaning in
the presence of another (second) object, i.e. by inter-action. Spatial
distance and ``directions'' are notions that emerge from the description
of this interaction, not vice versa.

Modern physics invented the notion of {\it background independence}:
``[...] a classical field theory is background-independent if the 
structure required to make sense of its 
equations is itself subject to dynamical evolution, rather than being 
imposed ab initio. [...] a theory is fully background-independent 
relative to an interpretation if each physical possibility corresponds 
to a distinct spacetime geometry; and it falls short of full 
background-independence to the extent that this condition fails''~\cite{Belot}.
Regarding these criteria we think that our physico-logical approach is,
though formally ``classical'', background independent. It directly and 
inevitably leads to the simplest possible physical objects and the 
first order interaction of these generates the 3+1 dimensional 
parameter space, that we human inhabitants of the constructed physical 
world call space-time.

The Lorentz transformations are not primarily required to describe
``coordinate'' transformations between ``inertial reference frames'', 
they are (also) active canonical transformations, changing the physical 
state of the system under consideration. The issues that many - also 
renowned - physicists had with special relativity~\cite{Wazeck}, might 
also be due to an unclear attitude (of Einstein, but also others) 
towards the ontological status of space-time. Though Einstein suggested 
in his theory of special relativity that the assumption of a material 
substance, an ether, is dispensable, he was not able or willing to 
fully dispense the hegemony of the Newtonian heritage of absolute 
space~\footnote{
For Einstein, {\it local causality} was a fundamental requirement. 
This (and not determinism) is the core of his concerns regarding 
QM~\cite{Fine,Becker}.}. 

According to that view it does not suffice to formulate a theory that 
allows for the derivation of geometrical notions, aka an emergent space-time, 
but spatio-temporal notions must inevitably be the most fundamental ones.
This philosophy might be called space-time fundamentalism (STF) and Einstein
frequently - though not always - appeared as a proponent of STF.
STF almost {\it requires} a reification of space, or space-time, respectively.

Mermin warned us that the reification of mathematical abstractions used in 
physics is a ``bad habit''~\cite{Habit}. He argues mainly from a pragmatic
point of view (which is welcome). However, viewed pragmatically:
Does the dogma of unreality of wave-functions enable students to understand 
QM or does it lead to unnecessary confusion? Is it {\it required} and 
{\it justified} to claim that ``the "orbit" is created by the fact that we 
observe it'' as Heisenberg claimed~\footnote{``Die "Bahn" entsteht erst 
dadurch, dass wir sie beobachten''~\cite{Heisenberg1927}.}? Do we really
need to accept an interpretation that questions object permanence, despite
the fact that it is, in the form of energy conservation, the fundament of 
physics?
Do these things at least correspond to any practice in physics? Do accelerator 
physicists provide any measure to ``observe'' particle beams in order to 
establish the existence of particle orbits? {\it Of course} they don't. 
{\it Of course} there is no need to do so. As far as we can tell, orbits 
are established by electromagnetic fields and not by observation.
We think it is time to stop paying lip service to a philosophy that even 
Heisenberg could not persevere in his uncertainty paper~\footnote{
A careful reading of that paper reveals that Heisenberg makes frequent
use of arguments that refer to supposed underlying physical processes, 
though he claims to have abandoned this kind of ``realistic'' and ``classical'' 
way of thinking.
}.

We believe that Heisenberg's dismissal of reality went beyond the 
requirements and his arguments suspiciously oscillate between the 
``uncertainty principle'', commutators of conjugate pairs and considerations 
about the disturbances by measurements, enriched with claims of
positivist nature~\cite{Heisenberg1927}. What he did, is exactly what 
Mermin (should have) criticized: He overrated mathematical abstractions, 
not in support of reification, but of un-reification. 
Heisenberg thus scrambled the quantum omelet with ingredients of a theory 
that wasn't finalized or established yet~\footnote{
This work is not supposed to be a blame game. But there are reasons for the
fact that I mostly refer to Heisenberg -- and not Bohr, Born, Pauli or others. 
Firstly, Bohr's writings are (said to be) obscure and elusive~\cite{Stapp} while 
Heisenberg expressed himself with sufficient clarity (and in my mother tongue).
Secondly, Heisenberg was the first who used the term {\it Copenhagen
 interpretation} (CI)~\cite{Howard}. Thirdly, Heisenberg agreed 
to authorize Stapp's presentation of the Copenhagen interpretation~\cite{Stapp}, 
which provides unambiguous evidence that he claimed (co-) authorship 
for the CI. And last but not least his writings leave no doubt that the 
struggle with interpretational issues (and Schr\"odinger's competing approach) 
deeply concerned him~\cite{HTG}. But while I considered his philosophy deep
and profound as a student, today I tend to agree with E.T. Jaynes, who wrote
that the CI ``constitutes a violent irrationality, that somewhere in this 
theory the distinction between reality and our knowledge of reality has 
become lost, and the result has more the character of medieval necromancy 
than of science.''~\cite{Jaynes5}. I think it is not contradiction to pay 
homage to Heisenberg as a physicist {\it and} to critically review his 
``philosophy'' at the same time.}.

But it would be completely wrong to conclude that emergence as such
necessarily implies a questionable reality status. Temperature is an 
emergent notion and there should be consensus that temperature is {\it real}.
Chemistry emerges from solutions of the Schr\"odinger equation, 
i.e. from wave-functions and orbitals. This alone does not suggest that 
chemical bonds are not real. Thus, even if our presentation of classical 
un-quantum mechanics refuses to regard space-time as apriori given, this 
does not mean that we regard space-time as being less {\it real}. 
We just doubt that it must inevitably be regarded as {\it fundamental}.
That is, we doubt that a fundamental theory of physics must invariably start
with the postulate of a specific space-time ``manifold'' in the first place.

Many, maybe most, of the alleged mysteries of QM have been debunked
before~\cite{Nicolic,RalstonBook}, or their non-classicality has been
critically reviewed~\cite{JenningsLeifer}. 
We suggest to go beyond a mere critic of the standard approach: as we 
have shown there is little in the {\it mathematical} formalism of quantum 
theory that can not be obtained from classical Hamiltonian methods.
We stress again that classicality is often misrepresented as some kind of
space-time meta-physics: If the objective is to describe motion, then, 
according to STF it must be motion in space and the classical canonical 
variables have to be understood literally as space-time coordinates and 
momenta. But this is not part of the classical {\it physical} but part 
of (Newtons) {\it metaphysical} presuppositions.

Classical analytical mechanics is but a mathematical framework. The 
applicability of the notion of {\it generalized} coordinates 
in the sense of dynamical variables in an abstract phase space is, 
in principle, unlimited. Abstract variables like those forming the wave 
function have not been invented by QM. We explained what exactly
distinguishes wave functions from observables and why this is a 
consequence of the fundamentality of $\psi$, using classical logic.

Though the Hamiltonian formalism has, with respect to Eq.~\ref{eq_lom3}, 
a perfect (skew-) symmetry between coordinates and momenta, the only
classical Hamiltonian that accounts for this symmetry, is the harmonic 
oscillator. In the Hamiltonian of a ``classical'' free particle, only 
the momenta appear, but there is no classical system, in which {\it only} 
coordinates are used. The derivation of Compton's scattering formula
demonstrates the irrelevance of coordinates in many practical problems.
It requires only the EMR, energy and momentum conservation and the 
de Broglie scaling relations $E=\hbar\,\w$ and $\vec p=\hbar\,\vec k$ 
for it's derivation. It is the same picture in many branches of physics: 
Spatial positions obtain their physical relevance exclusively from fields, 
i.e. from interactions, while the constraints that allow to draw physical 
conclusions and to make real calculations, are almost always be 
derived in energy-momentum-space. 
It is only the size of the human visual cortex that underlies our 
adherence to the idea that geometrical notions must somehow be the 
most fundamental ones. 

There is, within the standard presentation of classical physics, no 
explanation why the symmetry of coordinates and momenta {\it should} be 
broken. Max Born wrote\cite{BornRec}: 
\bquo
This lack of symmetry seems to me very strange and
rather improbable. There is strong formal evidence for
the hypothesis, which I have called {\it the principle of reciprocity}, 
that the laws of nature are symmetrical with regard to space-time and 
momentum-energy [...]
\equo
Born was absolutely right, but with respect to the fundamental level of 
Hamiltonian theory, namely the wave-function. We explained why and how 
the Dirac algebra breaks this symmetry between spatial coordinates and 
mechanical momenta.
Born spells out what we mentioned above: the principle of sufficient reason.
And his intuition was correct (as we think), but for spinorial phase space:
here the (skew-) symmetry of the canonical pair is fully valid. But yet again, 
this requires no independent principle: the PSR (and hence logic)
completely suffices. Our presentation of 
un-quantum mechanics is entirely based on simple math facts and 
straightforward logic. By construction it holds in the most general 
physical world. Hence, without commandments, the celebrated conjecture 
that many different physical worlds should be possible~\cite{Tegmark}, 
looses some of it's plausibility. The alleged infinity of mathematical
possibilities turns out to be a chimera. 

The intrinsic non-locality of un-quantum mechanics explains 
why it makes only limited sense to ask where an electron or a photon 
``really'' is in space: The electron itself is not located at some specific 
position in space at all. So what? Also billiard balls are not located at 
some unique position - only their center of gravity is.
Whence physical ontology is not primarily defined by spatial notions, 
it is meaningless to ask if it can simultaneously ``be'' at different 
positions. Surely it can, since projected into space-time, the electron 
has no definite unique location, but ``is'' a wave. However, the location 
of the constraining principles that {\it define} its dynamical character, 
is energy-momentum space. This energy-momentum space emerges from 
(auto-) correlations that originate elsewhere: in an abstract but
nonetheless 'classical' phase space.

\section{Prospects: Higher Dimensions}
\label{sec_prospects}

The distinction between Hamiltonian and skew-Hamiltonian elements and the
fact that skew-Hamiltonian elements have always, by definition, vanishing
expectation values, could be useful to discuss higher dimensional ``spaces''
as well. As mentioned above, our approach suggests to consider not only 
the Pauli and Dirac algebras alone, but two series of HCAs given in 
Tab.~\ref{tab_dim}.

Two CAs of this list, namely $Cl(9,1)$ and $Cl(25,1)$ are also regarded 
as interesting from the perspective of string theory~\cite{Kaku}.
Hence, by the way, we {\it already} succeeded to un-string these two 
algebras, unintentionally~\footnote{To un-string string theory means to
rationalize findings of the theory (if there are any), but without 
postulating ``strings''.}. 
As derived above, Hamilton's algebra of proper time allows, in principle, 
to consider many, arbitrarily large and complicated, algebras on the basis 
of classical phase space. But furthermore, HCAs have another feature, 
namely that they allow to ``hide'' dimensions. How that?

We found Clifford algebras from symmetries originating in Hamiltonian theory.
The analysis of the Hamiltonian Dirac algebra $Cl(3,1)$ generates a system 
of quantities and relations that precisely fits to relativistic electrodynamics. 
The formalism itself forced us to introduce a 3+1 dimensional parametric space-time. 
Without the use of Hamiltonian notions however, the Dirac matrices do not 
uniquely select a specific Clifford algebra: Instead of $Cl(3,1)$, 
the algebra of real $4\times 4$-matrices is also a ``representation'' of
$Cl(2,2)$. It was shown, that, {\it if} all Clifford generators are supposed
to be Hamiltonian, {\it then} we are restricted to the algebras listed in 
Tab.~\ref{tab_dim}, namely to a single time-like vector element and $N-1$
spatial elements. {\it Then} it seems, that $Cl(9,1)$ inevitably leads to
a 10-dimensional space-time. But as we argued in Ref.~\cite{qed_paper}, 
geometric spaces with more than $3$ dimensions, when derived from Clifford 
algebras, have some problematic features. Consider some $N$-dimensional 
space-time is supposed to emerge from a HCA. Then there are $N-1$ generators 
of boosts $\y_0\,\y_1,\y_0\,\y_2,\y_0\,\y_3,\dots$, 
but there are $\left( N-1\atop 2\right)=(N-1)(N-2)/2$ generators of 
rotations. While all generators of boosts mutually anti-commute, this 
does not hold for all rotators. $N-1=3$ is the largest number of spatial 
dimensions which is ``homogeneous'' in this respect~\cite{qed_paper}.
Hence, from the perspective of HCAs, $3$ spatial dimensions are a very
special case, a {\it unique} case.

This suggests to consider HCAs with the dimensionality $Cl(3,N-3)$.
From Bott's periodicity (Eq.~\ref{eq_bott0}) one obtains the condition
\begeq
p-q=6-N=0,\,2\,\rm{ mod }\,8\,.
\label{eq_bottr}
\endeq
which yields two sequences for $m\in\mathbb{N}$:
\begary{rcl}
N&=&6+8\,m\\
N&=&4+8\,m\\
\endary
The sequence $N=6+8\,m$ has no real representation in which all generators of 
the Clifford algebra are Hamiltonian, but HCAs of the Dirac type ($N=4+8\,m$) 
allow for a {\it reinterpretation} in $3$-dimensional space. In 
Ref.~\cite{qed_paper} we scetched an interpretation of the algebra of 
real $8\times 8$-matrices $Cl(3,3)$, which lead to the introducing 
an ``internal'' degree of freedom. 

The real representation of the $Cl(11,1)$ is based of the same set of 
matrices as $Cl(3,9)$ - they are just ordered and interpreted differently. 
All Dirac type HCAs with $m\ge 1$, i.e. $Cl(3,N-3)=Cl(3,1+8\,m)$, are algebras 
that can be obtained from multiple Kronecker products of Dirac matrices.
Indeed it has been claimed by Sogami that triple tensor products of Dirac 
spinors are able to reproduce much of the Standard model of particle 
physics~\cite{Sogami1,Sogami2,Sogami3}, an approach that fits seamlessly
to our presentation of un-quantum mechanics (see App.~\ref{sec_hm}).

As scetched in App.~\ref{sec_hm}, the Dirac sequence also emerges in a 
generalization with higher moments of degree $2\,M,\,M\in[3,5,7,\dots]$
or cross-correlations of multiple Dirac particles.
$Cl(3,9)$, for instance, has $3$ symmetric generators and $9$ skew-symmetric 
generators. One of the $9$ skew-symmetric generators is the symplectic unit 
matrix $\y_0$, which is Hamiltonian by construction. Then the remaining $8$ 
skew-symmetric Clifford generators are skew-Hamiltonian and neither represent 
observables nor are they generators of SSTs. They are in some sense ``hidden''.

Given $\y_a,\,a\ne 0$ is a skew-symmetric generator of $Cl(3,9)$, 
then $\y_a$ and $\y_0$ anti-commute (by definition of CA generators)
$\y_0\,\y_a=-\y_a\,\y_0$ so that $\y_a$ is skew-Hamiltonian:
\begary{rcl}
\y_0\,\y_a^T\,\y_0&=&-\y_0\,\y_a\,\y_0\\
                  &=&\y_0^2\,\y_a\\
                  &=&-\y_a\\
\endary
It is therefore neither a generator of a canonical transformation nor
an observable. Furthermore, $\y_a$ is skew-symplectic:
\begary{rcl}
\y_a\,\y_0\,\y_a^T&=&-\y_0\,\y_a\,\y_a^T\\
                  &=&-\y_0\\
\endary
since $\y_a\,\y_a^T={\bf 1}$. These are features known from the pseudo-scalar
of the Dirac-algebra. In context of the Dirac algebra, the pseudo-scalar
represents the charge conjugation operator. This suggests that the
skew-Hamiltonian $\y_a$ might be interpreted in a similar way, namely as 
representing discrete (instead of continuous) symmetry-transformations. 

\section{Conclusions and Outlook}
\label{sec_conclusion}

It was shown that it is the ``classical'' metaphysics of point particles
rather than classical mathematics that prevents from the insight that
both the Schr\"odinger and the Dirac theory (and hence QM)
are classical theories.

The SPQM attacks the problem by constructing a metaphysical rather 
than a physical paradox: It asks whether electrons {\it are} particles 
{\it or} waves, only in order to demonstrate that the electron can't 
{\it be} either. The typical conclusion however, that, since it is 
neither, QM can not possibly be understood, is untenable. We have shown
that classical mechanics emerges from the eigenvalues of a general 
Hamiltonian while quantum mechanics is mostly concerned with the 
eigenvectors of the same Hamiltonian. These are two elements which are 
both required for a complete description. Thus we have shown, that 
it is indeed possible to give a logical account of the ``wave-particle 
duality'', without the use of ad hoc metaphysical assumptions and without 
changing anything else but notation and presentation. Even if many
aspects of our presentation could not be discussed in all detail,
we believe to have shown that the math of CM and QM is not, contrary
to the usual assertions, fundamentally different.

\subsection{Approximations}

While there are examples of phase space distributions that are exclusively 
parametrized by their second moments, for instance (multivariate) Gaussians, 
there is no reason to presume that this is the only possible case. 
Hence we have no reason to believe that the second moments alone are 
sufficient to fully characterize the physical situation. 
The same holds with respect to the form of the Hamiltonian: 
While it is often possible and legitimate to use a truncated Taylor serie 
approximation as a simplifying assumption, this alone is no reason to 
believe that a limitation to second order is necessarily an intrinsic 
feature of nature. However we believe it is remarkable what can be obtained
on the basis of logic and some prelimenary simplifying assumptions.

The fact that we needed only second moments to derive many essentials 
of QM might be taken in support of the position that QM can not be
complete, but this does not mean that one has to presume ``hidden'' 
variables. It might suffice to consider higher moments and higher order
terms in the Hamiltonian. A first, very incomplete, 
look at possible generalizations for higher moments, represented 
by Kronecker products, is given in App.~\ref{sec_hm}.

\subsection{Classicality}

It was shown that Dirac's theory is {\it mathematically} classical.
A hint by Res Jost, pointing in this direction, even found 
its way into a celebrated paper of Dirac~\cite{Dirac63}: 
``It has been pointed out to me by R. Jost that this group is just 
the 4-dimensional simplectic group, which is equivalent to the 
3+2 de Sitter group.''
The mentioned group is the group of the Dirac matrices, and what Res Jost 
remarked is that this group is identical to a {\it classical} group, 
subject to classical Hamiltonian equations of motion. Furthermore it was
shown that QM altogether is {\it mathematically} classical: No anti-commutation
rules need to (and may) be presumed for the fundamental variables ($\psi$).

Since observables are, in our presentation, nothing but (auto-) correlations
of dynamical variables, followers of a purely information-theoretic 
approach of QM might feel confirmed. Though it is an intriguing idea
to think that ``correlations have physical reality; that which they correlate 
does not''~\cite{Mermin}, it is difficult to see, what exactly this claim
explains that cannot be obtained otherwise. We explained why our approach
makes it difficult to imagine that $\psi$ could be directly measured. But
we see no reason to conclude from this that $\psi$ could not be regarded
as physically real. 

Two methods to introduce ``physical'' space were considered: the first  
obtains the ``velocity'' directly by the use of Hamiltonian mechanics 
applied to the observables and secondly by a statistical description 
of moments based on the Fourier transform. 
The former case suggests properly defined trajectories and, given the
landscape of electric and magnetic fields is known, one can integrate
the trajectory of the particle (ignoring Heisenberg's claim that trajectories
don't exist) by the Lorentz force. There is little in this approach that
suggests quantum features. The second moments are correlations and they are in 
this sense mathematically exact~\footnote{
No one prevents us from interpreting
$\sqrt{\langle x^4\rangle-\langle x^2\rangle^2}$ as a measure
for the ``uncertainty'' of $x^2$. However, we doubt that 
the same quantity can be interpreted as an ``uncertainty'' of 
$\langle x^2\rangle$.}.
The matrix of second moments defines energy and momentum precisely.

The latter approach, the Fourier transform, is the basis of wave 
mechanics and it is required when the spatial extent of structures
in the vicinity of the particle's trajectory is in the same order of magnitude 
as the de-Broglie wavelength. Provided one has sympathy for the so-called 
``wave-particle duality'' or Bohr's ``complementarity principle'', 
one might take the duality of methods as a confirmation of these ideas.
However, if a ``principle'' can be mathematically derived, then it should
be called a theorem and there is few reason to call it a ``principle'' 
at all. And of course there is a bridge ``principle'' connecting both 
accounts, namely the relativistic energy momentum relation which can 
likewise be regarded as dispersion relation, where the ``group velocity'' 
is given by Eq.~\ref{eq_velocity}.

Is all this still classical physics? Maybe this depends on the point of view.
However, we have shown, as promised, that the difference between CM and QM
is not {\it mathematical}.
According to our reality condition, real physical objects (RPO) are characterized by 
their {\it permanence} which is, translated into the language of physics, 
symmetry in time. This implies not more and not less than a positive 
definite constant of motion (PDCOM). We have shown that the primary PDCOM 
of RPOs is mass, i.e. a form of energy. No commandments so far -- just a
single (reality) constraint.

Maybe this is what Feynman had in mind when he wrote
``...you know how it always is, every new idea, it takes a generation or two 
until it becomes obvious that there's no real problem''~\cite{Feynman82}.

\begin{acknowledgments}
Mathematica\textsuperscript{\textregistered} has been used for part of the
symbolic calculations. XFig 3.2.4 has been used to generate the figure, 
different versions of \LaTeX and GNU\textsuperscript{\copyright}-emacs 
for editing and layout.
\end{acknowledgments}

\begin{appendix}

\section{Unitary Motion is Symplectic}
\label{sec_unisym}

Linear symplectic motion is due to Eq.~\ref{eq_lom1} where ${\bf A}$ is 
real symmetric, $\psi$ is real and ${\bf J}$ is a symplectic unit matrix.
Linear unitary motion is given by
\begeq
i\,\dot\psi={\bf H}\,\psi\,,
\label{eq_lom_hermitian}
\endeq
where ${\bf H}$ is hermitian and $\psi$ complex.

If we split a Hermitian matrix ${\bf H}$ and a complex spinor $\psi$
into its respective real and imaginary parts ${\bf H}={\bf A}+i\,{\bf B}$,
such that ${\bf A}={\bf A}^T$ and $-{\bf B}={\bf B}^T$,
and $\psi=\phi+i\,\chi$, then Eqn.~\ref{eq_lom_hermitian}
can be written as follows:
\begary{rcl}
i\,(\dot\phi+i\,\dot\chi)&=&({\bf A}+i\,{\bf B})\,(\phi+i\,\chi)\\
i\,\dot\phi-\dot\chi&=&{\bf A}\,\phi+i\,{\bf B}\,\phi+i\,{\bf A}\chi-{\bf B}\,\chi\\
\dot\phi&=&{\bf B}\,\phi+{\bf A}\chi\\
\dot\chi&=&-{\bf A}\,\phi+{\bf B}\,\chi\\
\endary
Thus, if we compose a real spinor $\Psi=\bmtx{c}\phi\\\chi\emtx$ by the real and imaginary 
parts of the spinor $\psi$, then unitary motion has the form:
\begary{rcl}
\dot\Psi&=&\bmtx{cc}{\bf B}&{\bf A}\\-{\bf A}&{\bf B}\emtx\,\Psi\\
        &=&\bmtx{cc}0&{\bf 1}\\-{\bf 1}&0\emtx\,\bmtx{cc}{\bf A}&-{\bf B}\\{\bf B}&{\bf A}\emtx\,\Psi\\
        &=&\bmtx{cc}0&{\bf 1}\\-{\bf 1}&0\emtx\,\bmtx{cc}{\bf A}&{\bf B}^T\\{\bf B}&{\bf A}\emtx\,\Psi\\
        &=&\y_0\,{\cal A}\,\Psi\\
\label{eq_symplectic_dirac}
\endary
where $\y_0$ is a SUM and ${\cal A}$ is symmetric. In other words, {\it any}
unitary law of motion can always be expressed by symplectic motion 
with specific restrictions for the matrix ${\bf H}$ as given by 
Eq.~\ref{eq_symplectic_dirac}. 

\section{Periodic Time-Dependent Hamiltonian}
\label{sec_Tdep}

In the case of a time-dependent Hamiltonian matrix, the condition for 
a PDCOM Eq.~\ref{eq_com0} requires a modification:
\begary{rcl}
{d{\cal H}\over dt}&=&{\d{\cal H}\over\d t}+\nabla_{\psi}{\cal H}\,\dot\psi=0\\
                  0&=&\frac{1}{2}\psi^T\dot{\bf A}\psi+\psi^T\,{\bf
                    A}\,\dot\psi=0\\
\label{eq_Hoft}
\endary
We introduce an additional Hamiltonian matrix ${\bf G}$ and use the Ansatz
\begeq
\dot\psi=(\y_0\,{\bf A}+{\bf G})\,\psi\,,
\label{eq_transparency}
\endeq
Inserted into Eq.~\ref{eq_Hoft} this gives:
\begary{rcl}
0&=&\frac{1}{2}\psi^T\dot{\bf A}\psi+\psi^T\,{\bf A}\,(\y_0\,{\bf A}+{\bf G})\,\psi\\
0&=&\psi^T\left(\frac{1}{2}\dot{\bf A}+{\bf A}\,\y_0\,{\bf A}+{\bf A}\,{\bf G}\right)\,\psi\\
\endary
The due to the skew-symmetry of ${\bf A}\,\y_0\,{\bf A}$, it follows (as before)
$\psi^T{\bf A}\,\y_0\,{\bf A}\psi=0$. 
The remaining matrix $\frac{1}{2}\dot{\bf A}+{\bf A}\,{\bf G}$ must then also be
skew-symmetric to fulfill this condition. Since $\dot{\bf A}=\dot{\bf A}^T$ 
and ${\bf\dot F}=\y_0\,{\bf \dot A}$, the condition $\dot{\cal H}=0$
requires that~\cite{qed_paper}:
\begary{rcl}
0&=&(\frac{1}{2}\dot{\bf A}+{\bf A}\,{\bf G})^T+\frac{1}{2}\dot{\bf A}+{\bf A}\,{\bf G}\\
0&=&\dot{\bf A}+\y_0\,{\bf G}\,\y_0\,{\bf A}+{\bf A}\,{\bf G}\\
\endary
Multiplication with $\y_0$ from the left yields:
\begeq
0=\y_0\,\dot{\bf A}-{\bf G}\,\y_0\,{\bf A}+\y_0\,{\bf A}\,{\bf G}\\
\endeq
so that
\begeq
\dot{\bf F}={\bf G}\,{\bf F}-{\bf F}\,{\bf G}\,.
\endeq
Thus, if the time dependence of ${\bf F}$ can be obtained from Eq.~\ref{eq_enveq},
then then we have a kind of level transparency for the driving term ${\bf G}$: 
Concerning the original problem, Eq.~\ref{eq_transparency} suggests that ${\bf G}$ 
can be directly added to ${\bf F}$.

\section{(Multi-) Spinors in Electrodynamics?}

There are different possibilities to represent phase space densities.
One possibility has been used so far, namely a density function $\rho(\psi)$.
There is another approach, specifically useful in numerical simulations, 
namely phase space sampling. This implies to uses not a single spinor $\psi$,
but several, i.e. the column vector $\psi$ is replaced by a multi-column
vector with $m$ columns, aka matrix a $4\times m$-matrix. This approach can
also be used to impose a symmetry onto the phase space density~\cite{osc_paper}.
The matrix of second moments $\Sigma$ is then given in the form of Eq.~\ref{eq_sigma}.

Our approach so far concentrated on the description of the simplest RPOs,
i.e. {\it matter fields}.
Electromagnetic waves appeared only as terms that act on RPOs, but not
as objects in themselves. Even worse, we found that vector components
can not be generated from bi-vectors by Eq.~\ref{eq_enveq}. This still
holds, but raises the question of how to define electromagnetic energy
and momentum within our approach. The electromagnetic fields, 
written in the Dirac matrix formalism as a bi-vector, is given by
\begeq
{\bf F}=\y_0\,(\vec E\,\cdot\,\vec\y)+\y_{14}\,\y_0\,(\vec B\,\cdot\,\vec\y)\,.
\endeq
Eq.~\ref{eq_sigma} gives for an arbitrary Hamiltonian matrix:
\begeq
{\bf F}\,{\bf F}^T\,\y_0^T={\bf F}\,\y_0\,{\bf F}\,.
\endeq
For pure bi-vectors, this expression yields pure vector components for
free electromagnetic fields, aka ``photons'':
\begary{rcl}
\frac{1}{2}\,{\bf F}\,\y_0\,{\bf F}=\frac{1}{2}\,(\vec E^2+\vec B^2)\,\y_0+(\vec E\times\vec B)\cdot\vec\y\,.
\endary
This suggests that there is at least some {\it formal} similarity between 
the spinors (phase space coordinates) that we used to model RPOs and
electromagnetic fields.

The matrix ${\bf F}$ has,written explicitely, the form~\cite{rdm_paper}:
\begeq
{\bf F}=\bmtx{cccc}
-E_x&B_y+E_z&-B_z+E_y&B_x\\
-B_y+E_z&E_x&-B_x&-B_z-E_y\\
B_z+E_y&B_x&E_x&-B_y+E_z\\
-B_x&B_z-E_y&B_y+E_z&E_x\\
\emtx
\endeq
In case of free electromagnetic waves, we can choose a coordinate system such
that the wave propagates along the $z$-axis so that $E_z=B_z=0$ and hence:
\begeq
{\bf F}=\bmtx{cccc}
-E_x&B_y&E_y&B_x\\
-B_y&E_x&-B_x&-E_y\\
E_y&B_x&E_x&-B_y\\
-B_x&-E_y&B_y&E_x\\
\emtx
\endeq
If we define a spinor $\phi=(-E_x,-B_y,E_y,-B_x)^T$ as the first column (vector) of ${\bf F}$,
then the other columns are given by $-\y_6\,\phi$,  $-\y_9\,\phi$ and $\y_{14}\,\phi$,
respectively, so that ${\bf F}$ can be written as a ``multispinor''~\cite{osc_paper}:
\begeq
{\bf F}=(\phi,-\y_6\,\phi,-\y_9\,\phi,\y_{14}\,\phi)\,.
\endeq
Then one obtains a density matrix ${\bf F}\,{\bf F}^T\,\y_0$ of the form
\begeq
{\bf F}\,{\bf F}^T\,\y_0=2\,\bmtx{cc}1&0\\0&1\emtx\otimes\bmtx{cc} -P_z&{\cal E}\\-{\cal E}&P_z\emtx\,,
\endeq
with ${\cal E}=(B_x^2+B_y^2+E_x^2+E_y^2)/2$ and $P_z=E_x\,B_y-E_y\,B_x$.
That is, the matrix ${\bf F}$ is block-diagonal or {\it decoupled}. 

However, it must be kept in mind that ${\bf F}$ is {\it not} a multi-spinor, but
a Hamiltonian matrix. This becomes clear if one considers the transformation 
properties: Spinors transform (like conventional vectors) according to the rule
\begeq
\Psi'={\bf R}\,\Psi
\endeq
while Hamiltonian matrices transform according to
\begeq
{\bf F}'={\bf R}\,{\bf F}\,{\bf R}^{-1}
\endeq
so that 
\begeq
({\bf F}\y_0{\bf F})'={\bf R}\,{\bf F}\,({\bf R}^{-1}\,\y_0\,{\bf R})\,{\bf F}\,{\bf R}^{-1}\,.
\endeq
Then the matrix ${\bf F}\,{\bf F}^T\,\y_0$ is only a proper Hamiltonian,
if ${\bf R}^{-1}\,\y_0\,{\bf R}$ equals $\y_0$. 
According to Eq.~\ref{eq_symplectic}, this is the case, if the 
matrix ${\bf R}$ is not only symplectic but also orthogonal, i.e. in case 
of rotations. But it is not correct in case of boosts (see Eq.~\ref{eq_euler}). 
Hence the electromagnetic energy density is not a Minkowski 4-vector.
It is part of an object called ``stress-energy-tensor'' in classical
electrodynamics~\cite{Jackson}. 
This is due to the fact that the volume element in Minkowski space-time
is not an invariant quantity, while the volume element of a phase space 
{\it is} an invariant quantity.

The almost obscene complexity of many spatio-temporal descriptions in 
physics contrasts with the simplicity of the underlying phase 
space. Since equations should be the simpler the more fundamental, 
phase space and not space-time must be regarded as fundamental.

\section{Pseudoscalars}
\label{sec_PS}

The pseudo-scalar is defined by Eq.~\ref{eq_PS}. 
Is the pseudo-scalar Hamiltonian? The transpose of $\y_\pi$ is
\begeq
\y_\pi^T=\prod\limits_{\mu=N-1}^{0}\,\y_\mu^T=\y_{N-1}^T\,\y_{N-2}^T\dots\y_0^T\,.
\endeq
If all generators are Hamiltonian, this gives:
\begary{rcl}
\y_\pi^T&=&(\y_0\y_{N-1}\y_0)\,(\y_0\y_{N-2}\y_0)\dots(\y_0\y_0\y_0)\\
&=&(-1)^{N-1}\,\y_0(\y_{N-1}\y_{N-2}\dots\y_0)\,\y_0\\
\label{eq_PStrans}
\endary
A re-sorting of the order of the bracketed product requires a certain 
number of commutations of the factors and each commutation is accompanied
the a reversal of the sign. The number of permutations required to reverse 
the order of $N$ matrix factors is $N\,(N-1)/2$, so that 
\begeq
\y_\pi^T=(-1)^{N-1+N(N-1)/2}\,\y_0\,\y_\pi\,\y_0\,.
\endeq
The sign is hence positive and $\y_\pi$ is Hamiltonian, if the
exponent is an even integer. Since $N=2\,m$ is even, 
\begary{rcl}
N-1+N(N-1)/2=m-1+2\,m^2
\endary
the even term $2\,m^2$ can be skipped, so that $\y_\pi$ is Hamiltonian 
if $m-1=N/2-1$ is even, i.e. of the Pauli type (Eq.~\ref{eq_bott1}):
\begary{rcl}
N/2-1=4\,m\,.
\endary
The pseudo-scalar of Dirac type algebras (Eq.~\ref{eq_bott2}) 
is skew-Hamiltonian, since $N/2-1$ is odd:
\begeq
N/2-1=4\,m+1\,.
\endeq
This means that the Pauli algebra provides no criterium to distingush
between the even generator of $Cl(1,1)$ and the pseudo-scalar: The real
Pauli algebra is not uniquely defined by the Hamiltonian properties.

According to Eq.~\ref{eq_PStrans} general $k$-vectors are Hamiltonian if
the exponent $k-1+k(k-1)/2={k^2+k-1\over 2}$ is even. It is quickly
verified that this condition can
be written as
\begary{rcl}
{k^2+k-2\over 2}&=&2\,m\\
k^2+k&=&4\,m+2\\
\endary
and has a solution for integer $m$ for 
$k=4\,j+1$ and $k=4\,j+2$, but not for $k=4\,j$ and $k=4\,j+3$.
Hence in HCAs that are generated from Hamiltonian generators,
only $k$-vectors with $k=1,2,5,6,9,10,\dots$ are Hamiltonian
while $k$-vectors with $k=3,4,7,8,\dots$ are skew-Hamiltonian.

\section{Real Dirac Theory}
\label{sec_realDirac}

\subsection{Dirac Current Conservation}

The real Dirac equation (Eq.~\ref{eq_direac_real}) is:
\begary{rcl}
0&=&(\d_\mu\y_\mu\pm m)\,\psi\\
0&=&(\y_0\,\d_t+\vec\y\cdot\vec\nabla)\,\psi\pm m\,\psi\\
\endary
Since here we discuss spinors in ``physical space'' instead
of energy-momentum space, spinors are complex (Eq.~\ref{eq_ft})
and the ``adjunct'' spinor is $\bar\psi=\psi^t\y_0^t$ where
the superscript $t$ stands for the transposed complex conjugate.
Matrix transposition gives:
\begary{rcl}
0&=&(\d_\mu\psi^t)\,\y_\mu^t\pm m\,\psi^t\\
0&=&(\d_t\,\bar\psi)+(\vec\nabla\bar\psi)\,\cdot\,(\y_0\,\vec\y)\pm m\,\bar\psi\,\y_0\\
\endary
where we used the fact that $\y_k=\y_k^t$, $\y_0=-\y_0^t$ and the
anti-commutation rules.
Multiplication with $\y_0$ from the right yields
\begeq
0=(\d_t\bar\psi)\,\y_0+(\vec\nabla\bar\psi)\,\cdot\,\vec\y\mp m\,\bar\psi\,,
\endeq
so that, using $\bar\psi\psi=0$:
\begary{rcl}
0&=&\bar\psi(\y_0\,\d_t+\vec\y\cdot\vec\nabla)\psi\\
0&=&(\d_t\,\bar\psi)\,\y_0\psi+(\vec\nabla\bar\psi)\,\cdot\,\vec\y\,\psi\\
\endary
The sum of both equations then yields the conserved current:
\begeq
0=\d_t\,(\bar\psi\y_0\psi)+\vec\nabla\,\cdot\,(\bar\psi\vec\y\psi)\,.
\endeq
One obtains an electric 4-current $(\rho_e,\vec j_e)$ by multiplication with 
the scaling factor $\pm\,e$:
\begary{rcl}
\rho_e&=&\pm\,e\,\bar\psi\y_0\psi\\
\vec j_e&=&\pm\,e\,\bar\psi\vec\y\psi\\
\label{eq_DiracCurrent}
\endary

\subsection{Maxwell's Equations From Dirac Theory}
\label{sec_emeq}

This section provides evidence that the EMEQ (Eq.~\ref{eq_emeq1}) is consistent,
i.e. that it is sensible to identify the bi-vector elements of the Dirac
algebra with electric and magnetic fields, respectively.
We will show this by showing that the bi-vector elements $\vec E$ and $\vec B$
obey Maxwell's equations.

\subsubsection{Gauss Law}

In Ref.~\cite{qed_paper} we derived Maxwell's equations from the Hamiltonian
Dirac algebra. But it is also possible to use Dirac's equation in order to show, 
that the Dirac current (Eq.~\ref{eq_DiracCurrent}) is compatible with Maxwell's equations:
\begary{rcl}
0&=&-\d_t\,\psi+\y_0\,\vec\y\cdot\vec\nabla\,\psi\pm m\,\y_0\,\psi\\
0&=&\d_t\bar\psi+(\vec\nabla\bar\psi)\,\cdot\,\y_0\,\vec\y\pm\,m\,\bar\psi\,\y_0\,,
\endary
so that
\begary{rcl}
0&=&-\bar\psi\,\d_t\,\psi+\bar\psi\,\y_0\,\vec\y\cdot\vec\nabla\,\psi\pm m\,\bar\psi\,\y_0\,\psi\\
0&=&(\d_t\bar\psi)\,\psi+(\vec\nabla\bar\psi)\,\cdot\,\y_0\,\vec\y\,\psi\pm\,m\,\bar\psi\,\y_0\,\psi\,,
\endary
The sum yields
\begeq
0=(\d_t\bar\psi)\,\psi-\bar\psi\,\d_t\,\psi+\vec\nabla\cdot(\bar\psi\,\y_0\,\vec\y\,\psi)\pm 2\,m\,\bar\psi\,\y_0\,\psi\\
\endeq
From Eq.~\ref{eq_canonical} we take that the first term can be written as
$2\,i\,{\cal E}\,\bar\psi\,\psi$, which vanishes due to the algebraic identity 
$\bar\psi\psi=0$. Then one has
\begeq
0={e\over 2\,m}\,\vec\nabla\cdot(\bar\psi\,\y_0\,\vec\y\,\psi)\pm e\,\bar\psi\,\y_0\,\psi\,,
\endeq
which gives Gauss' law (for electron and positrons):
\begeq
\vec\nabla\,\cdot\vec E=\mp \rho_e\,,
\endeq
where $\vec E={e\over 2\,m}\,(\bar\psi\,\y_0\,\vec\y\,\psi)$ is the electric field.

\subsubsection{Gauss Law for Magnetism}

In order to show that the magnetic field is free of sources, we multiply with the
pseudo-scalar:
\begary{rcl}
0&=&-\y_{14}\,\d_t\,\psi+\y_{14}\,\y_0\,\vec\y\cdot\vec\nabla\,\psi\pm m\,\y_{14}\,\y_0\,\psi\\
0&=&\d_t\bar\psi\,\y_{14}+(\vec\nabla\bar\psi)\,\cdot\,\y_{14}\,\y_0\,\vec\y\mp\,m\,\bar\psi\,\y_{14}\,\y_0\,,
\endary
so that
\begary{rcl}
0&=&-\bar\psi\,\y_{14}\,(\d_t\,\psi)+\bar\psi\,\y_{14}\,\y_0\,\vec\y\cdot\vec\nabla\,\psi\pm m\,\bar\psi\,\y_{14}\,\y_0\,\psi\\
0&=&(\d_t\bar\psi)\,\y_{14}\,\psi+(\vec\nabla\bar\psi)\,\cdot\,\y_{14}\,\y_0\,\vec\y\,\psi\mp\,m\,\bar\psi\,\y_{14}\,\y_0\,\psi\,,
\endary
The terms containing the time derivatives again vanish by
Eq.~\ref{eq_canonical} as in case of Gauss Law. However, this time 
the ``mass term'' also vanishes when the equations are added:
\begeq
\vec\nabla\,\cdot\,(\bar\psi\,\y_{14}\,\y_0\,\vec\y\,\psi)=0\,,
\endeq
which gives 
\begeq
\vec\nabla\,\cdot\vec B=0\,.
\endeq
where $\vec B={e\over 2\,m}\,(\bar\psi\,\y_{14}\,\y_0\,\vec\y\,\psi)$ is the magnetic field.

\subsubsection{Ampere's Law}

If we regard the time derivative of the electric field, we obtain:
\begary{rcl}
{2\,m\over e}\,\d_t\,\vec E&=&(\d_t\,\bar\psi)\,\y_0\,\vec\y\,\psi+\bar\psi\,\y_0\,\vec\y\,(\d_t\,\psi)\\
&=&\left(\mp\,m\,\bar\psi\,\y_0-(\vec\nabla\,\bar\psi)\,\cdot\,\y_0\vec\y\,\right)\,\y_0\,\vec\y\,\psi\\
&+&\bar\psi\,\y_0\,\vec\y\,(\y_0\,\vec\y\cdot\vec\nabla\psi\pm\,m\,\y_0\,\psi)\\
&=&\pm\,{2\,m\over e}\,\vec j_e-[(\vec\nabla\,\bar\psi)\,\cdot\,\vec\y]\,\vec\y\,\psi\\
  &+&\bar\psi\,\vec\y\,(\vec\y\cdot\vec\nabla\psi)\\
\endary
Let us consider the $x$-component of the remaining terms of the right side:
\begary{rcl}
&&\bar\psi\,\y_1\,(\y_1\d_x+\y_2\d_y+\y_3\d_z)\psi\\
&-&[\d_x\bar\psi\y_1+\d_y\bar\psi\y_2+\d_z\bar\psi\y_3]\,\y_1\,\psi\\
&=&\bar\psi\,\d_x\psi+\bar\psi\,\y_1\y_2\d_y\psi+\bar\psi\,\y_1\y_3\d_z\,\psi\\
&-&(\d_x\bar\psi)\psi-(\d_y\bar\psi)\y_2\y_1\psi-(\d_z\bar\psi)\y_3\y_1\,\psi\\
&=&\bar\psi\,\d_x\psi-(\d_x\bar\psi)\psi+{2\,m\over e}\,(\d_y\,B_z-\d_z\,B_y)\\
\endary
Once again, it follows from Eq.~\ref{eq_canonical}, that the first 
term vanishes and one obtains Ampere's law:
\begeq
\vec\nabla\times\vec B-\d_t\,\vec E=\mp\,\vec j_e\,.
\endeq
We leave Faraday's Law as an exercise. 

\section{Higher Even Moments}
\label{sec_hm}

Second moments are averages of quadratic forms and can either be represented 
in the form of the $\Sigma$-matrix or alternatively by the use of Kronecker
products $\psi\otimes\psi$. The simplest spinor $\psi=(q,p)^T$ for instance 
generates a spinor $\psi \otimes\psi=(q^2,q\,p,p\,q,p^2)^T$, two different
spinors give $\psi_1 \otimes\psi_2=(q_1\,q_2,q_1\,p_2,p_1\,q_2,p_1\,p_2)^T$.

The rules for the Kronecker product ``$\otimes$'' are given in
Eq.~\ref{eq_Kronecker}.
If we define the second order spinor according to $\psi_2=\psi\otimes\psi$,
the 4th-order moments can be written in matrix form according to
\begeq
\Sigma_4=\langle\psi_2\psi_2^T\rangle\,.
\endeq
The spinor $\psi_2$ and the matrix $\Sigma_4$ are not free of redundancy, 
since $q$ and $p$ commute. However, the use of Kronecker products allows to 
stay within the algebraic framework as described for the case of 
simple spinors.

For (skew-) Hamiltonian matrices ${\bf S}$ (${\bf C}$) one finds:
\begary{rcl}
({\bf S}_1\otimes{\bf S}_2)^T&=&(\y_0\,{\bf S}_1\,\y_0)\otimes\,(\y_0\,{\bf S}_2\,\y_0)\\
&=&(\y_0\otimes\y_0)\,({\bf S}_1\,\otimes\,{\bf S}_2)\,(\y_0\,\otimes\,\y_0)\\
({\bf C}_1\otimes{\bf C}_2)^T&=&(-\y_0\,{\bf C}_1\,\y_0)\otimes\,(-\y_0\,{\bf C}_2\,\y_0)\\
&=&(\y_0\otimes\y_0)\,({\bf C}_1\,\otimes\,{\bf C}_2)\,(\y_0\,\otimes\,\y_0)\\
\endary
As mentioned above and explained in Refs.~\cite{qed_paper,osc_paper}, the 
constitutive properties of the SUM $\y_0$ are, that it must be skew-symmetric,
orthogonal and that it squares to $-{\bf 1}$, which is not fulfilled by
$\y_0\otimes\y_0$, but by $\y_0\otimes\y_0\otimes\y_0$, or more general:
The moments of order $D=2\,d$ for $d$ odd, lead automatically to symplectic 
motion, if the basic spinors is subject to symplectic EQOMs.

\subsection{Fourth Order Moments}
\label{sec_4moments}

Given that the fundamental EQOM are linear, e.g. are given by Eq.~\ref{eq_lom0}, 
one finds the simple generalization, starting with $\dot\psi_1={\bf F}\,\psi_1$
and $\dot\psi_2={\bf G}\,\psi_2$:
\begary{rcl}
\phi&\equiv&\psi_1\otimes\psi_2\\
\dot\phi&=&\dot\psi_1\otimes\psi_2+\psi_1\otimes\dot\psi_2\\
          &=&({\bf F}\otimes {\bf 1}+{\bf 1}\otimes{\bf G})\,(\psi_1\otimes\psi_2)\\
          &=&{\bf H}\,\phi\,,
\label{eq_eqom2}
\endary
such that the EQOM for the second moments are linear as well with the driving
matrix ${\bf H}$ given by 
\begeq 
{\bf H}={\bf F}\otimes {\bf 1}+{\bf 1}\otimes{\bf G}\equiv{\bf F} \oplus {\bf
  G}
\label{eq_H4th}
\endeq
which is called {\it Kronecker sum}. As well known from linear algebra, the
matrix exponential holds:
\begeq
\exp{({\bf F}\oplus{\bf G})}=\exp{({\bf F})}\,\otimes\,\exp({\bf G})\,.
\label{eq_expoplus}
\endeq

From Eq.~\ref{eq_enveq} one finds:
\begary{rcl}
{d\over dt}({\bf S}_1\otimes{\bf S}_2)&=&({\bf \dot S}_1\otimes{\bf S}_2)+({\bf S}_1\otimes{\bf\dot S}_2)\\
&=&({\bf F}\oplus{\bf G})\,({\bf S}_1\otimes{\bf S}_2)\\
&-&({\bf S}_1\otimes{\bf S}_2)\,({\bf F}\oplus{\bf G})
\endary
such that with ${\bf S}={\bf S}_1\otimes{\bf S}_2$ we may again write:
\begeq
{\bf\dot S}={\bf H}\,{\bf S}-{\bf S}\,{\bf H}\,.
\endeq
The transpose of the driving matrix ${\bf H}^T$ is given by:
\begary{rcl}
{\bf H}^T&=&{\bf F}^T\otimes{\bf 1}+{\bf 1}\otimes{\bf G}^T\\
               &=&\y_0\,{\bf F}\,\y_0\otimes{\bf 1}+{\bf 1}\otimes\y_0\,{\bf G}\,\y_0\\
               &=&-(\y_0\otimes\y_0)\,({\bf F}\otimes{\bf 1}+{\bf 1}\otimes {\bf G})\,(\y_0\otimes\y_0)\\
               &=&-(\y_0\otimes\y_0)\,{\bf H}\,(\y_0\otimes\y_0)\,.
\label{eq_symplex2}
\endary
Obviously ${\bf H}$ obeys a new criterium and is neither obviously Hamiltonian nor 
skew-Hamiltonian, since the matrix $(\y_0\otimes\y_0)$ is not skew-symmetric
and can hence not be interpreted as a symplectic unit matrix in the above
sense. Therefore ${\bf H}$ is not a (higher order) Hamiltonian matrix, 
though the trace of ${\bf H}$ is zero. Nevertheless the EQOM are of a form that 
constitutes a Lax pair. The corresponding constants of motion are then again
\begeq
\mathrm{Tr}({\bf S}^k)=\mathrm{Tr}\left(\langle {\bf S}_1\otimes{\bf S}_2\rangle^k\right)=\mathrm{const}\,.
\endeq

Also Eq.~\ref{eq_eqom2} can be derived within the framework
of Hamiltonian motion as we will show in the following.
If we write $\tilde\y_0=\y_0\otimes\y_0$ (where $\tilde\y_0^2={\bf 1}$ and $\tilde\y_0^T=\tilde\y_0$) 
and the Hamiltonian ${\cal H}(\phi)$ according to
\begeq
{\cal H}=\phi^T\,\tilde\y_0\,{\bf H}\,\phi\,,
\endeq
then we obtain
\begary{rcl}
\dot{\cal H}&=&\dot\phi^T\,\tilde\y_0\,{\bf H}\,\phi+\phi^T\,\tilde\y_0\,{\bf H}\,\dot\phi\\
            &=&\phi^T\,{\bf H}^T\,\tilde\y_0\,{\bf H}\,\phi+\phi^T\,\tilde\y_0\,{\bf H}^2\phi\\
            &=&\phi^T\,(-\tilde\y_0\,{\bf H}\,\tilde \y_0\,\tilde\y_0\,{\bf H}+\tilde\y_0\,{\bf H}^2)\phi\\
            &=&\phi^T\,(-\tilde\y_0\,{\bf H}^2+\tilde\y_0\,{\bf H}^2)\phi\\
            &=&0\\
\endary
So that ${\cal H}$ is conserved. However, we find that the product $\tilde\y_0\,{\bf H}$ is skew-symmetric 
\begary{rcl}
(\tilde\y_0\,{\bf H})^T&=&{\bf H}^T\,\tilde\y_0^T\\
(\tilde\y_0\,{\bf H})^T&=&-\tilde\y_0\,{\bf H}\,\tilde\y_0\,\tilde\y_0^T\\
(\tilde\y_0\,{\bf H})^T&=&-(\tilde\y_0\,{\bf H})\\
\endary
and the Hamiltonian function ${\cal H}$ of the fourth order moments vanishes, 
if the first order motion is symplectic.

The transfer matrix is given by Eq.~\ref{eq_expoplus} and is given by
\begeq
{\bf M}={\bf M}_1\,\otimes\,{\bf M}_2\,.
\endeq
Since ${\bf M}_1$ and ${\bf M}_2$ are symplectic, it follows that
\begary{rcl}
{\bf M}\,{\tilde\y}_0\,{\bf M}^T
&=&({\bf M}_1\,\otimes\,{\bf M}_2)(\y_0\,\otimes\,\y_0)({\bf M}_1^T\,\otimes\,{\bf M}_2^T)\\
&=&({\bf M}_1\,\y_0\,{\bf M}_1^T)\otimes\,({\bf M}_2\,\y_0\,{\bf M}_2^T)\\
&=&\y_0\,\otimes\,\y_0={\tilde\y}_0\\
\endary
Hence, though ${\tilde\y}_0$ is not a symplectic unit matrix (since $\tilde\y_0^2=+{\bf 1}$)
and though ${\bf M}$ is not symplectic, nonetheless ${\bf M}$ obeys an equation that
is equivalent to Eq.~\ref{eq_symplectic}.

\subsection{Eigenvalues of Kronecker Sums}

It is a known result in matrix analysis that the eigenvalues
of the Kronecker sum of ${\bf F}$ and ${\bf G}$ are sums
of eigenvalues of ${\bf F}$ and ${\bf G}$.
More precisely, if ${\bf f}$ is eigenvector of ${\bf F}$ with eigenvalue $f$
and ${\bf g}$ is eigenvector of ${\bf G}$ with eigenvalue $g$, then
${\bf f}\otimes{\bf g}$ is eigenvector of ${\bf F}\otimes{\bf G}$ with
eigenvalues $f+g$~\cite{MatrixAnalysis}.
${\bf F}$ and ${\bf G}$ are Hamiltonian matrices and for such matrices
it is known that if $f$ is an eigenvalue of ${\bf F}$, then $-f$, $\bar f$
and $-\bar f$ are also eigenvalue of ${\bf F}$~\cite{MHO}. Thus the 
$4$-th order moments contain the frequencies $f+g$ and $f-g$, $-f+g$ and
$-f-g$.

In case of two single degees of freedom, the normal forms are:
\begary{rcl}
{\bf F}&=&\w_1\,\y_0=\w_1\,\bmtx{cc}0&1\\-1&0\\\emtx\\
{\bf G}&=&\w_2\,\y_0\\
\endary
so that
\begary{rcl}
{\bf H}&=&{\bf F}\oplus{\bf G}=\bmtx{cccc}
0&\w_2&\w_1&0\\
-\w_2&0&0&\w_1\\
-\w_1&0&0&\w_2\\
0&-\w_1&-\w_2&0\\
\emtx
\endary
The eigenvalues of ${\bf H}$ are $\pm\vert\w_1+\w_2\vert$ and
$\pm\vert\w_1-\w_2\vert$.

Eq.~\ref{eq_symplex2} seemingly suggests the introduction of complex
numbers and of a symplectic unit matrix $(\y_0)_2=i\,(\y_0\otimes\y_0)$, but
there is no way to derive the EQOM from a non-zero real-valued Hamiltonian 
function.

\subsection{Sixth Order Moments}
\label{sec_6moments}

It is quite obvious that the next even order $\phi\equiv\psi_1\otimes\psi_2\otimes\psi_3$, 
based on the definitions
\begary{rclp{5mm}rcl}
\dot\phi_1&=&{\bf F}\,\psi_1&&\dot\phi_2&=&{\bf G}\,\psi_2\\
\dot\phi_3&=&{\bf H}\,\psi_3&&(\y_0)_3&\equiv&\y_0\otimes\y_0\otimes\y_0\\
\endary
again leads to
\begary{rcl}
{\bf J}&=&{\bf F}\otimes{\bf 1}\otimes{\bf 1}+{\bf 1}\otimes{\bf G}\otimes{\bf
  1}+{\bf 1}\otimes{\bf 1}\otimes{\bf H}\\
&=&{\bf F}\oplus{\bf G}\oplus{\bf H}\\
{\bf S}&=&{\bf S}_1\otimes{\bf S}_2\otimes{\bf S}_3\\
\endary
which again are symplectic laws of motion
\begary{rcl}
\dot\phi&=&{\bf J}\,\phi\\
(\y_0)_3^2&=&(-)^3\,{\bf 1}\otimes{\bf 1}\otimes{\bf 1}=-{\bf 1}_3\\
(\y_0)_3^T&=&-(\y_0)_3\\
{\bf J}^T&=&(\y_0)_3\,{\bf J}\,(\y_0)_3\\
{\bf\dot S}&=&{\bf J}\,{\bf S}-{\bf S}\,{\bf J}\,,
\endary
with the Lax pair ${\bf S}$ and ${\bf J}$ and the respective
constants of motion. The generalization of these findings is obvious:
All spinors $\psi_k=\prod\limits_{i=0}^k\,\otimes\psi_i$ with $k$ odd that
are composed of equal sized spinors $\psi_i$, each of which subject to 
symplectic motion, are again subject to symplectic motion. Spinors with 
$k$ even produce constants of motion, but the linearized Hamiltonian 
from which they can be derived, is identically zero.

\subsection{Symplectic High Order Moments}
\label{sec_symplectic_moments}

As we argued above, the simplest non-trivial and hence fundamental algebra
is the real Dirac algebra and the size of the corresponding spinor is
$2\,n=4$. Hence if spinors for higher moments are composed as a Kronecker
product from an uneven number $k=2\,m+1$ of simple spinors, then they fulfill 
the constraints for symplectic motion, if all individual spinors do.
For the fundamental spinor size of $2\,n=4$ this means that spinors composed from 
Kronecker-products corresponding to these moments have the size $4^k=4^{2\,m+1}$ and 
hence the corresponding matrices have the size $(4^{k})^2=4^{4\,m+2}=2^{8\,m+4}$. 
This matrix size corresponds to Clifford algebras $Cl(N-1,1)$ with $N=8\,m+4$, 
i.e. HCAs or the Dirac type.

Real Dirac spinors, that are constructed from even Kronecker products with an
even number $2\,m$ of Dirac spinors, have the size $4^{2\,m}=2^{4\,m}$, which 
corresponds to real matrix reps with $2^{8\,m}$ independent elements. However
there exists no real Clifford algebra $Cl(N-1,1)$ of size $N=8\,m$. And vice
versa, the case $N=2+8\,m$ has no correspondence in higher order moments or 
higher order correlations. One could say that HCAs of the Pauli type can not
be Kronecker-decomposed.

\subsection{Higher Order Hamiltonian}
\label{sec_ho_hamiltonian}

Given a Hamiltonian is of higher than 2nd order, then it can be written as a Taylor series according to
\begeq
{\cal H}(\psi)=\frac{1}{2!}\,\psi^T\,{\bf A}\,\psi+\frac{1}{3!}\,{\bf B}_{ijk}\,\psi_i\,\psi_j\,\psi_k+\dots
\label{eq_hamiltonian_general}
\endeq
One can argue that in stable (static) systems all terms of odd order vanish, 
i.e. that the Hamiltonian is invariant under $\psi\to\,-\psi$, then:
\begeq
{\cal H}(\psi)=\frac{1}{2!}\,\psi^T\,{\bf A}\,\psi+\frac{1}{4!}\,{\bf C}_{ijkl}\,\psi_i\,\psi_j\,\psi_k\,\psi_l+\dots
\label{eq_hamiltonian_even}
\endeq
Since the SUM is orthogonal $\y_0^T\,\y_0={\bf 1}$, the second order term ${\cal H}_2$ can be written as
\begeq
{\cal H}_2\,(\psi)=\frac{1}{2!}\,\bar\psi^T\,{\bf F}\,\psi
\endeq
where $\bar\psi^T\equiv\psi^T\,\y_0^T$. Using the Kronecker product we define $\y_2=\psi\otimes\psi$ and
$\bar\y_2^T=\psi^T\,\y_0^T\otimes\psi^T\,\y_0^T$ such that the fourth order term is
\begeq
{\cal H}_4(\psi_2)=\frac{1}{4!}\,\bar\psi_2^T\,{\bf F}_{2}\,\psi_2
\endeq
where ${\bf F}_{2}=(\y_0\otimes\y_0)\,{\bf A}_2$ with a symmetric matrix ${\bf A}_2$. The sixth order term
is then written as
\begeq
{\cal H}_6(\psi_3)=\frac{1}{6!}\,\bar\psi_3^T\,{\bf F}_{3}\,\psi_3\,,
\endeq
and so on. If the spinor $\psi$ is of Dirac size $2\,n=4$, (or, more generally, if $(2\,n)^2=2^N$), then there exists
a complete real matrix system that represents some Clifford algebra with elements $\Gamma_k\,,k\in\,[0,\dots,2^N-1]$ such
that any matrix ${\bf F}$ can by written as ${\bf F}=\sum_k\,f_k\,\Gamma_k$ and the tensor products can be written as
\begary{rcl}
{\bf F}_2&=&\sum_{jk}\,f_{jk}\,\Gamma_j\otimes\Gamma_k\\
{\bf F}_3&=&\sum_{ijk}\,f_{ijk}\,\Gamma_i\otimes\Gamma_j\otimes\Gamma_k\,,
\endary
so that
\begary{rcl}
{\bf H}_4&=&{1\over 4!}\,\sum_{jk}\,f_{jk}\,(\bar\psi^T\otimes\bar\psi^T)\,(\Gamma_j\otimes\Gamma_k)\,(\psi\otimes\psi)\\
&=&{1\over 4!}\,\sum_{jk}\,f_{jk}\,(\bar\psi^T\Gamma_j\psi)\,(\bar\psi^T\Gamma_k\psi)\\
{\bf H}_6&=&{1\over 6!}\,\sum_{ijk}\,f_{ijk}\,(\bar\psi^T\otimes\bar\psi^T\otimes\bar\psi^T)\,\\
&\times&(\Gamma_i\otimes\Gamma_j\otimes\Gamma_k)\,(\psi\otimes\psi\otimes\psi)\\
{\bf H}_6&=&{1\over 6!}\,\sum_{ijk}\,f_{ijk}\,(\bar\psi^T\Gamma_i\psi)\,(\bar\psi^T\Gamma_j\psi)\,(\bar\psi^T\Gamma_k\psi)\\
\endary
such that all terms (of the last line) vanish unless ($\y_0^T\,\Gamma_i$) and $\y_0^T\,\Gamma_j$ and $\y_0^T\,\Gamma_k$ are symmetric
matrices, respectively.

The number of symmetric matrices of size $2\,n\times 2\,n$ is $2\,n\,(2\,n+1)/2$, so that
the number of non-vanishing terms in the Hamiltonian ${\bf H}_{2k}$ has an upper limit of
\begeq
[n\,(2\,n+1)]^k
\endeq

\end{appendix}

%\section*{References}
\bibliography{uqm_paper.bib}{}
\bibliographystyle{unsrt}

\end{document}